\documentclass[final,3p,12pt,times]{elsarticle}
\usepackage{graphicx}
\usepackage{epstopdf}
\usepackage{color}
\usepackage{amsmath}
\usepackage{amsfonts}
\usepackage{hyperref}
\usepackage{amssymb}
\usepackage{subfigure}
\usepackage{multirow}
\usepackage{array}
\usepackage{bm}
\usepackage{breqn}

\journal{}

\begin{document}

\begin{frontmatter}

\title{A NURBS-based Inverse Analysis for Reconstruction of Nonlinear Deformations of Thin Shell Structures}

\author[WM]{N. Vu-Bac}
\author[AA]{T.X. Duong}
\author[WM]{T. Lahmer}
\author[HN]{X. Zhuang}
\author[AA]{R.A. Sauer}
\author[BU]{H.S. Park}
\author[DT,WM]{T. Rabczuk\corref{spb}}

\cortext[spb]{Corresponding Author. Tel.: +49 (0)3643 58 4511. Email: timon.rabczuk@uni-weimar.de}
\address[DT]{Institute of Research and Development, Duy Tan University, Da Nang, Vietnam}
\address[WM]{Institute of Structural Mechanics, Bauhaus-Universit\"{a}t Weimar, Marienstr. 15, D-99423 Weimar, Germany}
\address[AA]{Aachen Institute for Advanced Study in Computational Engineering Science (AICES), RWTH Aachen University, Templergraben 55, 52056 Aachen, Germany}
\address[HN]{Institut f\"{u}r Kontinuumsmechanik, Gottfried Wilhelm Leibniz Universit\"{a}t Hannover, Appelstra{\ss}e 11, 30167 Hannover, Germany}
\address[BU]{Department of Mechanical Engineering, Boston University, Boston, MA 02215, USA}

\address[]{}
\address[]{}

\address[]{\normalfont Published \footnote{This pdf is the personal version of an article whose final publication is available at \href{url}{https://www.sciencedirect.com/science/article/pii/S0045782517306849}} in Computer Methods in Applied Mechanics and Engineering, \href{url}{DOI:10.1016/j.cma.2017.09.034}}

\address[]{\normalfont Submitted on 20 May 2017, Revised on 23 September 2017, Accepted on 27 September 2017}

\begin{abstract}

This article presents original work combining a NURBS-based inverse analysis with both kinematic and constitutive nonlinearities to recover the applied loads and deformations of thin shell structures. The inverse formulation is tackled by gradient-based optimization algorithms based on computed and measured displacements at a number of discrete locations. The proposed method allows accurately recovering the target shape of shell structures such that instabilities due to snapping and buckling are captured. The results obtained show good performance and applicability of the proposed algorithms to computer-aided manufacturing of shell structures.


\end{abstract}


\end{frontmatter}

\section{Introduction}

There exist many examples in the natural world of shape changes of slender structures driven by external sources. For example, biological structures can change shape due to stimuli such as nonuniform heating, local swelling and growth of thin sheets. These geometric modifications formed by such morphological changes can have the benefit of enhancing biological functions\cite{Boudaoud:2010,Mirabet:2011}. For instance, osmotic swelling leads to snap closing of the leaves of the Venus flytrap, with the resultant configuration being crucial for its nutrition \cite{Forterre:2005}. Since thin structures tend to bend to reduce their stretching energy when subjected to dramatic growth-induced deformations as occurs in growing leaves \cite{Marder:2003}, wrinkling skin \cite{Kuecken:2005} and the writhing of tendril-bearing climbers \cite{McMillen:2002}, they often morph into nontrivial three dimensional (3D) shapes \cite{Klein:2007}. Furthermore, large deformations occurring in a growing body result in an interesting shape-altering mechanism, i.e. if specific areas within a thin material are deformed, the entire structure will morph into a new shape. Hence, understanding and controlling the shape changes that result in thin objects subject to external stimuli such as external loads, swelling or temperature-induced thermal strain is crucial for biomimetic engineering. Various studies have been performed to understand the shape change mechanism from swelling or growing of thin, soft bodies \cite{Pezzulla:2015,Nardinochi:2015}.

In order to obtain a target shape, the growth-like morphing of a 2D flat sheet into curved, 3D shapes under external stimuli needs to be understood. In particular, the stimuli that is required to produce a target shape needs to be determined and then the resultant configuration of the sheet subject to these stimuli can be reconstructed. Inverse analysis is a widely used approach to determining the external stimuli and the reconstruction of the displacement vector at every material point of the structure \cite{Constantinescu:1995,Marin:2002,Geymonat:2003,Bonnet:2005}. For instance, an inverse method is presented in \cite{Lucantonio:2014} to determine the active strain needed to deform a bio-hybrid system to a desired curvature that is obtained by appropriate experiments. Inverse techniques are also widely used in aerospace applications. For instance, control-surface morphing usually used in commercial aircraft, employs piezo-electric actuators integrated with shape memory alloys \cite{Bogert:2003}. The method enable identifying a detailed state of structural deformations from a set of experimentally measured displacements. Consequently, other fundamental response quantities such as stress, material and structural failure can be estimated.


Various inverse problems and their applications have been discussed in literature, see \cite{Maniatty:1989,Schnur:1990,Maniatty:1994,Liu:1996}. However, very few studies address the reconstruction of three-dimensional deformations of bending structures \cite{Bletzinger:2005,Wuechner:2005,Bletzinger:2010}. Even fewer studies on inverse analysis deal with high-fidelity plate or shell structures as indicated by Tessler et al. \cite{Tessler:2003}. As studied by Jones et al. \cite{Jones:1998}, an inverse approach employing least-squares approach was used to reconstruct deformations of a cantilever plate from a set of strain measurements. Later, Shkarayev \cite{Shkarayev:2001} developed an inverse approach employing finite element-based parametrization that uses measured surface strains to determine the applied loads, stresses, which then leads to reconstructing structural deformations in an aerospace vehicle. It should be noted that the existing inverse methods are limited to solving small strain elasticity problems, and have not been utilized to account for nonlinear kinematics and large deformations.


For the reconstruction of structural deformations, numerical methods require the solution of inverse problems. Finite element (FE) analysis is commonly used to discretize the spatial domain in combination with the methods of mathematical optimization. Inverse problems may be ill-posed, particularly when considering uniqueness of the solution. In order to obtain a solution, iterative methods of nonlinear optimization are applied. The optimization problem can be solved by using gradient-free or gradient-based methods \cite{Nanthakumar:2015,Nanthakumar:2016a,Nanthakumar:2016b}. In gradient-free methods, only the relationship between an objective and any associated constraint functions and the structural problem are required whereas the gradient-based optimizers request derivatives of the objective and constraint functions. Those can be computed via analytical or semi-analytical sensitivities. Since gradient-free optimization strategies, like genetic algorithms \cite{Waisman:2010} or multilevel coordinate search \cite{Nanthakumar:2013} are often computationally expensive in finding the optimal solution, particularly when the number of parameters to be identified increases \cite{Nanthakumar:2014a}, gradient-based optimization algorithms are employed in this study.


Kirchhoff-Love shell theory is suitable for thin shells. However, since $C^1$-continuity within the computational domain are required, it is rarely used in FE analysis. Meshfree \cite{Rabczuk:2006,Rabczuk:2007,Rabczuk:2010} and NURBS-based formulations \cite{Kiendl:2009,Benson:2010,Benson:2011,Benson:2013,Nhon:2011} on the other hand fulfill the $C^1$-continuity requirement of Kirchhoff-Love shell theory and allow to avoid rotational degrees of freedom. Due to the lower computational cost and the better opportunity to model complex geometries, we will pursue the NURBS-based approach. NURBS have also been recently used in shape optimization \cite{Wall:2008,Kiendl:2014} and topology optimization \cite{Seo:2010a,Ghasemi:2017}. In this article, we will present a NURBS-based inverse analysis using analytical and semi-analytical sensitivities for shell structures.



To the best of our knowledge, this is the first time an inverse approach is presented for thin shell structures that accounts for both geometric and material nonlinearities. The problem of interest is the determination of the external applied loads and the reconstruction of structural deformations from given data where shape changes due to instabilities like snap-through, snap-back and buckling are allowed.

The article is divided into three major sections. The next section presents a brief description of the theory of thin shells. The governing equation, weak form, and constitutive equations are discussed. The FE formulation is also addressed in this section. The third section describes the inverse analysis based on the gradient-based optimization using analytical and semi-analytical sensitivity approaches. The proposed method is verified and illustrated by various numerical examples in section \ref{sec:numerical_examples}. In particular, elastic instabilities like snap-through, snap-back and buckling will be taken into account.


\section{A brief description of rotational-free thin shell theory} \label{sec:thin_shell_theory}

\subsection{Thin shell kinematics} \label{subsec:kinematics}

\begin{figure}[htbp] \centering
	\includegraphics[width = 0.6\textwidth]{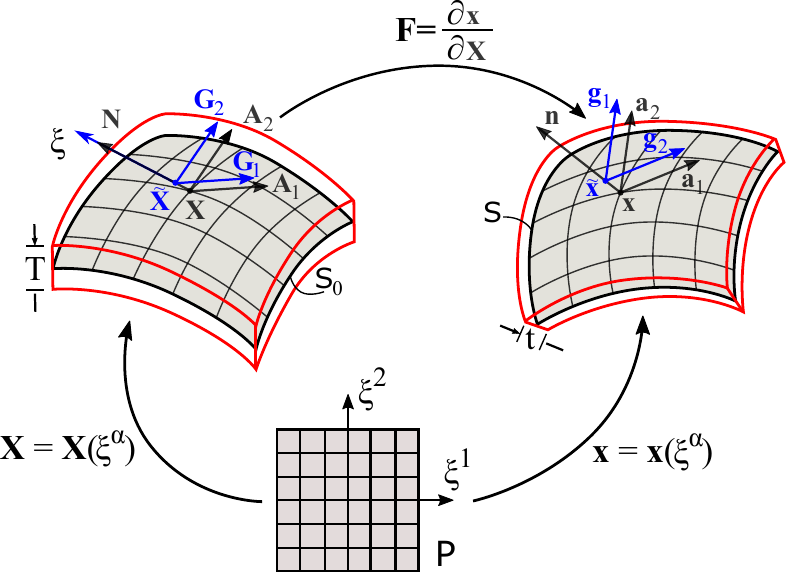}
	\caption{Mapping between parameter domain $\mathcal{P}$, reference surface $\mathcal{S}_0$ and current surface $\mathcal{S}$. $\tilde{\bm{X}}$ and $\tilde{\bm{x}}$ are the respective reference and current position of a material point of the 3D continuum. $G_{\alpha}, ~G_3$ are covariant tangent and normal vectors of a shell layer in reference configuration and $g_{\alpha}, ~g_3$ are the corresponding ones in current configuration. The figure is adopted from \cite{Duong:2017}}
	\label{fig:kinematics}
\end{figure}

Consider a general surface $\mathcal{S}$ of the shell in Figure \ref{fig:kinematics} on which a point $\bm{x}$ can be described by the parametric mapping

\begin{equation} \label{eq:covariant_tangent}
	\bm{x} = \bm{x} (\xi^{\alpha}), \quad \alpha = 1,2,
\end{equation}

\noindent where $\xi^{\alpha}$ denotes coordinates in the parameter domain $\mathcal{P}$. The deformation of $\mathcal{S}$ is described by introducing the reference configuration $\mathcal{S}_0$. It is characterized by the parametric description $\bm{X} = \bm{X} (\xi^{\alpha})$. From these mappings, the covariant tangent vectors $\bm{A}_{\alpha} = \frac{ \partial \bm{X} }{ \partial \xi^{\alpha} }$ and $\bm{a}_{\alpha} = \frac{ \partial \bm{x} }{ \partial \xi^{\alpha} }$ can be determined, respectively. The respective surface normal vectors of the reference and current configurations are then defined by $\bm{N} = \frac{ \bm{A}_1 \times \bm{A}_2 }{ \parallel \bm{A}_1 \times \bm{A}_2 \parallel }$ and $\bm{n} = \frac{ \bm{a}_1 \times \bm{a}_2 }{ \parallel \bm{a}_1 \times \bm{a}_2 \parallel }$.

Given the covariant tangent vectors, the components of the covariant metric tensor of the midsurface

\begin{equation} \label{eq:covariant_metric_tensor}
\begin{aligned}
	A_{\alpha  \beta} &= \bm{A}_{\alpha} \cdot \bm{A}_{\beta}, \quad \alpha, \beta = 1,2 \\
	a_{\alpha  \beta} &= \bm{a}_{\alpha} \cdot \bm{a}_{\beta},
\end{aligned}
\end{equation}

\noindent and the components of the contravariant metric tensors $[A^{\alpha \beta}] = [A_{\alpha \beta}]^{-1}$ and $[a^{\alpha \beta}] = [a_{\alpha \beta}]^{-1}$ can be evaluated. From the contravariant metric tensors, the contravariant tangent vectors are defined as $\bm{A}^{\alpha} = A^{\alpha \beta} \bm{A}_{\beta}$ and $\bm{a}^{\alpha} = a^{\alpha \beta} \bm{a}_{\beta}$. The covariant components of the curvature tensor are defined as

\begin{equation} \label{eq:curvature_tensor}
\begin{aligned}
	B_{\alpha  \beta} &= \bm{N} \cdot \bm{A}_{\alpha, \beta}, \\
	b_{\alpha  \beta} &= \bm{n} \cdot \bm{a}_{\alpha, \beta}.
\end{aligned}
\end{equation}

The mean, Gaussian and principal curvatures of surface $\mathcal{S}$ are then given by

\begin{equation} \label{eq:curvature_properties}
	H = \frac{1}{2} a^{\alpha \beta} b_{\alpha \beta}, \quad \kappa = \frac{det[b_{\alpha \beta}]}{det[a_{\alpha \beta}]}, \quad \kappa_{1/2} = H \pm \sqrt{H^2 - \kappa}
\end{equation}

The surface deformation gradient tensor $\bm{F}$ describing the mapping from $\mathcal{S}_0$ to $\mathcal{S}$ is defined as $\bm{F} = \bm{a}_{\alpha} \otimes \bm{A}^{\alpha}$. Given $\bm{F}$, the right and left Cauchy-Green surface tensors are introduced by:

\begin{equation} \label{eq:right_Cauchy_Green}
	\bm{C} = \bm{F}^T \bm{F} = a_{\alpha \beta} \bm{A}^{\alpha} \otimes \bm{A}^{\beta}, \quad \bm{B} = \bm{F} \bm{F}^T = A^{\alpha \beta} \bm{a}_{\alpha} \otimes \bm{a}_{\beta}
\end{equation}

The surface Green-Lagrange strain tensor and the curvature tensor can then be expressed as follows \cite{Steigmann:1999b}

\begin{equation} \label{eq:Green_Lagrange_strain_and_curvature}
\begin{aligned}
	\bm{E} &= E_{\alpha \beta} \bm{A}^{\alpha} \otimes \bm{A}^{\beta} = \frac{1}{2} \left( a_{\alpha \beta} - A_{\alpha \beta} \right) \bm{A}^{\alpha} \otimes \bm{A}^{\beta}, \\
	\bm{K} &= K_{\alpha \beta} \bm{A}^{\alpha} \otimes \bm{A}^{\beta} = \left( b_{\alpha \beta} - B_{\alpha \beta} \right) \bm{A}^{\alpha} \otimes \bm{A}^{\beta}.
\end{aligned}
\end{equation}

\subsection{Balance laws}

Given a surface $\mathcal{S}$ subjected to a prescribed body force $\bm{f}$, the quasi-static equilibrium of the thin shell is described by

\begin{equation} \label{eq:governing_equation}
	\bm{T}^{\alpha}_{; \alpha} + \bm{f} = \bm{0}.
\end{equation}

\noindent Here, $\bm{T}^{\alpha}_{; \beta}$ denotes the covariant derivative of $\bm{T}^{\alpha}$ defined by $\bm{T}^{\alpha}_{; \beta} = \bm{T}^{\alpha}_{, \beta} + \Gamma^{\alpha}_{\beta \gamma} \bm{T}^{\gamma}$, where $\Gamma^{\alpha}_{\beta \gamma}$ are the Christoffel symbols defined by $\Gamma^{\gamma}_{\alpha \beta} = \bm{a}^{\gamma} \cdot \bm{a}_{\alpha, \beta}$. $\bm{T}^{\alpha}$ denotes the traction defined on the surface normal to $\bm{a}^{\alpha}$, given by


\begin{equation} \label{eq:traction_vector}
	\bm{T}^{\alpha} = N^{\alpha \beta} \bm{a}_{\beta} + S^{\alpha} \bm{n}.
\end{equation}

\noindent Here the respective in plane and shear stress components of the distributed sectional force are defined by

\begin{equation} \label{eq:inplane_and_shear_stresses}
\begin{aligned}
	N^{\alpha \beta} &= \sigma^{\alpha \beta} + b^{\alpha}_{\gamma} M^{\gamma \beta}, \\
	S^{\alpha} &= -M^{\beta \alpha}_{; \beta},
\end{aligned}
\end{equation}

\noindent where $\sigma^{\alpha \beta}$ and $M^{\alpha \beta}$ are the respective membrane stress and bending moment components \cite{Sauer:2017a}. The boundary conditions at the boundary $\partial \mathcal{S}$ are expressed as

\begin{equation} \label{eq:boundary_conditions}
\begin{aligned}
	\bm{u} &= \bar{\bm{u}}, \quad on ~ \partial_u \mathcal{S} \\
	\bm{n} &= \bar{\bm{n}}, \quad on ~ \partial_n \mathcal{S} \\
	\bm{t} &= \bar{\bm{t}}, \quad on ~ \partial_t \mathcal{S} \\
	m_{\tau} &= \bar{m_{\tau}}, \quad on ~ \partial_m \mathcal{S}
\end{aligned}
\end{equation}

\noindent with $\bar{\bm{u}}$, $\bar{\bm{n}}$, $\bar{\bm{t}} = \bar{t}^{\alpha} \bm{a}_{\alpha}$, and $\bar{m_{\tau}}$ being the respective prescribed displacement, rotation, boundary traction and bending moment. $m_{\tau}$ and $m_{\nu}$ are the bending moment components parallel and perpendicular to boundary $\partial \mathcal{S}$, respectively.

\subsection{Weak form}

It can be shown that the weak form of Equation (\ref{eq:governing_equation}) is given by the statement \cite{Sauer:2017b}: Find $\bm{x} \in \mathcal{V}$ $\forall \delta \bm{x} \in \mathcal{V}_0$ such that

\begin{equation} \label{eq:weak_form}
	\delta \Pi = \delta \Pi_{\mathrm{int}} - \delta \Pi_{\mathrm{ext}} =	0 \quad	\forall \delta \bm{x} \in \mathcal{V},
\end{equation}


\noindent with the internal and external virtual work given by

\begin{align} 
	\delta \Pi_{\mathrm{int}} &= \int_{\mathcal{S}} \frac{1}{2} \delta a_{\alpha \beta} \sigma^{\alpha \beta} ~da + \int_{\mathcal{S}} \delta b_{\alpha \beta} M^{\alpha \beta} ~da, \label{eq:weak_form_term1}\\
	\delta \Pi_{\mathrm{ext}} &= \int_{\mathcal{S}} \delta \bm{x} \cdot \bm{f} ~da + \int_{\partial_t \mathcal{S}} \delta \bm{x} \cdot \bm{t} ~ds + \int_{\partial_m \mathcal{S}} \delta \bm{n} \cdot m_{\tau} \bm{\nu} ~ds + [ \delta \bm{x} \cdot m_{\nu} \bm{n} ], \label{eq:weak_form_term2}
\end{align}




\noindent where the last term in Equation (\ref{eq:weak_form_term2}) refers to the virtual work of the point loads $m_{\nu} \bm{n}$ applied at corners on the boundary $\partial_m \mathcal{S}$ (in case $m_{\nu} \neq 0$). It can be shown that after linearization the external and the internal virtual work are characterized by the increments

\begin{equation} \label{eq:Gext_linearization}
	\Delta \delta \Pi_{\mathrm{ext}} = \int_{\partial \mathcal{S}} m_{\tau} \delta \bm{a}_{\alpha} \cdot \left( \nu^{\beta} \bm{n} \otimes \bm{a}^{\alpha} + \nu^{\alpha} \bm{a}^{\beta} \otimes \bm{n} \right) \Delta \bm{a}_{\beta} ~ds
\end{equation}

\begin{dmath} \label{eq:Gint_linearization}
	\Delta \delta \Pi_{\mathrm{int}} = \int_{\mathcal{S}_0} \left( c^{\alpha \beta \gamma \delta} \frac{1}{2} \delta a_{\alpha \beta} \frac{1}{2} \Delta a_{\gamma \delta} + d^{\alpha \beta \gamma \delta} \frac{1}{2} \delta a_{\alpha \beta} \Delta b_{\gamma \delta} + \tau^{\alpha \beta} \frac{1}{2} \Delta \delta a_{\alpha \beta} + e^{\alpha \beta \gamma \delta} \frac{1}{2} \delta b_{\alpha \beta} \Delta a_{\gamma \delta} + f^{\alpha \beta \gamma \delta} \delta b_{\alpha \beta} \Delta b_{\gamma \delta} + M^{\alpha \beta}_0 \Delta \delta b_{\alpha \beta} \right) ~dA
\end{dmath}

\noindent in which $c^{\alpha \beta \gamma \delta}$, $d^{\alpha \beta \gamma \delta}$, $e^{\alpha \beta \gamma \delta}$, and $f^{\alpha \beta \gamma \delta}$ are defined by

\begin{equation} \label{eq:material_tangent_matrices}
\begin{aligned}
	& c^{\alpha \beta \gamma \delta} := 2 \frac{ \partial \tau^{\alpha \beta} }{ \partial a_{\gamma \delta} }, \quad d^{\alpha \beta \gamma \delta} := \frac{ \partial \tau^{\alpha \beta} }{ \partial b_{\gamma \delta} }, \\
	& e^{\alpha \beta \gamma \delta} := 2 \frac{ \partial M^{\alpha \beta}_0 }{ \partial a_{\gamma \delta} }, \quad f^{\alpha \beta \gamma \delta} := \frac{ \partial M^{\alpha \beta}_0 }{ \partial b_{\gamma \delta} },
\end{aligned}
\end{equation}

\noindent and where $\tau^{\alpha \beta} = J \sigma^{\alpha \beta}$, $M^{\alpha \beta}_0 = J M^{\alpha \beta}$, and $da = J ~dA$. The Newton-Raphson and the arc length methods \cite{Crisfield:1981} are employed to solve the linearized system of equations.

\subsection{Constitutive equations} \label{subsec:constitutive_equations}

A Neo-Hookean material model is used to describe nonlinear stress-strain relationship of the materials under large deformations. This model is non-dissipative. For hyperelastic material models, constitutive equations for stretching and bending are derived from a strain energy function $W$, i.e.

\begin{equation} \label{eq:Cauchy_stress_and_moment}
	\tau^{\alpha \beta} = 2 \frac{\partial W}{\partial a_{\alpha \beta}}, \quad M^{\alpha \beta}_0 = \frac{\partial W}{\partial b_{\alpha \beta}}.
\end{equation}


%

\subsection{Koiter shell material model} \label{subsec:Koiter_model}

According to the Koiter material model \cite{Ciarlet:2005,Steigmann:2013}, the surface strain energy for initially curved shells is defined as

\begin{equation} \label{eq:strain_energy_Koiter}
	W (\bm{E}, \bm{K}) = \frac{1}{2} \bm{E} : \mathbb{C} : \bm{E} + \frac{1}{2} \bm{K} : \mathbb{F} : \bm{K},
\end{equation}


\noindent in which $\mathbb{C} = \Lambda \bm{I} \odot \bm{I} + 2 \mu \left( \bm{I} \otimes \bm{I} \right)$ and $\mathbb{F} = \frac{T^2}{12} \mathbb{C}$. Consequently, the Kirchhoff stress and moment in Equation (\ref{eq:Cauchy_stress_and_moment}) are

\begin{equation} \label{eq:KHstress_and_moment_Koiter}
\begin{aligned}
	\tau^{\alpha \beta} &= \Lambda tr (\bm{E}) A^{\alpha \beta} + 2 \mu E^{\alpha \beta}, \\
	M^{\alpha \beta}_0 &= \frac{T^2}{12} \left( \Lambda tr (\bm{K}) A^{\alpha \beta} + 2 \mu K^{\alpha \beta} \right),
\end{aligned}
\end{equation}

\noindent where $\Lambda$ and $\mu$ can be determined in different ways. As presented in \cite{Duong:2017}, we can integrate analytically over the thickness of the 3D Saint Vernant-Kirchhoff material model to obtain

\begin{equation} \label{eq:Lame_constants}
	\Lambda := T \frac{ 2 \tilde{\Lambda} \tilde{\mu} }{ \tilde{\Lambda} + 2 \tilde{\mu} }, \quad \mu := T \tilde{\mu},
\end{equation}

\noindent with $\tilde{\Lambda}$ and $\tilde{\mu}$ being the classical 3D Lam\'{e} constants in linear elasticity.

\subsection{Shell constitutive model derived from 3D material model} \label{subsec:proj_model}

In this section, a Kirchhoff-Love shell constitutive model is derived from a three-dimensional continuum mechanics. The 3D model is projected onto the surface $\mathcal{S}$ to obtain the 2D surface shell model as shown in Figure \ref{fig:kinematics}. The actual metric at an arbitrary material point $P$ in the shell has to be mapped onto the midsurface in the respective reference and deformed configuration \cite{Wriggers:2008} as

\begin{equation} \label{eq:coords_projection}
\begin{aligned}
	\tilde{\bm{X}} (\xi^{\alpha}, \xi) &= \bm{x} (\xi^{\alpha}) + \xi \bm{N} (\xi^{\alpha}) \\
	\tilde{\bm{x}} (\xi^{\alpha}, \xi) &= \bm{X} (\xi^{\alpha}) + \xi \bm{d} (\xi^{\alpha}),
\end{aligned}
\end{equation}

\noindent where $\xi \in [-T/2 ~ T/2]$ is the coordinate along the shell thickness and $\bm{d} := \lambda_3 \bm{n}$ with $\lambda_3$ being the stretch in the normal direction for the Kirchhoff-Love shell model. The components of the projected Kirchhoff stress and moment in Equation (\ref{eq:Cauchy_stress_and_moment}) can be expressed as


\begin{equation} \label{eq:KHstress_and_moment_projection}
\begin{aligned}
	\tau^{\alpha \beta} &= \int^{\frac{T}{2}}_{- \frac{T}{2}} \tilde{\tau}^{\alpha \beta} d \xi, \\	
	M^{\alpha \beta}_0 &= - \int^{\frac{T}{2}}_{- \frac{T}{2}} \xi \tilde{\tau}^{\alpha \beta} d \xi,
\end{aligned}
\end{equation}

\noindent where $\tilde{\tau}^{\alpha \beta}$ denote the 3D Kirchhoff stress components. The detailed projection procedure can be found in \cite{Duong:2017}.


%
%
%
%
%
%

\subsection{Compressible Neo-Hookean materials} \label{subsec:comp_NeoHookean_model}

The 3D compressible Neo-Hookean strain energy per reference configuration is provided as follows:

\begin{equation} \label{eq:strain_energy_compressible_NH}
	\tilde{W} (\tilde{I}_1, \tilde{J}) = \frac{\tilde{\Lambda}}{4} \left( \tilde{J}^2 -1 - 2ln \tilde{J} \right) + \frac{\tilde{\mu}}{2} \left( \tilde{I}_1 - 3 - 2ln \tilde{J} \right),
\end{equation}

\noindent with $\tilde{I}_1$ and $\tilde{J}$ being invariants of the 3D Cauchy-Green tensor $\tilde{\bm{C}}$. The 3D Kirchhoff stress is then given by \cite{Duong:2017} as




\begin{equation} \label{eq:KHstress_compressible_NH}
	\tilde{\tau}^{\alpha \beta}	= \tilde{\mu} G^{\alpha \beta} - \tilde{\mu} \frac{ \tilde{\Lambda} + 2\tilde{\mu} }{\tilde{\Lambda} {J^*}^2 + 2\tilde{\mu}} g^{\alpha \beta}.
\end{equation}

\noindent where the respective covariant and contravariant metric tensors are defined as $g_{\alpha \beta} := \bm{g}_{\alpha} \cdot \bm{g}_{\beta}$ and $g^{\alpha \beta} := \bm{g}^{\alpha} \cdot \bm{g}^{\beta}$, where $\bm{g}_{\bullet}$ and $\bm{g}^{\bullet}$ denote the covariant and contravariant tangent vectors at P, respectively. Further $J^* = \sqrt{\frac{g}{G}}$ with $G := \mathrm{det} [G_{\alpha \beta}]$ and $g := \mathrm{det} [g_{\alpha \beta}]$.

By substituting the stress in Equation (\ref{eq:KHstress_compressible_NH}), the stress and moment shown in Equation (\ref{eq:KHstress_and_moment_projection}) can be evaluated. In this article, we will consider two constitutive material models: (1) the Koiter material model based on Equation (\ref{eq:KHstress_and_moment_Koiter}) and (2) the projected material model based on Equations (\ref{eq:KHstress_and_moment_projection}) and (\ref{eq:KHstress_compressible_NH}).

\subsection{FE approximation} \label{subsec:FE_approx}

Because the Kirchoff-Love shell formulations involve second derivatives, $C^1$-continuity of the surface is required. Hence, NURBS-based shape functions proposed by \cite{Borden:2011} are used to discretize the surface $\mathcal{S}$ and to solve Equation (\ref{eq:weak_form}). The NURBS basis functions are defined by

\begin{equation} \label{eq:NURBS_basis func}
	\mathrm{N}_A (\xi, \eta) = \frac{ \omega_A \hat{\mathrm{N}}^e_A (\xi, \eta) }{ \sum^n_{A=1} \omega_A \hat{\mathrm{N}}^e_A (\xi, \eta) },
\end{equation}

\noindent in which $\{ \hat{\mathrm{N}}^e_A \}^n_{A=1}$ is the B-spline basis function written in terms of Bernstein polynomials as

\begin{equation} \label{eq:Bernstein_polynomials}
	\hat{\mathbf{N}}^e (\xi, \eta) = \mathbf{C}^e_{\xi} \mathbf{B} (\xi) \otimes \mathbf{C}^e_{\eta} \mathbf{B} (\eta),
\end{equation}

\noindent with $\hat{\mathrm{N}}^e_A$ being entries of matrix $\hat{\mathbf{N}}^e$. The geometry of the respective reference and deformed surfaces are approximated by interpolation of the control point coordinates $\bm{X}_e$ and $\bm{x}_e$ using the FE approach as follows:

\begin{equation} \label{eq:interpolation}
	\bm{X} = \mathbf{N} \bm{X}_e, \quad \bm{x} = \mathbf{N} \bm{x}_e,
\end{equation}

\noindent where $\mathbf{N} (\xi, \eta) := [N_1 \bm{1}, ~N_2 \bm{1}, ~..., ~N_n \bm{1}]$ contains the nodal shape functions of element $\Omega^e$ shown in Equation (\ref{eq:NURBS_basis func}). The covariant tangent vectors of the surface are then determined by
 
\begin{equation} \label{eq:discretized_covariant_vectors}
	\bm{A}_{\alpha} = \frac{\partial \bm{X}}{\partial \xi^{\alpha}} \approx \mathbf{N}_{,\alpha} \bm{X}_e, \quad \bm{a}_{\alpha} = \frac{\partial \bm{x}}{\partial \xi^{\alpha}} \approx \mathbf{N}_{,\alpha} \bm{x}_e.
\end{equation}

\noindent Variation of $\bm{x}$ and $\bm{a}_{\alpha}$ are expressed as

\begin{equation} \label{eq:discretized_variation_quantities}
	\delta \bm{x} \approx \mathbf{N} \delta \bm{X}_e, \quad \delta \bm{a}_{\alpha} \approx \mathbf{N}_{, \alpha} \delta \bm{x}_e.
\end{equation}

\noindent Consequently, all kinematical quantities presented in section \ref{subsec:kinematics} and their variation can be determined, see \cite{Sauer:2017b,Duong:2017} for details.

\subsection{Discretized weak form} \label{subsec:discretized_weak_form}

The weak form in Equation (\ref{eq:weak_form}) is discretized using the above interpolations. This leads to the approximation

\begin{equation} \label{eq:discretized_weak_form}
	\delta \Pi \approx \sum^{n_{el}}_{e=1} \delta \Pi^e = \sum^{n_{el}}_{e=1} \left( \delta \Pi^e_{\mathrm{int}} - \delta \Pi^e_{\mathrm{ext}} \right).
\end{equation}

\noindent with $n_{el}$ being the number of elements. The internal and external virtual work contributions of element $e$ are given by

\begin{equation} \label{eq:Gint_and_Gext}
\begin{aligned}
	\delta \Pi^e_{\mathrm{int}} &= \delta \bm{x}^T_e \left( \bm{f}^e_{\mathrm{int} \tau} + \bm{f}^e_{\mathrm{int} M} \right), \\
	\delta \Pi^e_{\mathrm{ext}} &= \delta \bm{x}^T_e \left( \bm{f}^e_{\mathrm{ext} 0} + \bm{f}^e_{\mathrm{ext} p} + \bm{f}^e_{\mathrm{ext} t} + \bm{f}^e_{\mathrm{ext} m} \right),
\end{aligned}
\end{equation}

\noindent in which the internal FE force vectors due to the membrane stress $\tau^{\alpha \beta}$ and the bending moment $M^{\alpha \beta}_0$ are defined by \cite{Duong:2017}

\begin{equation} \label{eq:int_force}
\begin{aligned}
	\bm{f}^e_{\mathrm{int} \tau} &:= \int_{\Omega^e_0} \tau^{\alpha \beta} \mathbf{N}^T_{, \alpha} \bm{a}_{\beta} ~dA, \\
	\bm{f}^e_{\mathrm{int} M} &:= \int_{\Omega^e_0} M^{\alpha \beta}_0 \mathbf{N}^T_{; \alpha \beta} \bm{n} ~dA,
\end{aligned}
\end{equation}

\noindent and the external force FE vectors due to a constant body force $\bm{f}_0$, external pressure perpendicular to the surface $\mathcal{S}$, boundary traction $\bm{t}$ and boundary moment $m_{\tau}$ are defined by \cite{Duong:2017}

\begin{equation} \label{eq:ext_force}
\begin{aligned}
	\bm{f}^e_{\mathrm{ext} 0} &:= \int_{\Omega^e_0} \mathbf{N}^T \bm{f}_0 ~dA \\
	\bm{f}^e_{\mathrm{ext} p} &:= \int_{\Omega^e_0} \mathbf{N}^T p \bm{n} ~dA \\
	\bm{f}^e_{\mathrm{ext} t} &:= \int_{\partial_t \Omega^e} \mathbf{N}^T \bm{t} ~ds \\
	\bm{f}^e_{\mathrm{ext} m} &:= \int_{\partial_m \Omega^e} \mathbf{N}^T_{, \alpha} \nu^{\alpha} m_{\tau} \bm{n} ~ds
\end{aligned}
\end{equation}

\noindent The linearization of $\delta \Pi^e_{\mathrm{int}}$ and $\delta \Pi^e_{\mathrm{ext}}$ is shown in \ref{appendix:FE_tangent_matrices}.

\subsection{Lagrange multiplier method for rotational constraints} \label{subsec:Lagrange_multiplier_constraints}

This section presents an approach used to enforce $G^1$-continuity between patches. The rotational constrains is introduced by adding the constraint potential, see \cite{Duong:2017}

\begin{equation} \label{eq:Lagrange_multiplier_potential}
	\Pi_n = \int_{\mathcal{L}_0} p(\bar{g}_c + \bar{g}_s) ~dS,
\end{equation}


\noindent to the shell formulation, with $p$ being the Lagrange multiplier and

\begin{align} \label{eq:gcbar_and_gsbar}
	\bar{g}_c &:= 1-\cos (\theta - \theta_0), \\
	\bar{g}_s &:= \sin (\theta - \theta_0),
\end{align}

\noindent where $\cos \theta_0 := \bm{N} \cdot \bar{\bm{N}}$ and $\cos \theta := \bm{n} \cdot \bar{\bm{n}}$. Here $\bm{N}$, $\bar{\bm{N}}$ are the two surface normals of neighboring patches in the reference configuration, and $\bm{n}$, $\bar{\bm{n}}$ are the corresponding normals in the deformed configuration. Details of the variation, linearization and FE discretization of $\Pi_n$ can be found in \cite{Duong:2017}. The particular patch boundary denoted as $\mathcal{L}_0$ in Equation (\ref{eq:Lagrange_multiplier_potential}) refers to the reference configuration.

\section{Inverse analysis} \label{sec:inverse}

This section presents an inverse analysis procedure for the isogeometric rotation-free thin shell considering hyperelastic material behavior. Nonlinear kinematics and material response are considered. The boundary conditions, i.e. prescribed boundary displacements, tractions and moments, as well as surface loads like out-of-plane pressure, that are applied to produce a desired shape of thin shell structures, can be identified. The considered technique is based on gradient-based strategies, where the required derivatives are determined either analytically, by the finite difference method or by the adjoint approach. Inverse problems are often solved iteratively in form of the optimization problem \cite{Firl:2012}

\begin{equation} \label{eq:optimization_formulation}
\begin{aligned}
	\text{minimize} & \quad \mathrm{J} (s, u (s)), s \in \mathbf{R}^n \\
	\text{such that} & \quad g_j (s, u(s)) \leq 0, \quad j = \{ 1, ..., n^g \}, \\
	~ & \quad s_l \leq s \leq s_u
\end{aligned}
\end{equation}

\noindent in which $s$ are the design variables; $u$ are the state variables describing the structural response, e.g. the displacements; $\mathrm{J}$ is the objective function; $g_j$ are inequality constraints imposed on these variables to restrict quantities describing the shell response; $s_l$ and $s_u$ are the respective side constraints which restrict the design variables. In this study, the Dirichlet boundary conditions - the parameters $s$ - can be determined based on the measured displacements of the surface. The forward operator which maps those parameters to measurements is shown by \cite{Nanthakumar:2014a} as follows:

\begin{align} \label{eq:inverse_problem}
	& F : X \rightarrow Y, \\
	& s \mapsto u
\end{align}

\noindent where $X$ refers to the parameter space, i.e. the range of Neumann or Dirichlet boundary conditions (external loads, prescribed displacements etc.) and $Y$ refers to the measurement space. Given the measurements with noise $u^{meas}$, the inverse problem corresponds to determining $s$ from the following equation

\begin{equation} \label{eq:inverse_equation}
	F(s) = u^{meas}
\end{equation}

\noindent which is rewritten in terms of a least-squares functional form

\begin{equation} \label{eq:least_squares_form}
	\mathrm{J} (s,u (s)) = \frac{1}{2} \int_{\Omega} | u^{meas} - u (s) |^2 d \Omega
\end{equation}

The inverse analysis will be employed to identify unknown applied loads and reconstruct nonlinear deformations of the shell structure where the shape changes due to instabilities (i.e. snapping or buckling) are allowed. The above objective function in Equation (\ref{eq:least_squares_form}) is expressed in terms of the FE discretization as

\begin{equation} \label{eq:obj_func_discret}
	\mathrm{J} = \frac{1}{2} \Vert \bm{\mathrm{u}}^{meas} - \bm{\mathrm{u}} \Vert^2,
\end{equation}

\noindent where $\bm{\mathrm{u}}$ is the vector of nodal displacements on the surface; $\bm{\mathrm{u}}^{meas}$ is the vector of the displacements measured on the target shape. The design variables are chosen as the applied loads. The state variables are the nodal displacements. The gradient based method of moving asymptotes (MMA) \cite{Svanberg:1987} is employed together with analytically derived sensitivities. Thus, the differentiation of the objective function $\mathrm{J}$ and constraint $g$ with respect to (w.r.t.) design variables $\bm{s}$ is required. Using the chain rule of differentiation, the total derivative of the objective in Equation (\ref{eq:obj_func_discret}) w.r.t. the design variable $s_i$ is expressed as


\begin{equation} \label{eq:derv_obj_func}
	\frac{d \mathrm{J}}{d s_i} = \frac{\partial \mathrm{J}}{\partial s_i} + \frac{\partial \mathrm{J}}{\partial \bm{u}} \frac{\partial \bm{u}}{\partial s_i}
\end{equation}

The partial derivative w.r.t. the design variable $\frac{\partial \mathrm{J}}{\partial \bm{s}}$ is zero according to Equation (\ref{eq:obj_func_discret}). The equilibrium equations for nonlinear FE analysis is given by

\begin{equation} \label{eq:equil_cond}
	\bm{r} (\bm{u}, \bm{s}) = \bm{f}_{\mathrm{int}} (\bm{u}, \bm{s}) - \bm{f}_{\mathrm{ext}} (\bm{u}, \bm{s}) = \bm{0},
\end{equation}

\noindent where $\bm{f}_{\mathrm{int}}$ and $\bm{f}_{\mathrm{ext}}$ denote the internal force and the external applied load, respectively. Differentiating Equation (\ref{eq:equil_cond}) w.r.t. the design variables $\bm{s}$ gives

\begin{equation} \label{eq:der_equil_cond}
	\frac{d \bm{r}}{d \mathrm{s}_i} = \frac{\partial \bm{f}_{\mathrm{int}}}{\partial \mathrm{s}_i} + \frac{\partial \bm{f}_{\mathrm{int}}}{\partial \bm{u}} \frac{\partial \bm{u}}{\partial s_i} - \left( \frac{\partial \bm{f}_{\mathrm{ext}}}{\partial s_i} + \frac{\partial \bm{f}_{\mathrm{ext}}}{\partial \bm{u}} \frac{\partial \bm{u}}{\partial s_i} \right) = \bm{0}.
\end{equation}

\noindent Rewriting this equation gives the nonlinear state derivative w.r.t. the design variables $\frac{\partial \bm{u}}{\partial \bm{s}}$ as

\begin{equation} \label{eq:du_dsi}
	\frac{\partial \bm{u}}{\partial s_i} = - \Biggl( \underbrace{ \frac{\partial \bm{f}_{\mathrm{int}}}{\partial \bm{u}} - \frac{\partial \bm{f}_{\mathrm{ext}}}{\partial \bm{u}} }_{\bm{K}_T} \Biggr)^{-1} \Biggl( \frac{\partial \bm{f}_{\mathrm{int}}}{\partial s_i} - \frac{\partial \bm{f}_{\mathrm{ext}}}{\partial s_i} \Biggr) = - \bm{K}^{-1}_T \underbrace{ \Biggl( \frac{\partial \bm{f}_{\mathrm{int}}}{\partial s_i} - \frac{\partial \bm{f}_{\mathrm{ext}}}{\partial s_i} \Biggr) }_{\bm{f}^*_{\mathrm{nln}}},
\end{equation}

\noindent where $\bm{K}_T$ is the tangent stiffness matrix; $\bm{f}^*_{\mathrm{nln}}$ is known as the nonlinear pseudo load vector \cite{Firl:2012}. In optimization algorithms, the derivatives of the displacement $\frac{\partial \bm{u}}{\partial s_i}$ are not computed explicitly. Instead, they are obtained by using the \emph{adjoint method} that defines an adjoint vector $\bm{\lambda}$ \cite{Haftka:2012} which is the solution of the equation

\begin{equation} \label{eq:adjoint}
	\bm{K}_T \bm{\lambda} = \bm{z},
\end{equation}

\noindent where $\bm{z}$, with $z_i = \frac{\partial J}{\partial u_i}$, is the vector of derivatives of the objective function w.r.t. the displacement. The solution $\bm{\lambda}$ of Equation (\ref{eq:adjoint}) is a FE solution for displacements and is obtained by solving the governing equation (\ref{eq:governing_equation}) with Neumann boundary condition defined by

\begin{equation} \label{eq:dummy_load} 
	z_i = u^{meas}_i - u_i ~ \text{on} ~ \Omega,
\end{equation}

\noindent and the Dirichlet boundary conditions are the same as those of the actual problem shown in Equations (\ref{eq:boundary_conditions}.1). Then, Equation (\ref{eq:derv_obj_func}) is written as

\begin{equation} \label{eq:derv_obj_func_adj}
	\frac{dJ}{ds_i} =  \frac{\partial J}{\partial s_i} - \bm{\lambda}^T \left( \frac{\partial \bm{f}_{\mathrm{int}}}{\partial s_i} - \frac{\partial \bm{f}_{\mathrm{ext}}}{\partial s_i} \right).
\end{equation}

As observed, the derivatives of $\bm{f}_{\mathrm{int}}$ and $\bm{f}_{\mathrm{ext}}$ w.r.t. the design variables $s_i$ in Equation (\ref{eq:derv_obj_func_adj}) are required in the adjoint method. While it is possible to use numerical sensitivities, it is more efficient to use analytical sensitivities. The analytical sensitivities will be presented in the following section \ref{appendix:analsa} whilst the semi-analytical sensitivities will be described in section \ref{appendix:semianalsa}.

\subsection{Analytical sensitivities} \label{appendix:analsa}

In this section, the analytical sensitivities of the quantities in Equation (\ref{eq:derv_obj_func_adj}) w.r.t. the respective prescribed displacement ($\bm{u}$), traction ($\bm{t}$), point load ($\bm{f}_0$), moment ($m_{\tau}$) and lateral pressure ($q$) are presented.

\subsubsection{Prescribed displacement} \label{appendix:sensitivities_displacement}

Assume a vector of displacements is prescribed at $\bm{X}_A$ along $\bm{e}_i$, the design variable is defined as $\bm{s} = \bm{u}_A$ such that $\bm{u}_A = u ~\delta(\bm{X}_A) ~\bm{e}_i$. The nonlinear pseudo load vector $\bm{f}^*_{\mathrm{nln}}$ is given by

\begin{equation} \label{eq:deriv_fint_wrt_disp}
	\bm{f}^*_{\mathrm{nln}} = \frac{\partial \bm{f}_{\mathrm{int}}}{\partial \bm{s}} = \bm{K}_{T,A}.
\end{equation}

\noindent where $\bm{K}_{T,A}$ refers to all columns of the tangent stiffness matrix $\bm{K}_T$ defined in Equation (\ref{eq:du_dsi}) that correspond to the degrees of freedom (dofs) in the design vector $\bm{s}$. Note that here $\delta$ is the Dirac-Delta distribution. Implementation should be done at the global level (i.e. after elemental assembly). By doing so, the sensitivity for Equation (\ref{eq:derv_obj_func_adj}) is obtained.

\subsubsection{Traction} \label{appendix:sensitivities_traction}

For an unknown edge traction along direction $\bm{e}_i$ the design variable $s$ is

\begin{equation} \label{eq:design_var_traction}
	s = \bm{t} \cdot \bm{e}_i.
\end{equation}

\noindent The nonlinear pseudo load vector $\bm{f}^*_{\mathrm{nln}}$ is obtained by

\begin{equation} \label{eq:deriv_fext_wrt_tract}
	\bm{f}^*_{\mathrm{nln}} = - \frac{\partial \bm{f}^e_{\mathrm{ext}}}{\partial s},
\end{equation}

\noindent where $\bm{f}^e_{\mathrm{ext}}$ is given by Equation (\ref{eq:ext_force}.3). Since now $\bm{t} = s \bm{e}_i$, we find

\begin{equation} \label{eq:fstartnln_tract}
	\bm{f}^*_{\mathrm{nln}} = - \int_{\partial_t \Omega^e} \mathbf{N}^T \bm{e}_i ~ds.
\end{equation}

\noindent More general, if $\bm{s} = \bm{t}$, the pseudo load vector is

\begin{equation} \label{eq:fstartnln_tract_general}
	\bm{f}^*_{\mathrm{nln}} = - \int_{\partial_t \Omega^e} \mathbf{N}^T ~ds.
\end{equation}

\noindent And if $s = \bm{t}_0$, where $\bm{t}_0 = \bm{t} \frac{ds}{dS}$, we find

\begin{equation} \label{eq:fstartnln_tract_general1}
	\bm{f}^*_{\mathrm{nln}} = - \int_{\partial_t \Omega^e_0} \mathbf{N}^T ~dS.
\end{equation}

\subsubsection{Point load} \label{appendix:sensitivities_pointload}

Considering a shell structure under a point load $\bm{F}$, at $\bm{X}_A$ along $\bm{e}_i$, i.e. $\bm{F} = F ~\delta (\bm{X}_A) ~\bm{e}_i$, the design variable $s$ is equal to $F$. The nonlinear pseudo load vector $\bm{f}^*_{\mathrm{nln}}$ is obtained by

\begin{equation} \label{eq:deriv_fext_wrt_pointload}
	\bm{f}^*_{\mathrm{nln}} = - \frac{\partial \bm{f}^e_{\mathrm{ext}}}{\partial s},
\end{equation}

\noindent where $\bm{f}^e_{\mathrm{ext}}$ is given by Equation (\ref{eq:ext_force}.1). The only non-zero entry within Equation (\ref{eq:ext_force}.1) is $\bm{f}_{\mathrm{ext} A} = \mathrm{N}_A ~F ~\bm{e}_i$ where $A$ denotes the FE node at $\bm{X}_A$. Hence, Equation (\ref{eq:deriv_fext_wrt_pointload}) becomes

\begin{equation} \label{eq:fstartnln_pointload}
	\bm{f}^*_{\mathrm{nln}} = - \mathrm{N}_A \bm{e}_i.
\end{equation}

\noindent More general, in case $\bm{s} = \bm{f}_0$, $\bm{f}^*_{\mathrm{nln}}$ is given by

\begin{equation} \label{eq:fstartnln_pointload_general}
	\bm{f}^*_{\mathrm{nln}} = - \int_{\Omega^e_0} \mathbf{N}^T ~dA.
\end{equation}

\subsubsection{Uniaxial load and bending moment} \label{appendix:sensitivities_moment}

Let consider a shell structure subjected to uniaxial traction and bending moment as shown in example \ref{subsec:ex3}. The two design variables are $s_1 = \bm{t} \cdot \bm{e}_1$ and $s_2 = m_{\tau}$. For the uniaxial traction, the derivative $\bm{f}^*_{\mathrm{nln1}} = \frac{\partial \bm{f}^e_{\mathrm{ext}}}{\partial s_1}$ is given in Equation (\ref{eq:fstartnln_tract}). Taking the derivative of $\bm{f}^e_{\mathrm{ext}}$ w.r.t. $s_2$, we obtain

\begin{equation} \label{eq:fstartnln_eccentric_load}
		\bm{f}^*_{\mathrm{nln2}} = - \frac{\partial \bm{f}^e_{\mathrm{ext}}}{\partial s_2} = - \frac{\partial \bm{f}^e_{\mathrm{ext}}}{\partial m_{\tau}} = - \int_{\partial_m \Omega^e} \mathbf{N}^T_{, \alpha} \nu^{\alpha} \bm{n} ~ds,
\end{equation}

\noindent where $m_{\tau}$ denotes the distributed moment per current length. Assuming that $m_{\tau_0}$ = distributed moment per reference length is the design variable, then $\bm{f}^e_{\mathrm{ext}} = \int_{\partial_m \Omega^e_0} \mathbf{N}^T_{, \alpha} \nu^{\alpha} m_{\tau_0} \bm{n} ~dS$ and

\begin{equation} \label{eq:fstartnln_eccentric_load_reference}
	\bm{f}^*_{\mathrm{nln2}} = - \int_{\partial_m \Omega^e_0} \mathbf{N}^T_{, \alpha} \nu^{\alpha} \bm{n} ~dS
\end{equation}

\subsubsection{Uniaxial load and lateral pressure} \label{appendix:sensitivies_pressure}

Let us consider a shell structure that is subjected to uniaxial traction and lateral pressure as shown in example \ref{subsec:ex4}. In this case, the two design variables will be $s_1 = \bm{t} \cdot \bm{e}_1$ and $s_2 = q$. The derivatives $\bm{f}^*_{\mathrm{nln1}} = \frac{\partial \bm{f}^e_{\mathrm{ext}}}{\partial s_1}$ is given in Equation (\ref{eq:fstartnln_tract}) and $\bm{f}^*_{\mathrm{nln2}} = \frac{ \partial \bm{f}^e_{\mathrm{ext}} }{ \partial s_2 }$ is given by

\begin{equation} \label{eq:fstartnln_lateral_pressure}
		\bm{f}^*_{\mathrm{nln2}} = - \frac{\partial \bm{f}^e_{\mathrm{ext}}}{\partial p} = - \int_{\Omega^e} \mathbf{N}^T \bm{n} ~da.
\end{equation}

\subsection{Semi-analytical sensitivities} \label{appendix:semianalsa}

The semi-analytical sensitivities of the quantities in Equation (\ref{eq:derv_obj_func_adj}) are approximated by the finite difference method as follows:

\begin{equation} \label{eq:semianal_sens_anal}
\begin{aligned}
	\frac{\partial \bm{f}_{\mathrm{int}}}{\partial s_i} &\approx \frac{ \bm{f}_{\mathrm{int}} (\bm{u},\bm{s} + \Delta \bm{s}_i) - \bm{f}_{\mathrm{int}} (\bm{u},\bm{s}) }{\Delta s_i}, \\
	\frac{\partial \bm{f}_{\mathrm{ext}}}{\partial s_i} &\approx \frac{ \bm{f}_{\mathrm{ext}} (\bm{u},\bm{s} + \Delta \bm{s}_i) - \bm{f}_{\mathrm{ext}} (\bm{u},\bm{s}) }{\Delta s_i}
\end{aligned}
\end{equation}

\noindent where $f_{\bullet} (\bm{u}, \bm{s} + \Delta \bm{s}_i)$ refers to perturbed quantities in direction of variable $s_i$. The small step size $\Delta s_i = 0.001 \times s_i$ is selected to reduce the truncation error to acceptable value \cite{Haftka:2012}. The details of the selection of the step size can be found in \cite{Gill:1983,Iott:1985}.

\section{Numerical examples} \label{sec:numerical_examples}

Since experimental measurements are not available for this study, the proposed approach is validated using a two-step procedure. First, a FE model on the basis of the isogeometric approach \cite{Hughes:2005} is developed for a given shell structure under the applied loads. Both nonlinear kinematics and nonlinear material constitutive laws are accounted for. Surface displacements $\bm{\mathrm{u}}^{iga}$ are then computed at discrete locations. Random noise representing small disturbances is added to the displacement vector $\bm{\mathrm{u}}^{iga}$, giving $\bm{\mathrm{u}}^{meas} := \bm{\mathrm{u}}^{iga} (1 + 0.01 \gamma)$, with $\gamma$ being a random number in the interval $[-1, 1]$. Since this disturbed data \emph{could} originate from an experiment we call it \emph{experiment-like}. Secondly, using this experiment-like data, we apply gradient-based optimization methods on Equation (\ref{eq:obj_func_discret}). The approximation of the applied loads are inversely solved and the structural deformations are reconstructed. Solution of the inverse analysis is achieved when insignificant changes occur in the configuration between iterations. Particularly, the algorithm is completed where the convergence criterion, $\epsilon_{obj. func.} = \vert \frac{J^I - J^{I-1}}{J^0} \vert \leqslant 10^{-3}$, is satisfied with $I$ being the iteration number and $J^0$ is the objective value computed at the initial input.

Four numerical examples on cantilevered and hinged cylindrical shells governed by geometrically nonlinear mechanics are presented. Hyperelastic constitutive laws which account for stretching and bending are considered. The first example is a cantilever shell subjected to prescribed end displacement (also end shear traction), the second is the hinged cylindrical shell subjected to a prescribed center displacement. In the two last examples, imperfections, which are essential for instability shape changes are considered. The inverse analysis of nonlinear instabilities of shell structures under compression using both the Koiter material model and the projected shell material model using the compressible Neo-Hookean formulation are examined. Note that hereafter we use the \emph{Koiter model} to refer to the former and the \emph{projected model} for the latter.


\subsection{Cantilever subjected to prescribed end displacement and end shear traction} \label{subsec:ex1}

\begin{figure}[htbp]
\begin{center}
	\subfigure[]{\includegraphics[width = 0.475\textwidth]{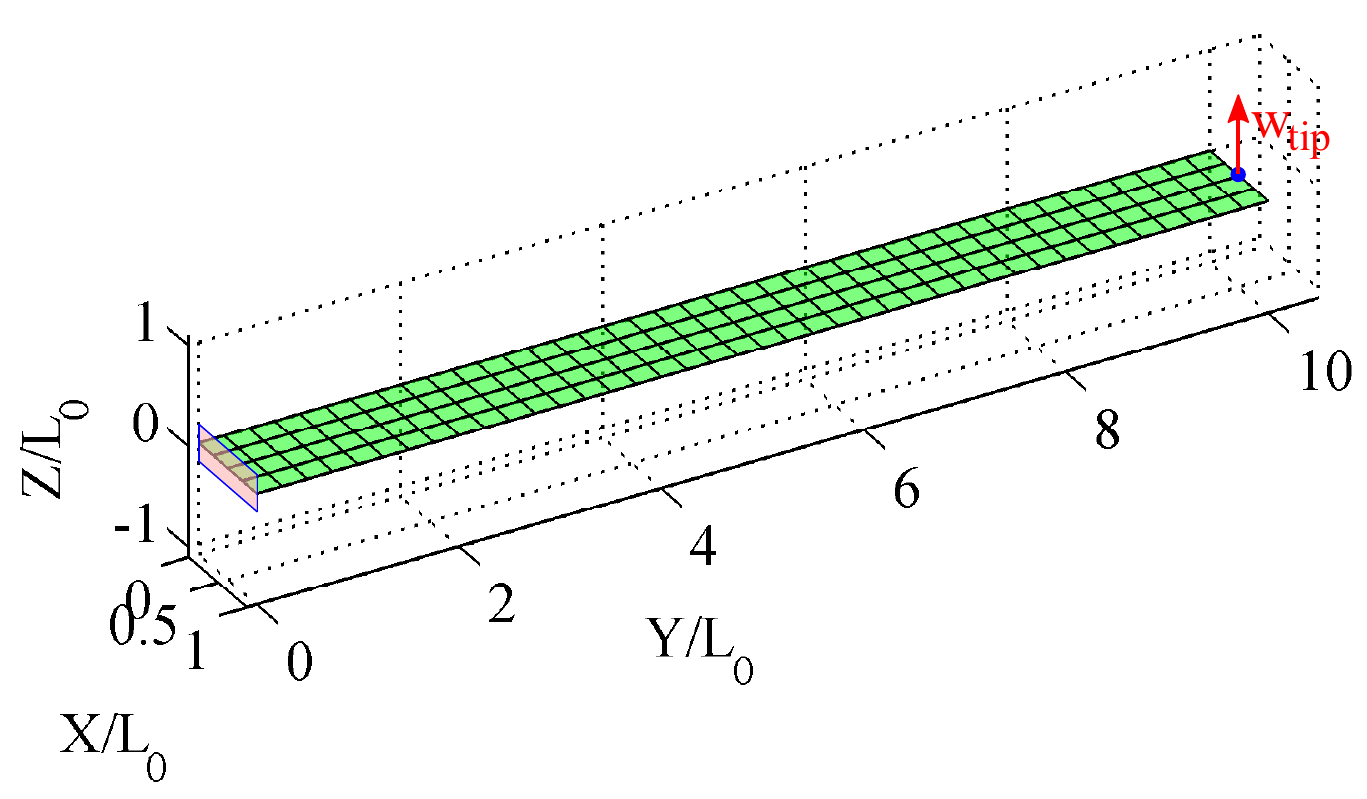}}
	\subfigure[]{\includegraphics[width = 0.475\textwidth]{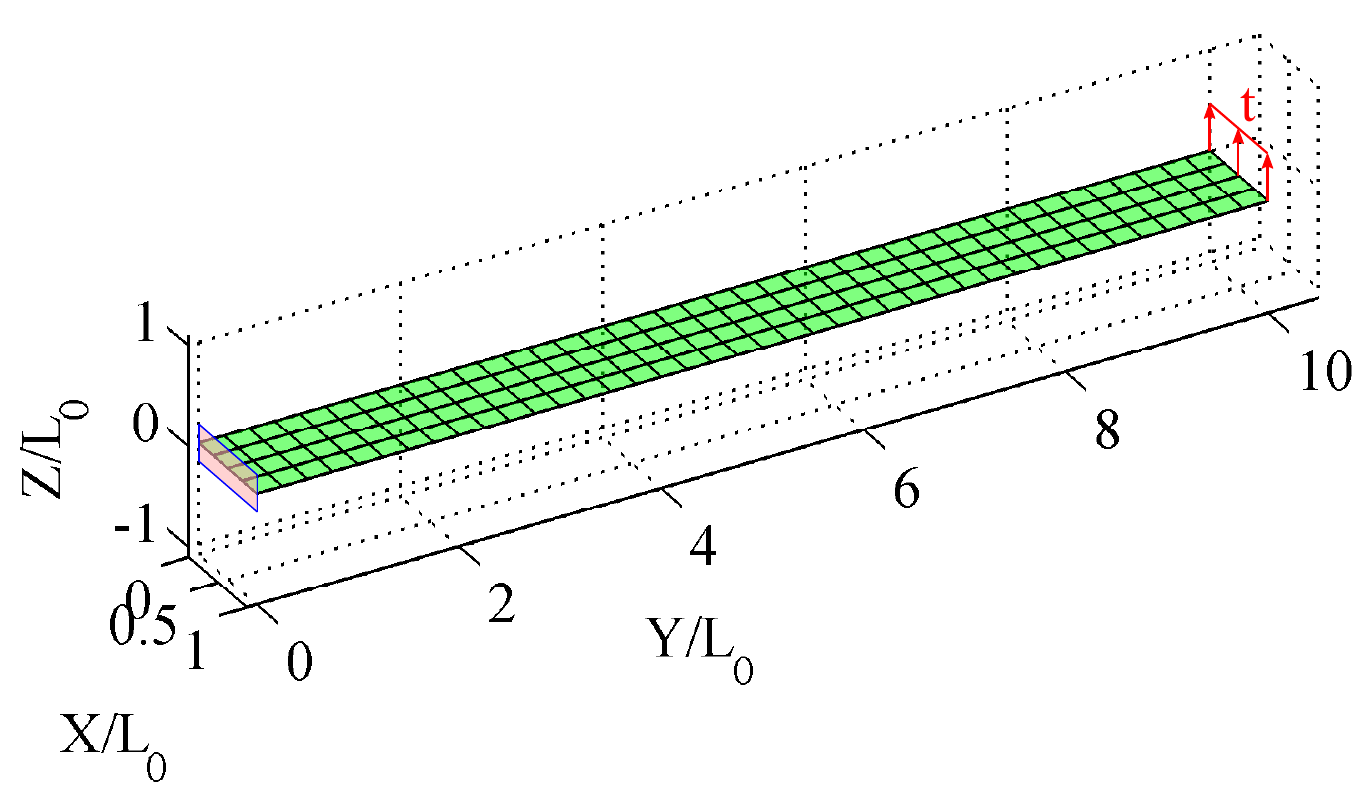}}
	\subfigure[]{\includegraphics[width = 0.45\textwidth]{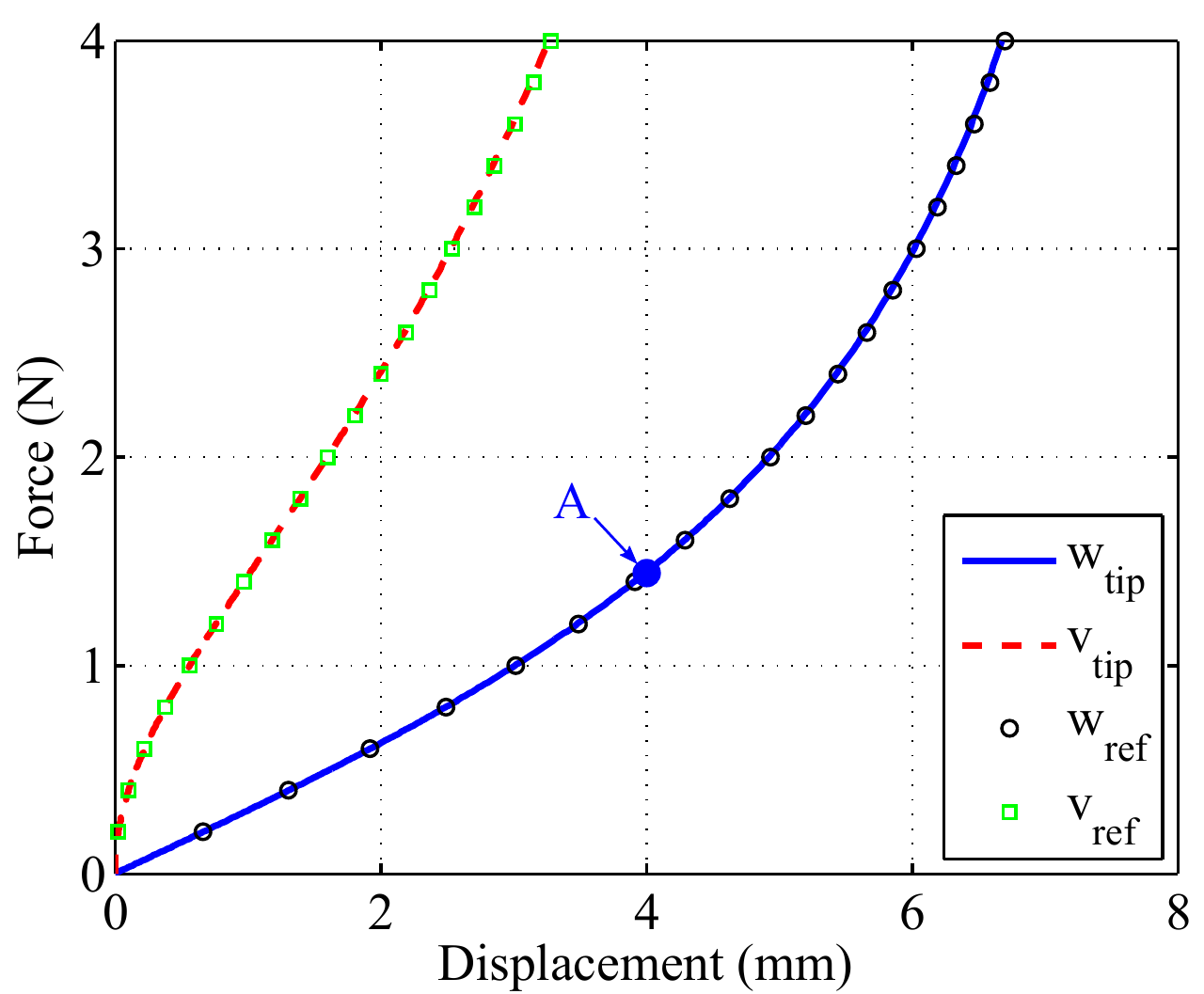}}
	\subfigure[]{\includegraphics[width = 0.5\textwidth]{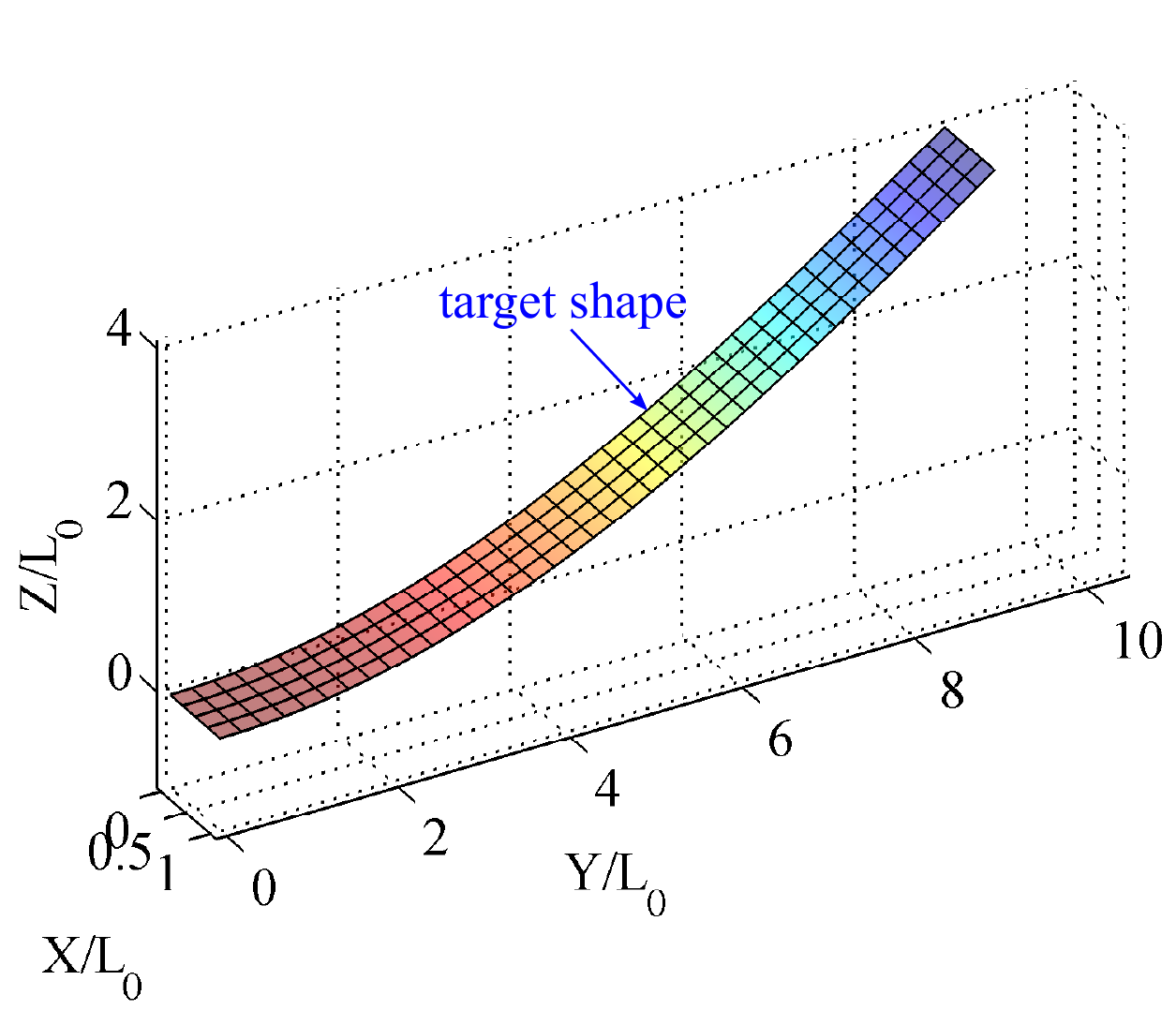}}
	\caption{Cantilever shell: (a) undeformed shape under the prescribed end displacement $w_{tip}$, (b) undeformed shape under the end shear traction $t$, (c) vertical tip forces versus vertical ($w_{tip}$) and horizontal ($v_{tip}$) tip displacements. The force-displacement responses are in good agreement with the ones shown in \cite{Sze:2004}, (d) deformed shape at displacement $w^{meas}_{tip}$ corresponding to point \textcolor{blue}{A} in (c). This current shape is used as target shape and nodal displacements represent the experiment-like data.}
	\label{fig:cantilever_shell_model}
\end{center}
\end{figure}

\begin{figure}[htbp]
\begin{center}
	\subfigure[]{\includegraphics[width = 0.45\textwidth]{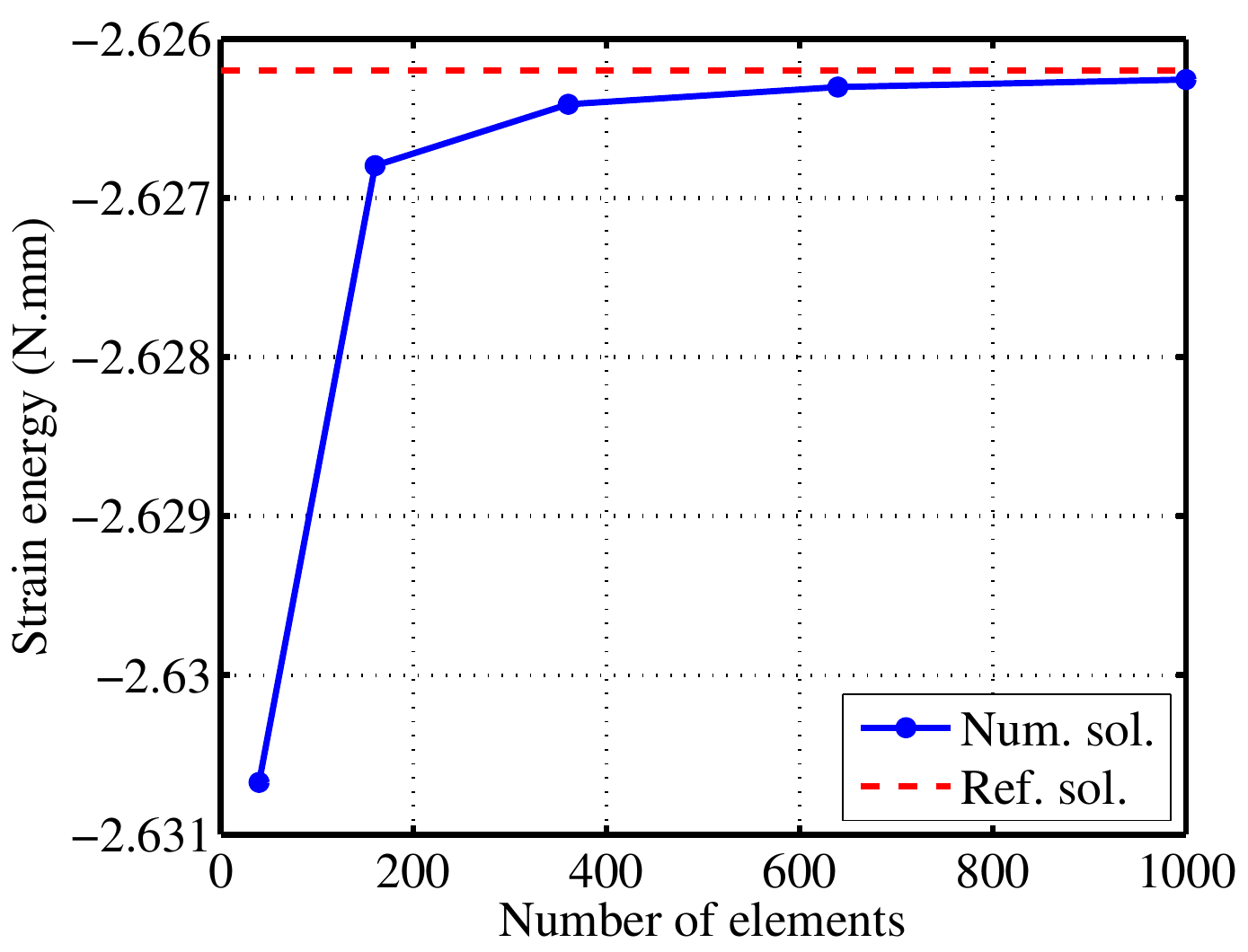}}
	\subfigure[]{\includegraphics[width = 0.45\textwidth]{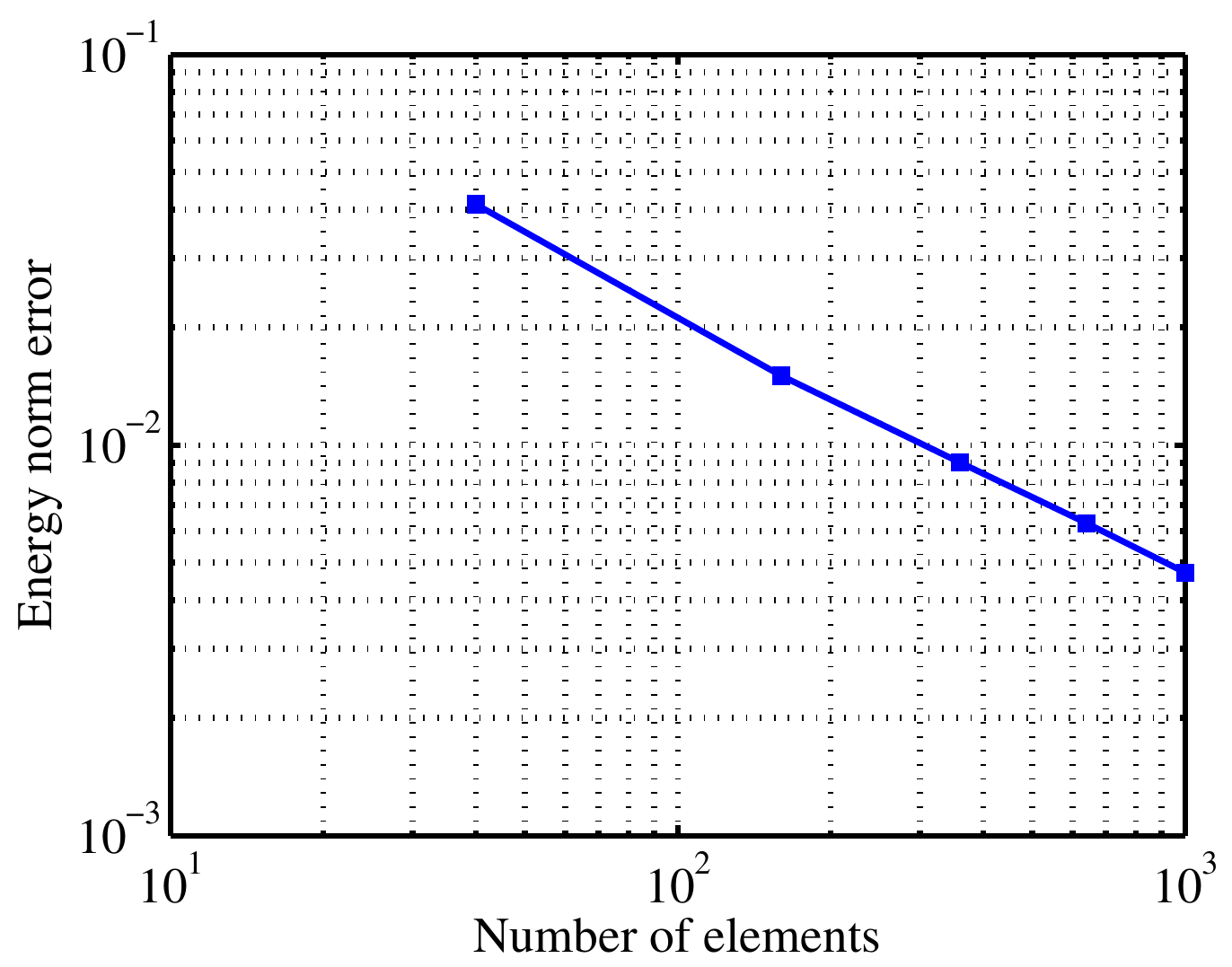}}
	\caption{Cantilever shell subjected to prescribed end displacement $w_{tip}$: (a) strain energy, (b) convergence of the energy norm error w.r.t. the number of elements.}
	\label{fig:energy_norm}
\end{center}
\end{figure}

In this example, the prescribed displacement $w_{tip}$ or the shear traction $t$ applied to the free end of a cantilever shell is identified using experiment-like data, see Figures \ref{fig:cantilever_shell_model}(a) and \ref{fig:cantilever_shell_model}(b). The experiment-like data is generated by considering the large deflection of a cantilever shell, with dimensions $L \times W \times T = 1 \times 10 \times 0.1$ mm, under the prescribed displacement and the end shear traction as shown in Figure \ref{fig:cantilever_shell_model}. The shell is discretized  by $4 \times 40$ NURBS elements. The projected model using the compressible Neo-Hookean formulation, according to sections \ref{subsec:proj_model} and \ref{subsec:comp_NeoHookean_model} is used with $E = 1.2 \times 10^6 ~N/mm^2$ and $\nu = 0$. The force-displacement responses obtained from the forward problem are compared to that reported by Sze et al. \cite{Sze:2004}. Convergence of the energy norm error determined by Equation (\ref{eq:energy_norm_error}) is shown in Figure \ref{fig:energy_norm} to assess the FE results. The error $e_{E}$ is calculated as

\begin{equation} \label{eq:energy_norm_error}
	e_E = \sqrt{ \left\lvert \frac{E^{num} - E^{ref}}{E^{ref}} \right\rvert }
\end{equation}

\noindent where $E^{num}$ and $E^{ref}$ are the numerical and the reference strain energies, respectively. It should be noted that since the analytical solution is not available, the reference strain energy obtained from a fine mesh of $20 \times 200$ NURBS-based elements is used. The experiment-like data, i.e. the nodal displacements, is obtained by using NURBS-based finite elements. The target configuration corresponds to point A on the solid blue curve in Figure \ref{fig:cantilever_shell_model}(c) where $w^{meas}_{tip} = 4.0$ or $t^{meas} = 1.5 ~N/mm$.


\begin{figure}[htbp]
\begin{center}
	\subfigure[]{\includegraphics[width = 0.5\textwidth]{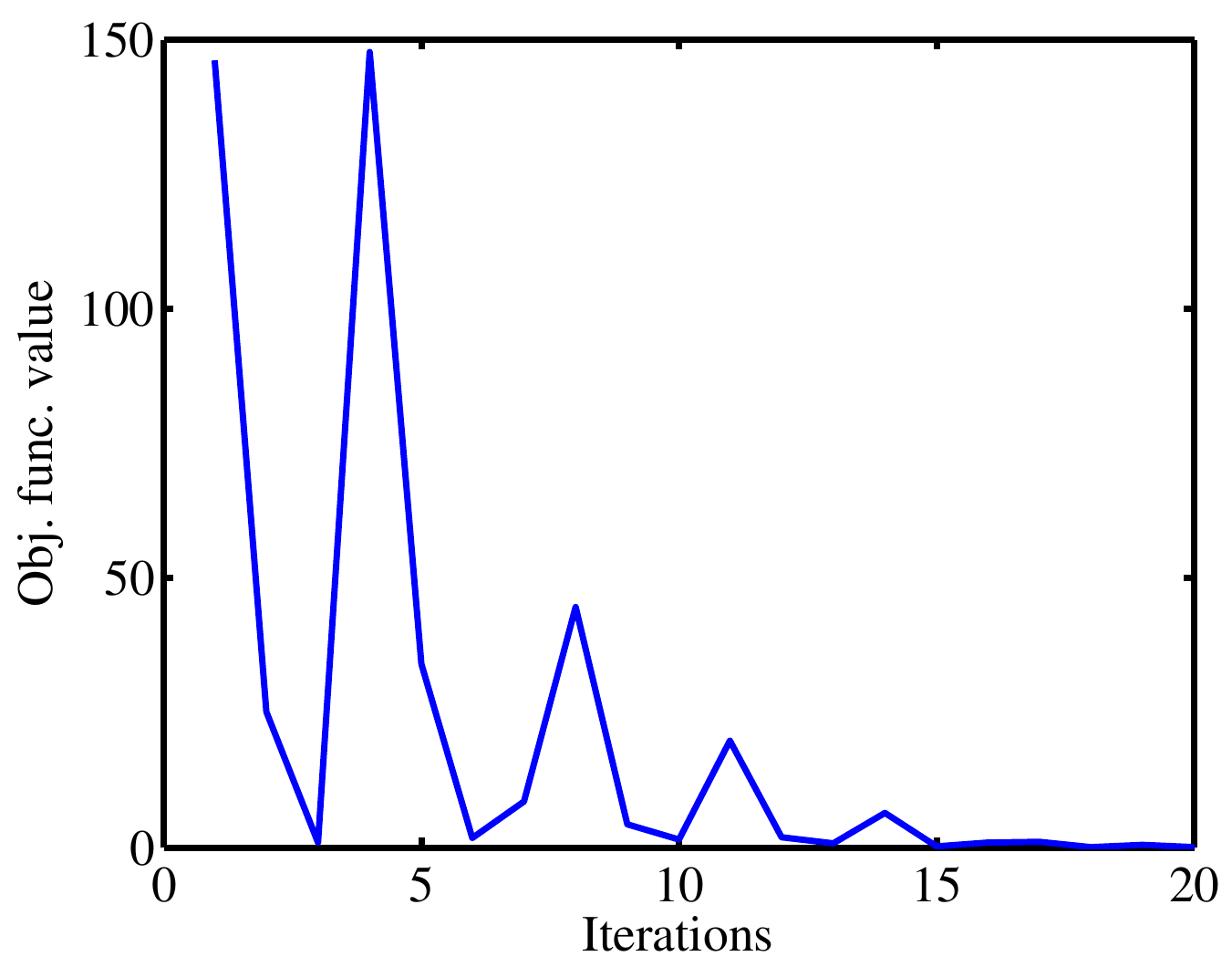}}
	\subfigure[]{\includegraphics[width = 0.45\textwidth]{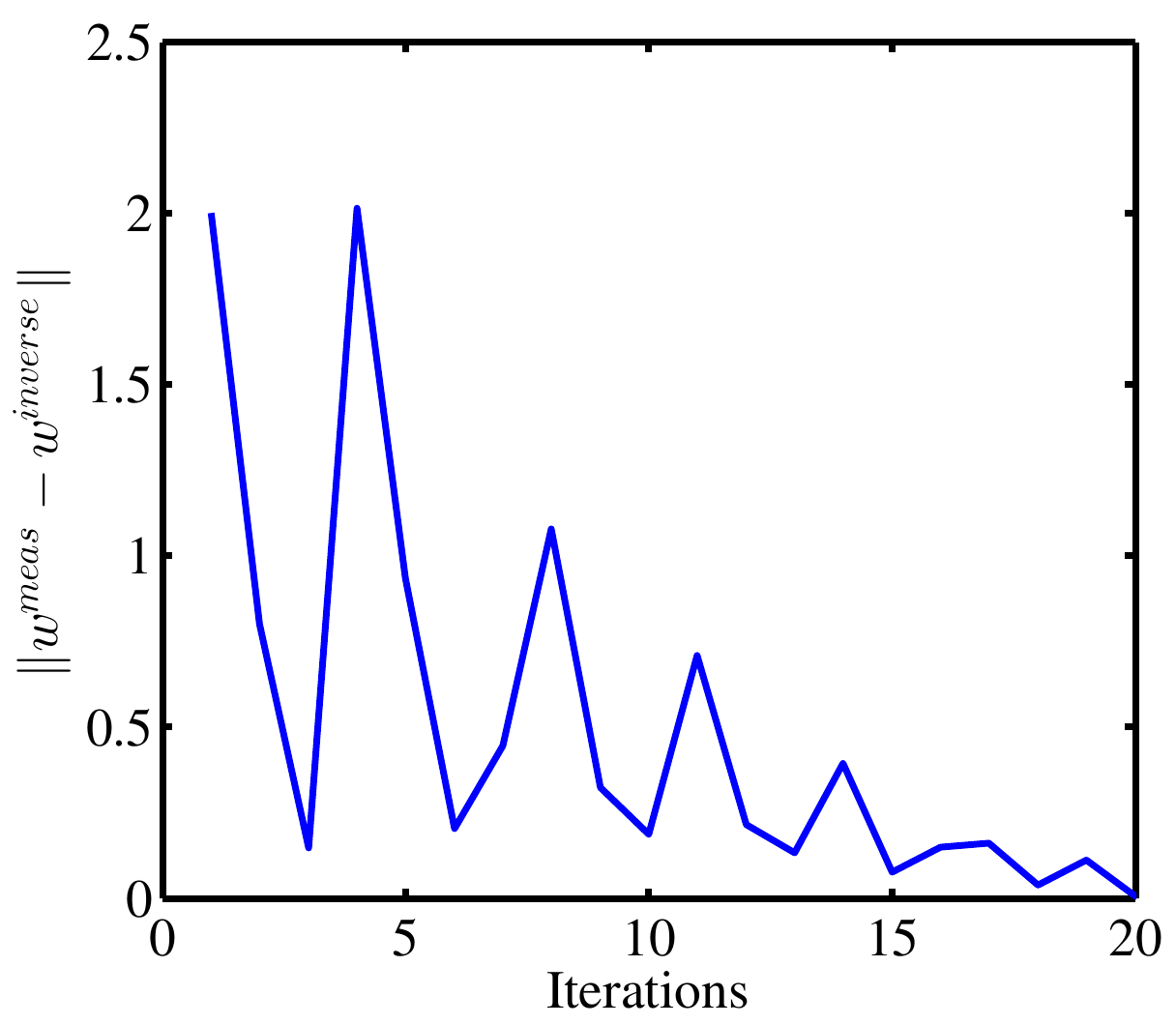}}
	\caption{Cantilever shell subjected to the displacement $w_{tip}$: (a) the objective function versus the number of iterations, (b) convergence of $L^2$ error norm in parameter space $X$ versus the number of iterations. The analytical sensitivity presented in Equation (\ref{eq:deriv_fint_wrt_disp}) is employed to compute the element derivatives in Equation (\ref{eq:derv_obj_func_adj}).}
	\label{fig:cantilever_shell_inverse_diranalsen}
\end{center}
\end{figure}

\begin{figure}[htbp]
\begin{center}
	\subfigure[]{\includegraphics[width = 0.475\textwidth]{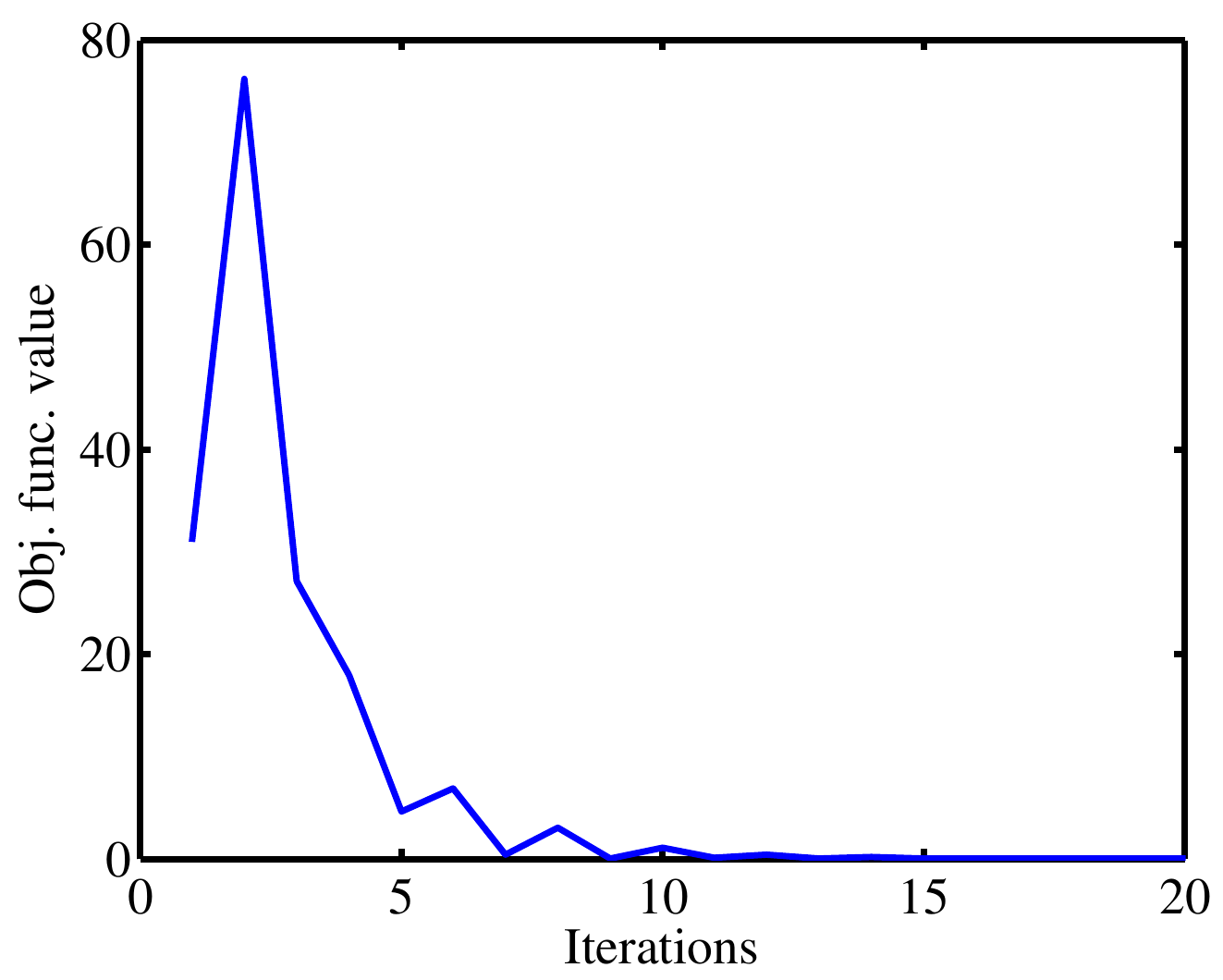}}
	\subfigure[]{\includegraphics[width = 0.45\textwidth]{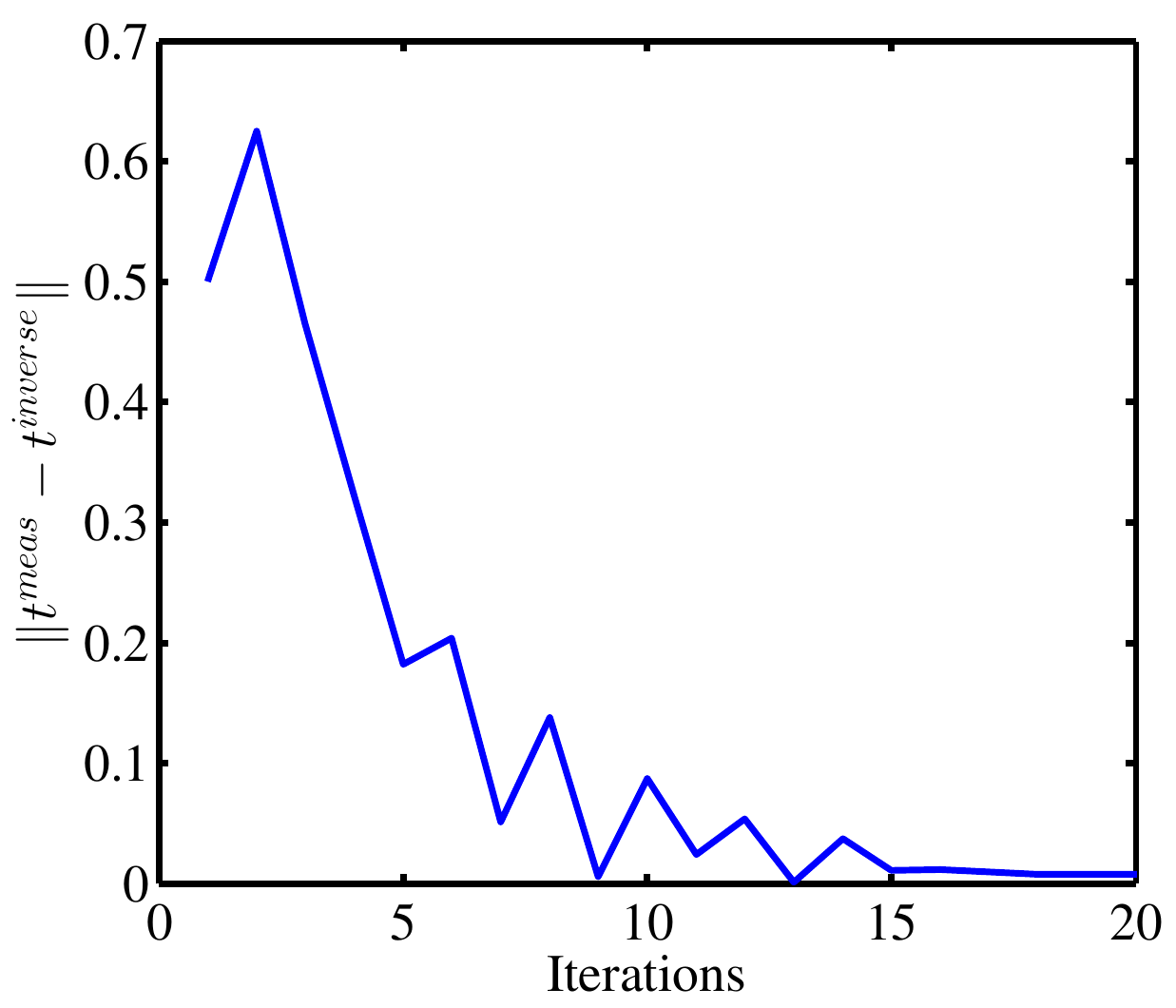}}
	\caption{Cantilever shell subjected to the end shear traction $t$: (a) the objective function versus the number of iterations, (b) convergence of $L^2$ error norm in parameter space $X$ over the number of iterations. The analytical sensitivity presented in Equation (\ref{eq:fstartnln_tract}) is employed to compute the element derivatives in Equation (\ref{eq:derv_obj_func_adj}).}
	\label{fig:cantilever_shell_inverse_neuanalsen}
\end{center}
\end{figure}

Inverse analysis based on the target configuration is then carried out corresponding to Neumann (traction $t$) and Dirichlet (prescribed displacement $w_{tip}$) boundary conditions (BCs). The element derivatives in Equation(\ref{eq:derv_obj_func_adj}) are obtained by using analytical and semi-analytical sensitivities. Detailed implementation for the analytical sensitivities when the cantilever shell subjected to the displacement $w_{tip}$ or the end shear traction $t$ can be found in sections \ref{appendix:sensitivities_displacement} and \ref{appendix:sensitivities_traction}, respectively. Accordingly, the convergence of the objective functions and $L^2$ error norms in spaces $Y$ and $X$, see Equation (\ref{eq:inverse_problem}), w.r.t. iterations is shown in Figures \ref{fig:cantilever_shell_inverse_diranalsen} and \ref{fig:cantilever_shell_inverse_neuanalsen}. The inverse solutions $t^{inverse} = 1.4927 ~N/mm$ (for Neumann BCs) and $w^{inverse}_{tip} = 4.0077$ (for Dirichlet BCs) are obtained respectively. It can be seen that they are in good agreement with $t^{meas} = 1.5 ~N/mm$ $w^{inverse}_{tip} = 4$.

\begin{figure}[htbp]
\begin{center}
\vspace*{-2.0cm}
	\subfigure[]{\includegraphics[width = 0.45\textwidth]{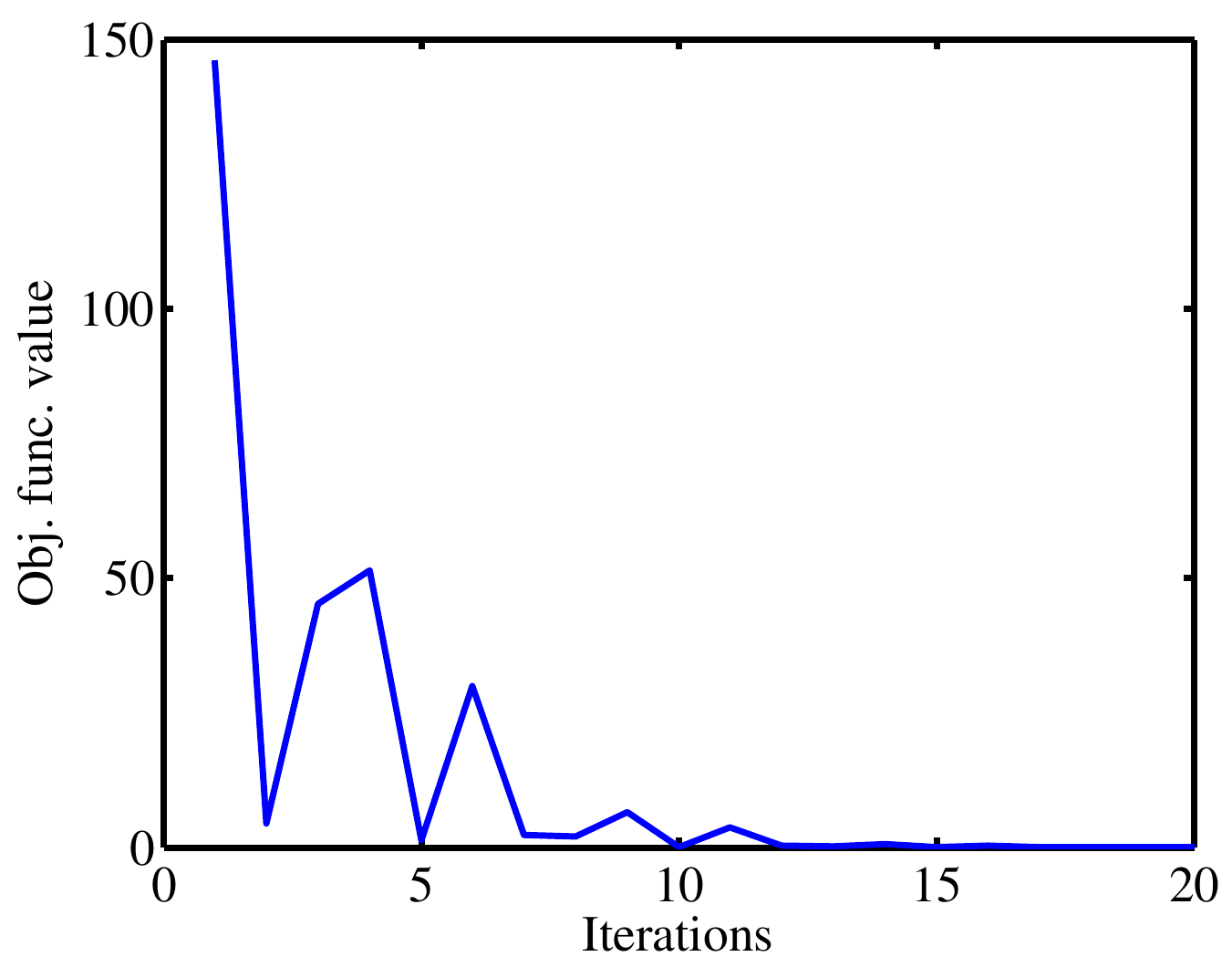}}
	\subfigure[]{\includegraphics[width = 0.4\textwidth]{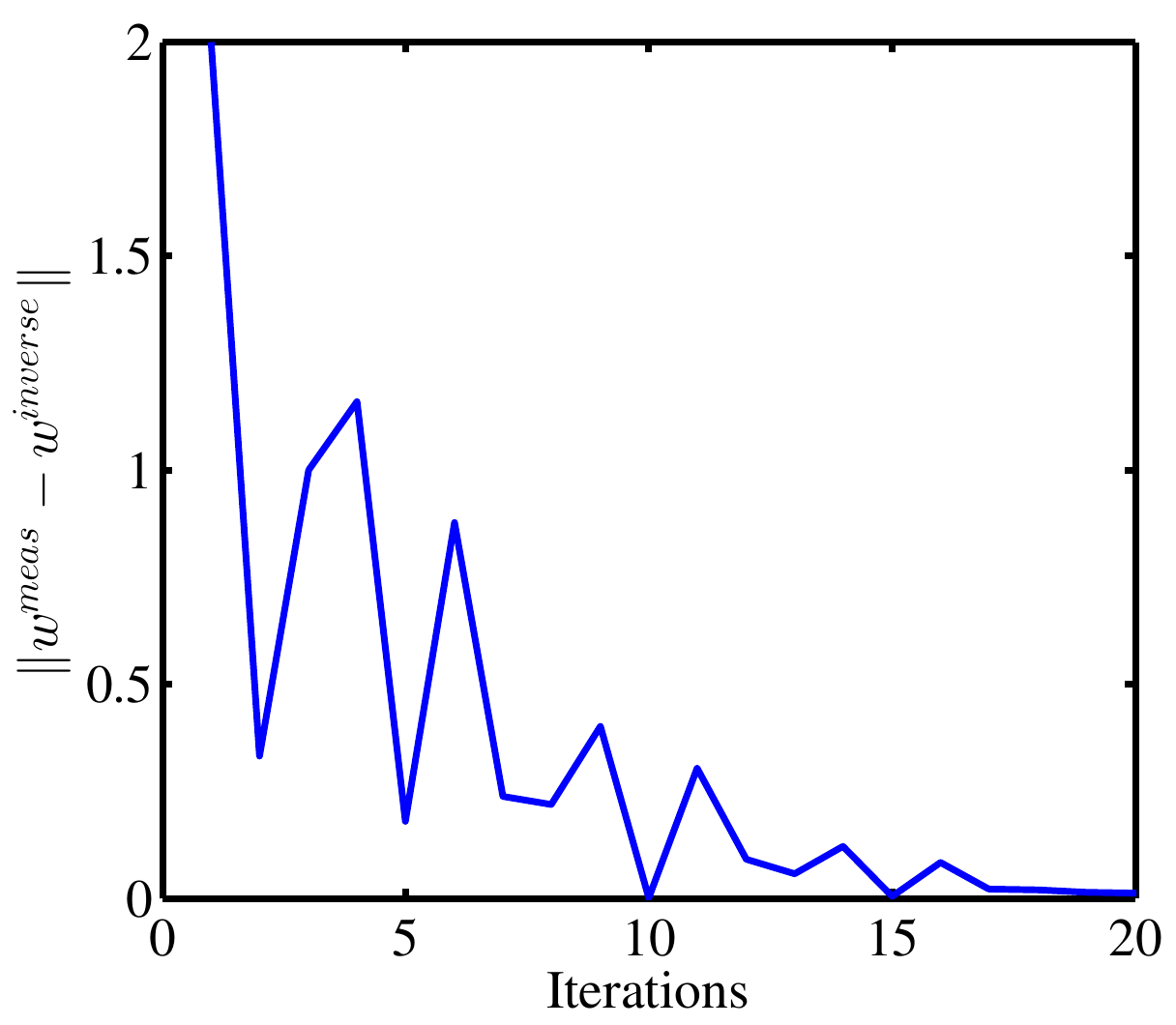}} \\
\vspace*{-0.25cm}
	\subfigure[]{\includegraphics[width = 0.45\textwidth]{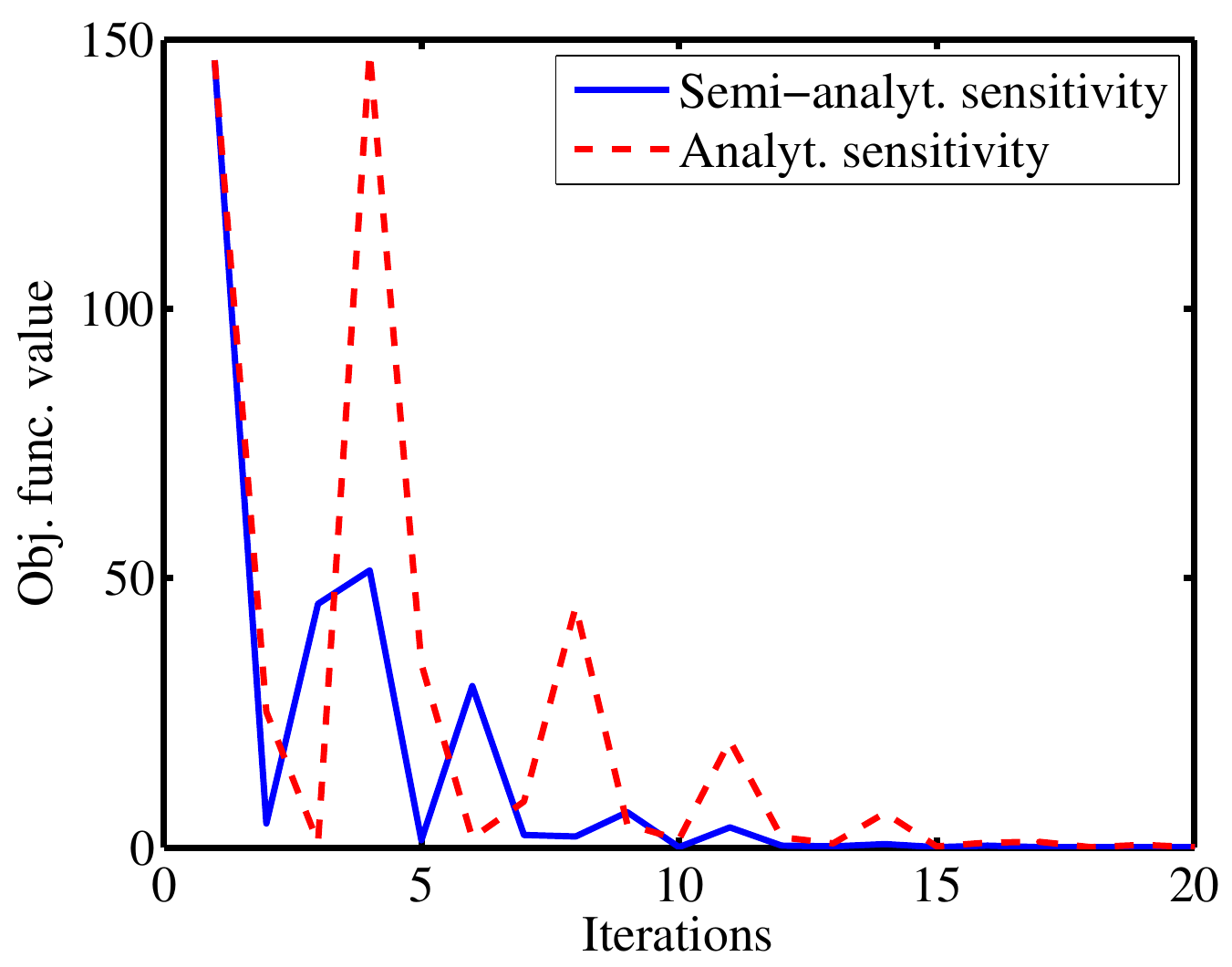}}
	\subfigure[]{\includegraphics[width = 0.45\textwidth]{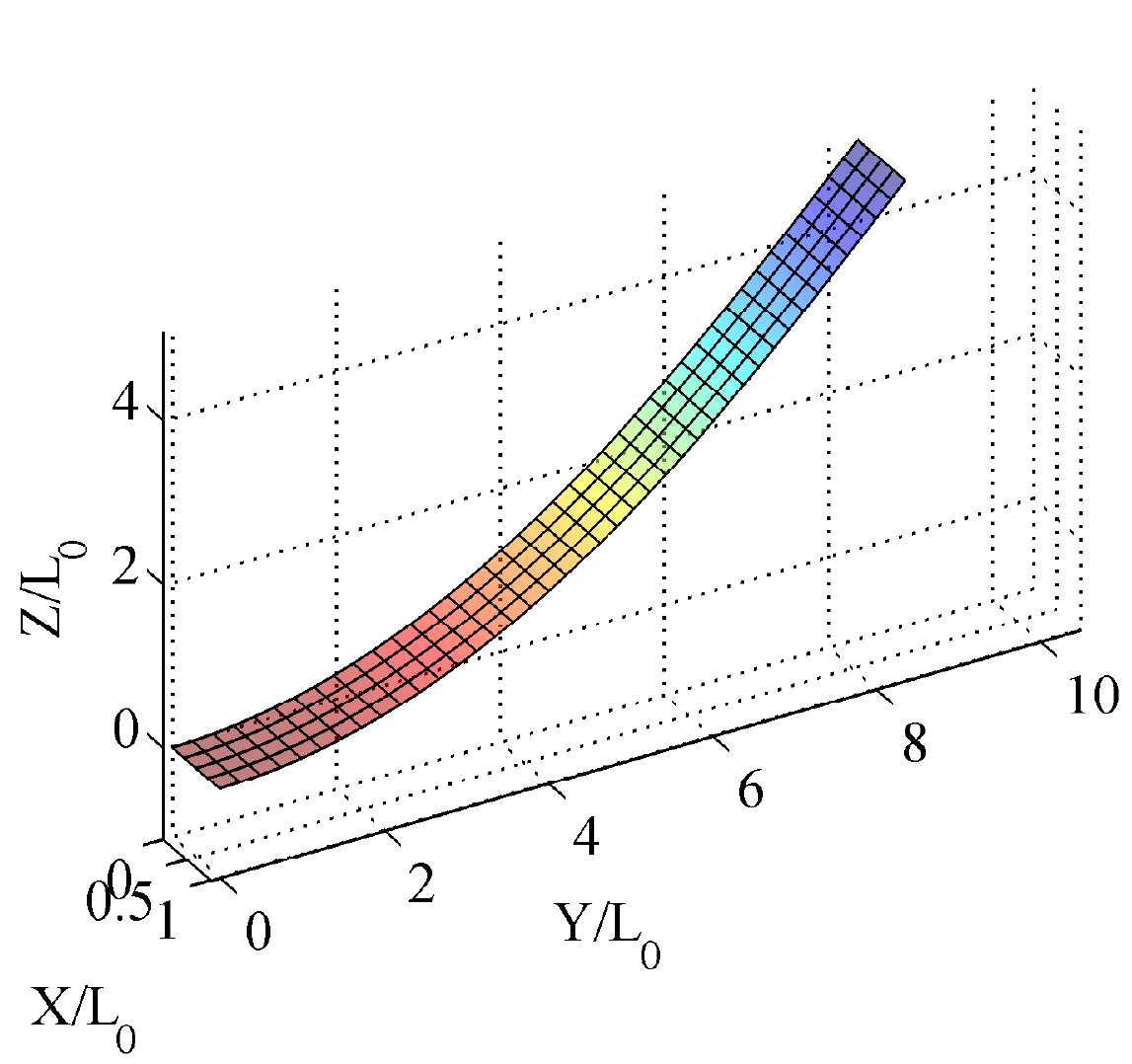}} \\
\vspace*{-0.25cm}
	\subfigure[]{\includegraphics[width = 0.45\textwidth]{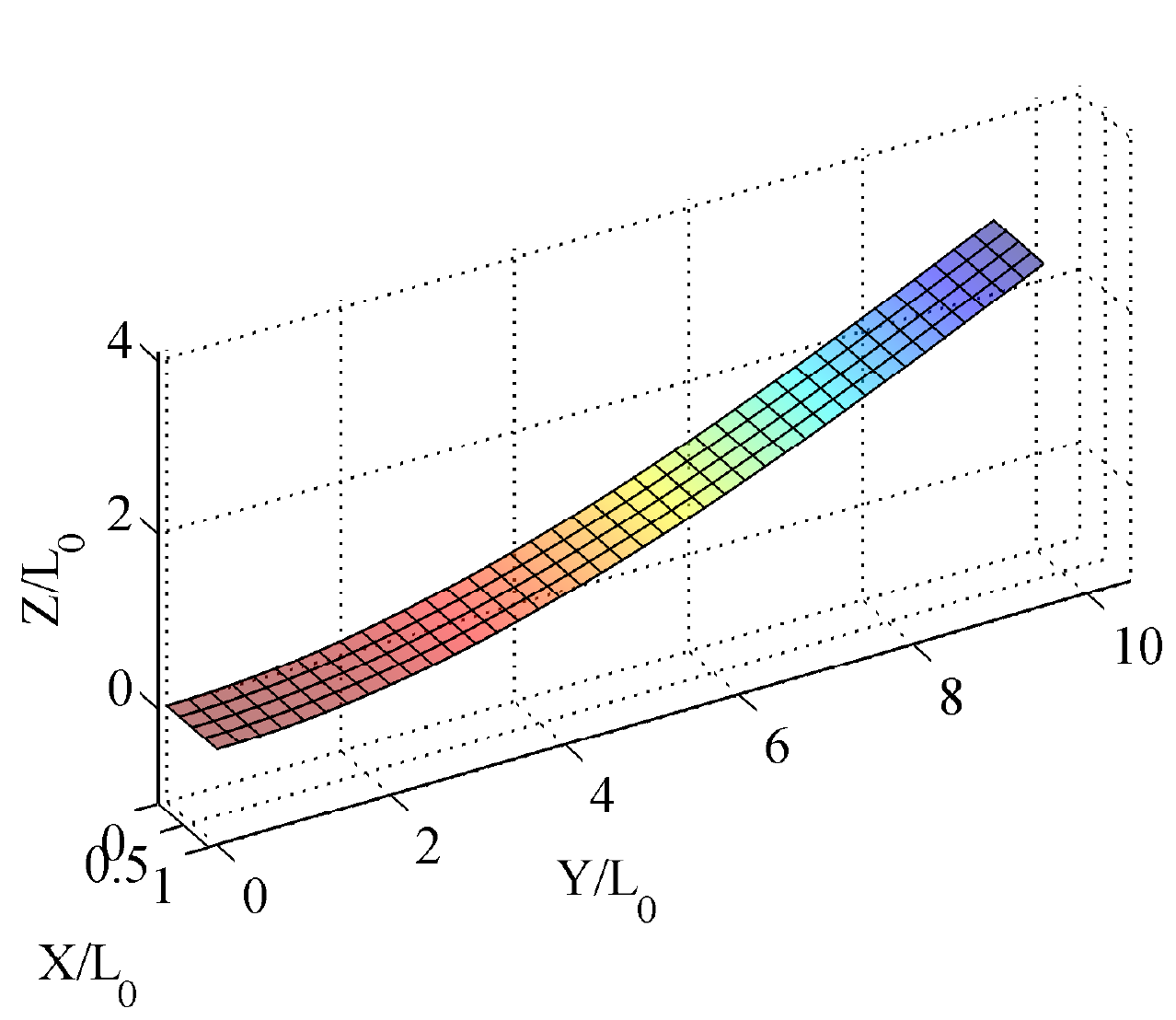}}
	\subfigure[]{\includegraphics[width = 0.45\textwidth]{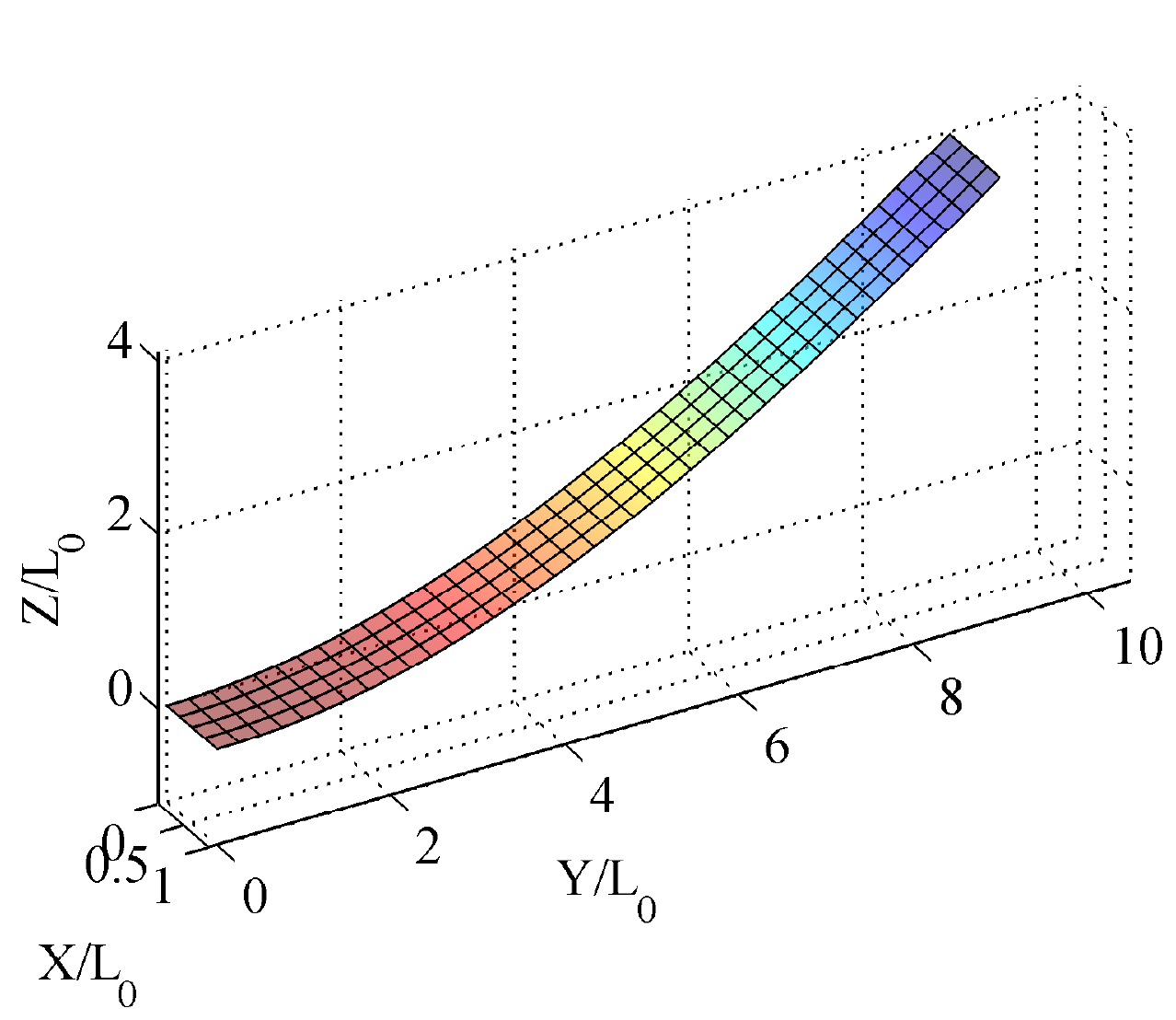}}
	\caption{Cantilever shell subjected to the displacement $w_{tip}$: (a) the objective function versus the number of iterations, (b) convergence of $L^2$ error norm in parameter space $X$ over the number of iterations, (c) convergence of the objective function over the number of iterations using the analytical and semi-analytical sensitivities, (d) the reconstructed deformation at the second iteration, (e) the reconstructed deformation after $3$ iterations, (f) the reconstructed deformation after $20$ iterations. The semi-analytical sensitivity is employed to compute the element derivatives in Equation (\ref{eq:derv_obj_func_adj}).}
	\label{fig:cantilever_shell_inverse}
\end{center}
\end{figure}

In addition, inverse approach using the semi-analytical sensitivity is performed to reconstruct the target configuration of the cantilever shell. Figure \ref{fig:cantilever_shell_inverse} shows the excellent accuracy of the proposed inverse method. The convergence of the objective function and $L^2$ error norm over iterations is illustrated in Figures \ref{fig:cantilever_shell_inverse}(a+b), respectively. Very good agreement of the convergence of the objective function w.r.t. the number of iterations is obtained between the analytical and semi-analytical sensitivities, see Figure \ref{fig:cantilever_shell_inverse}(c). The deformed configuration under the inverse solution $w^{inverse}_{tip} = 3.9696$ shows excellent agreement with the target configuration in Figure \ref{fig:cantilever_shell_model}(d). As observed, the maximal error ($err_{\bm{w}} = \frac{4 - 3.9696}{4} = 0.76 \%$) between two configurations occurs at the nodes along the free end.

\begin{figure}[htbp]
\begin{center}
	\subfigure{\includegraphics[width = 0.45\textwidth]{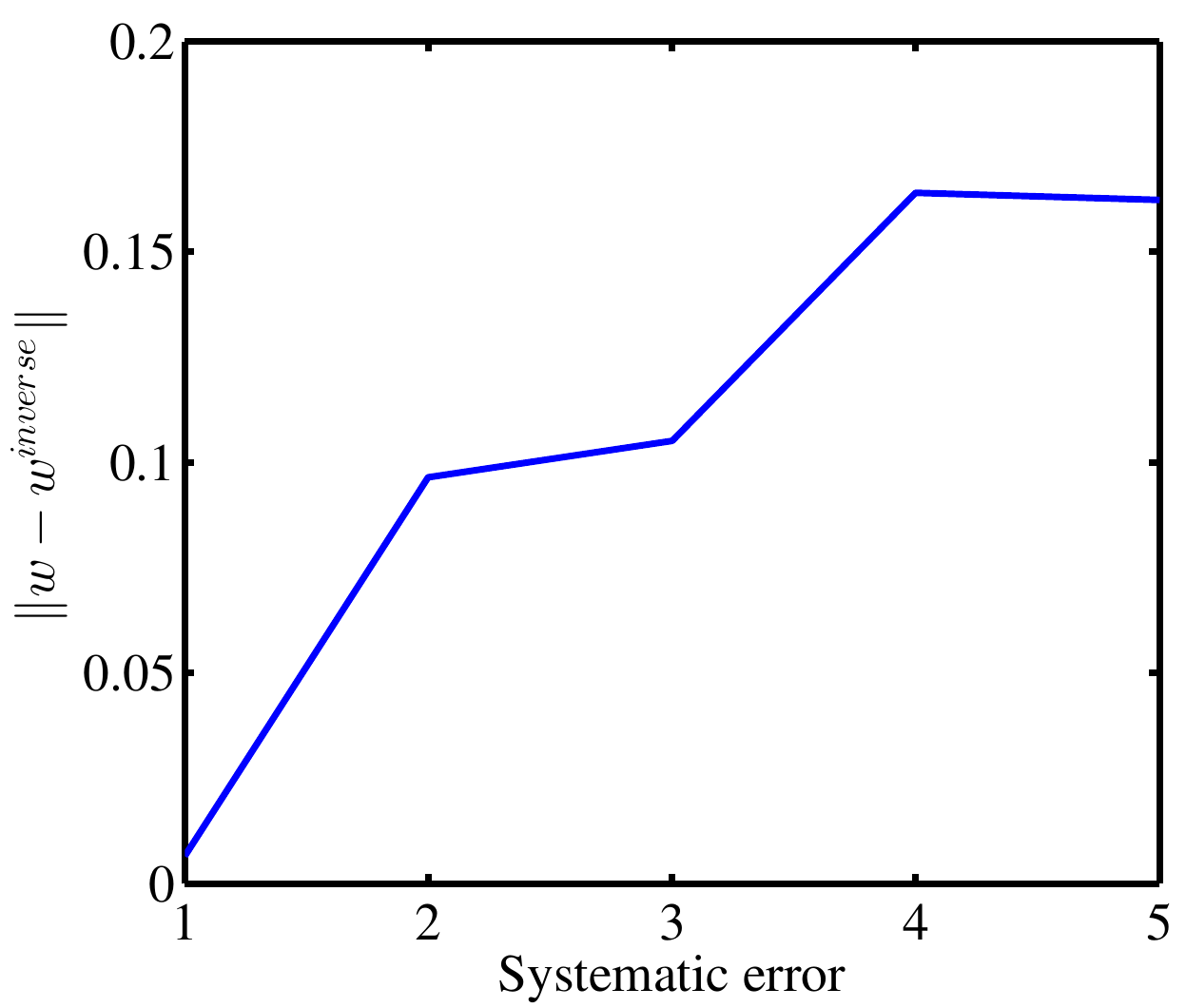}}
	\subfigure{\includegraphics[width = 0.45\textwidth]{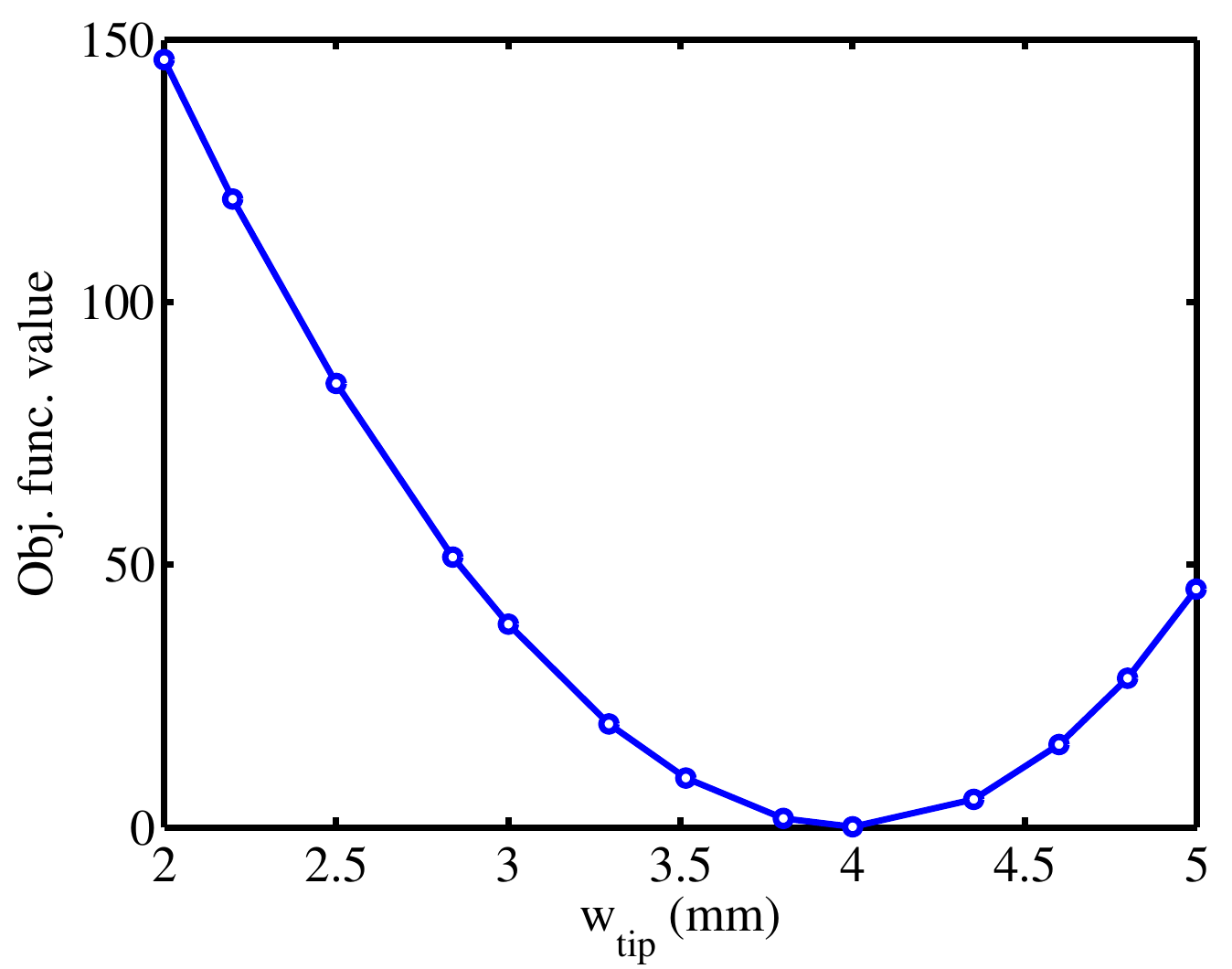}}
	\caption{Cantilever shell subjected to the displacement $w_{tip}$: (a) the effect of noise (with changing $\gamma$) on accuracy of the algorithm, (b) the convexity of the objective function w.r.t. the end displacement $w_{tip}$.}
	\label{fig:systematic_error_and_convexity}
\end{center}
\end{figure}

In order to estimate the effect of the noise on accuracy of the algorithm, systematic noise ($\gamma$ is taken as deterministic numbers) is examined. As shown in Figure \ref{fig:systematic_error_and_convexity}(a), the accuracy of the algorithm decreases with an increase in systematic noise $\gamma$. The convexity of the objective function w.r.t. the end displacement $w_{tip}$ can be seen in Figure \ref{fig:systematic_error_and_convexity}(b).

Generally, inverse problems may be ill-posed, i.e., that one of the following conditions is violated: Existence, uniqueness and stable dependency on the data. As we do not aim to solve the inverse problem directly (Equation (\ref{eq:inverse_equation})) but make use of a sum of least-squares approach (Equation (\ref{eq:least_squares_form})) the existence is guaranteed, at least in the least-squares sense. Proving uniqueness analytically is challenging and most probable not possible for the complex nonlinear forward problem. So consequently, we followed a numerical approach where values of the objective are plotted over different values for the excitations, see Figure \ref{fig:systematic_error_and_convexity}(b). As a clear minimal point is indicated, we can assume uniqueness at least for the simplified problems. For problems with more degrees of freedom a generalization is unfortunately impossible.

To solve now the inverse problem, iterative methods are applied. Regarding error norms in $X$ and $Y$ space we see good convergence and no strong tendency for divergence, even for higher error amounts in the data (here the prescribed shape), see Figures \ref{fig:cantilever_shell_inverse_diranalsen}, \ref{fig:cantilever_shell_inverse_neuanalsen} and \ref{fig:systematic_error_and_convexity}(a).

\subsection{Hinged cylinderical shell subjected to prescribed center displacement and central point force} \label{subsec:ex2}

\begin{figure}[htbp]
\begin{center}
	\subfigure[]{\includegraphics[width = 0.6\textwidth]{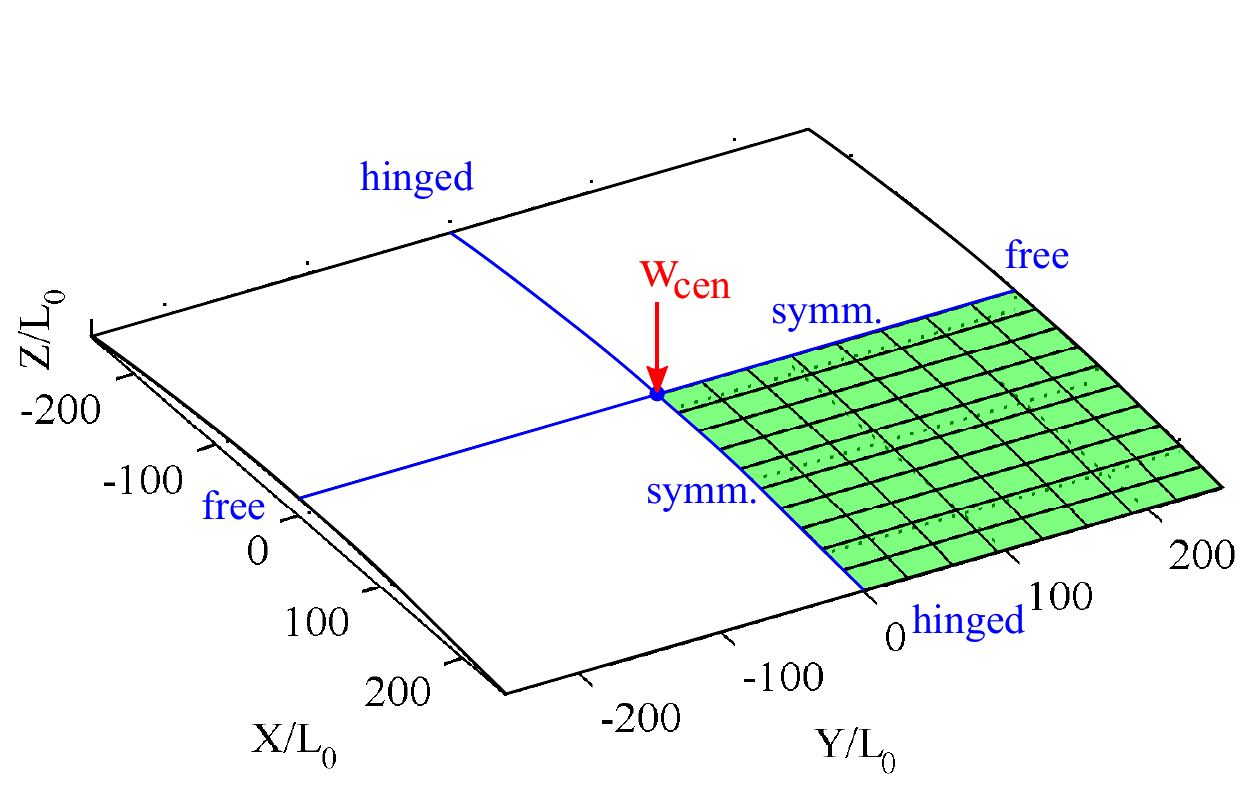}} \\
	\subfigure[]{\includegraphics[width = 0.5\textwidth]{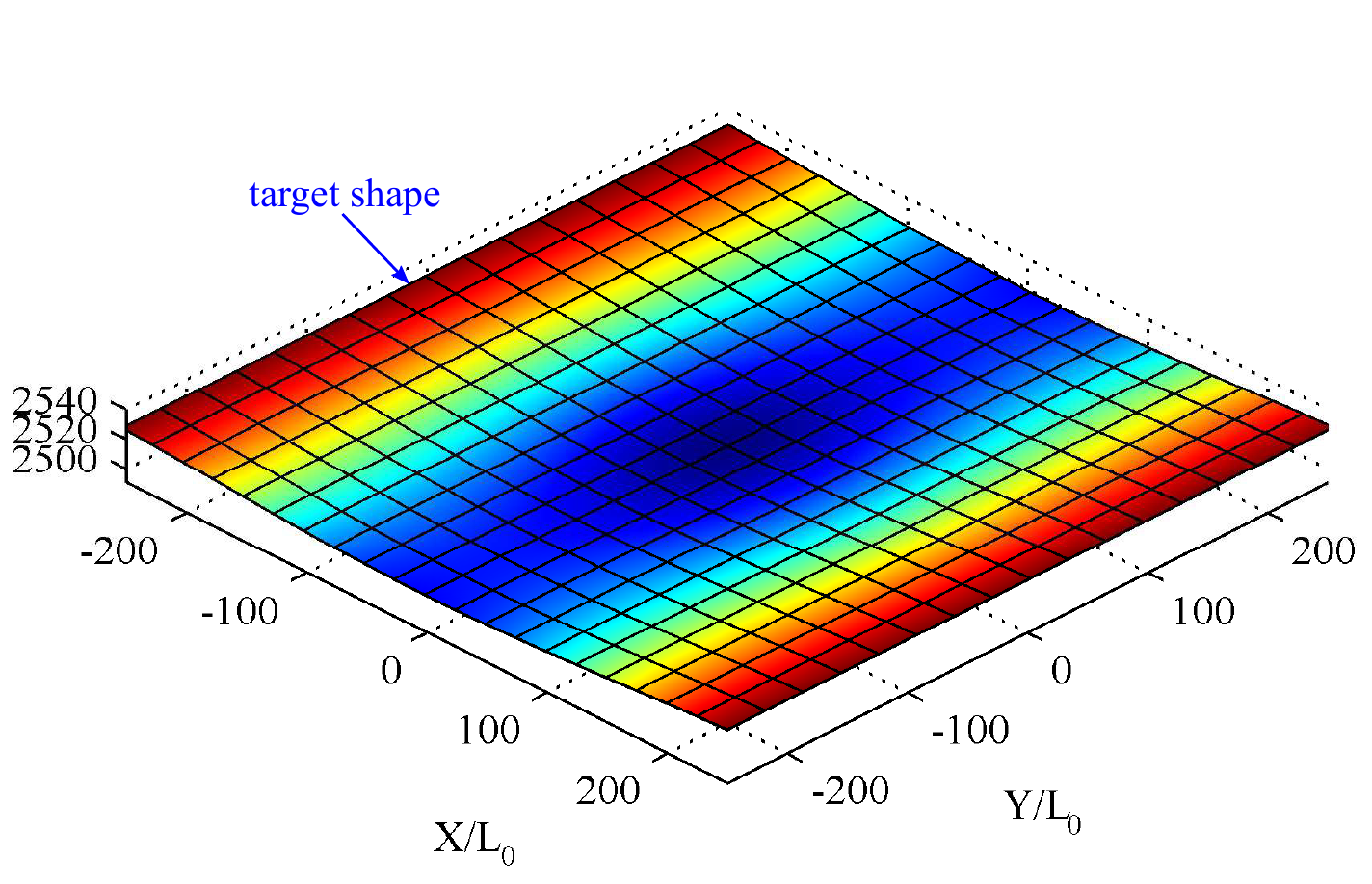}}
	\subfigure[]{\includegraphics[width = 0.45\textwidth]{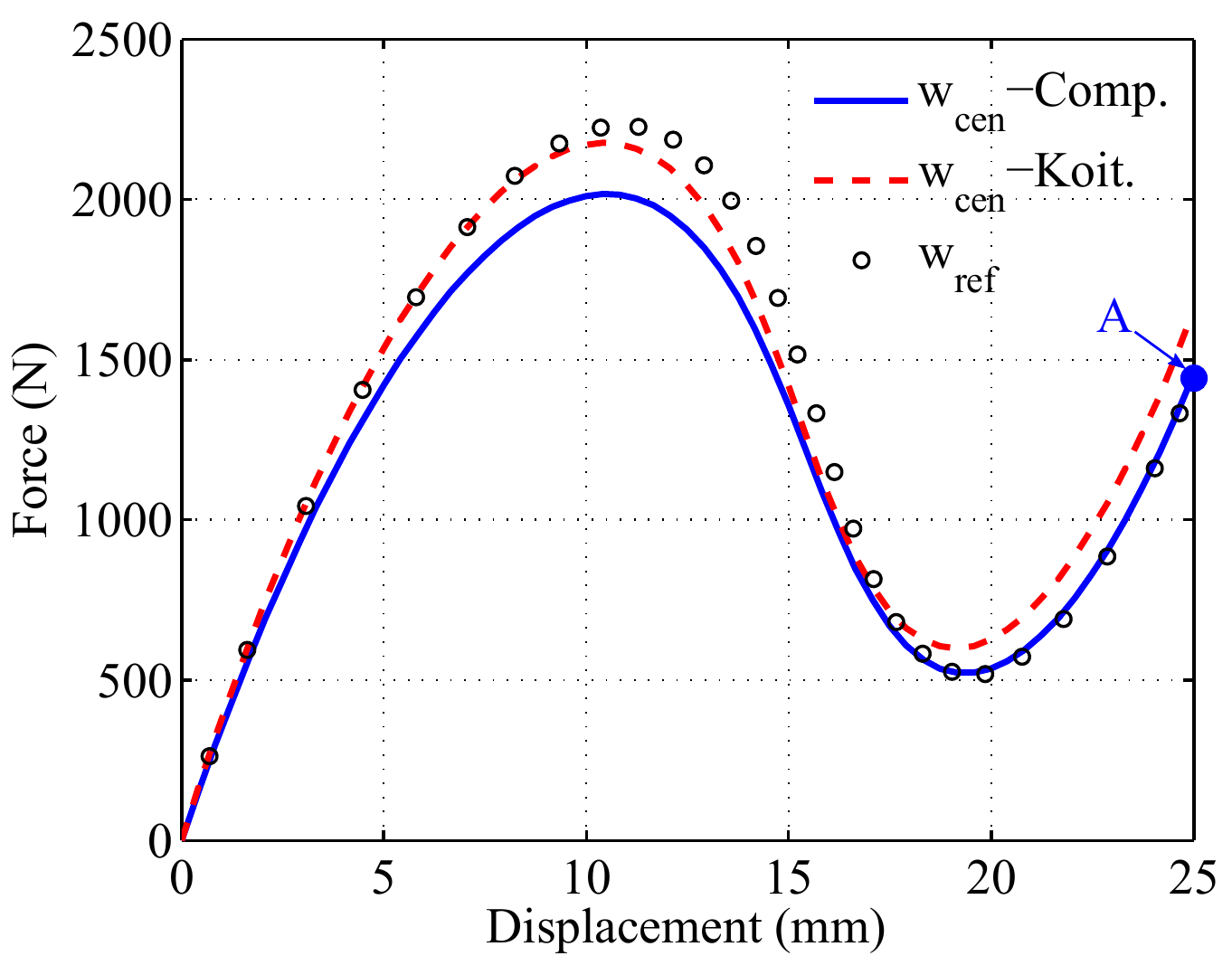}}
	\caption{Hinged cylindrical shell subject to prescribed center displacement $w_{cen}$: (a) undeformed shape, (b) deformed shape at displacement $w^{meas}_{cen} = -24.98$ using the projected shell model. This shape corresponds to point \textcolor{blue}{A} in (c), (c) reaction forces versus vertical displacement ($w$) in $z$ direction for the Koiter shell model (red dashed line) and the projected shell model (blue solid line) using the compressible Neo-Hookean material. The force-displacement curve obtained from the Koiter material model is in good agreement with the one, using isotropic Saint Venant-Kirchhoff model, shown in \cite{Sze:2004}.}
	\label{fig:Scoderlis_Lo_roof_model}
\end{center}
\end{figure}

The second benchmark example consists of a cylindrical shell of dimensions $R = 2540 ~mm$, $L = B = 504 ~mm$ and thickness $T = 12.7 ~mm$. The longitudinal edges of the shell are pinned, its curved edges are free and the shell is subjected to a prescribed center displacement $w_{cen}$. Due to the symmetry, a quarter of the shell is modeled using $8 \times 10$ quadratic NURBS elements as shown in Figure \ref{fig:Scoderlis_Lo_roof_model}(a). The symmetry boundary conditions are enforced by the Lagrange multiplier method presented in section \ref{subsec:Lagrange_multiplier_constraints}.

The forward problem is first solved based on the isogeometric analysis approach. The material is described by the Koiter model and the projected model with compressible Neo-Hookean formulation using the Young's modulus $E = 3105 ~N/mm^2$ and the Poisson's ratio $\nu = 0.3$. The force-displacement response is compared in Figure \ref{fig:Scoderlis_Lo_roof_model}(c) to the one reported by Sze et al. \cite{Sze:2004}, which considered the Saint Venant-Kirchhoff formulation with the same material properties. Good agreement between the response obtained by using the Koiter model with the reference result demonstrates the accuracy of the model. The deformed configuration corresponding to point A on the solid blue curve in Figure \ref{fig:Scoderlis_Lo_roof_model}(c) is then taken as the target shape.

\begin{figure}[htbp]
\begin{center}
	\subfigure[]{\includegraphics[width = 0.475\textwidth]{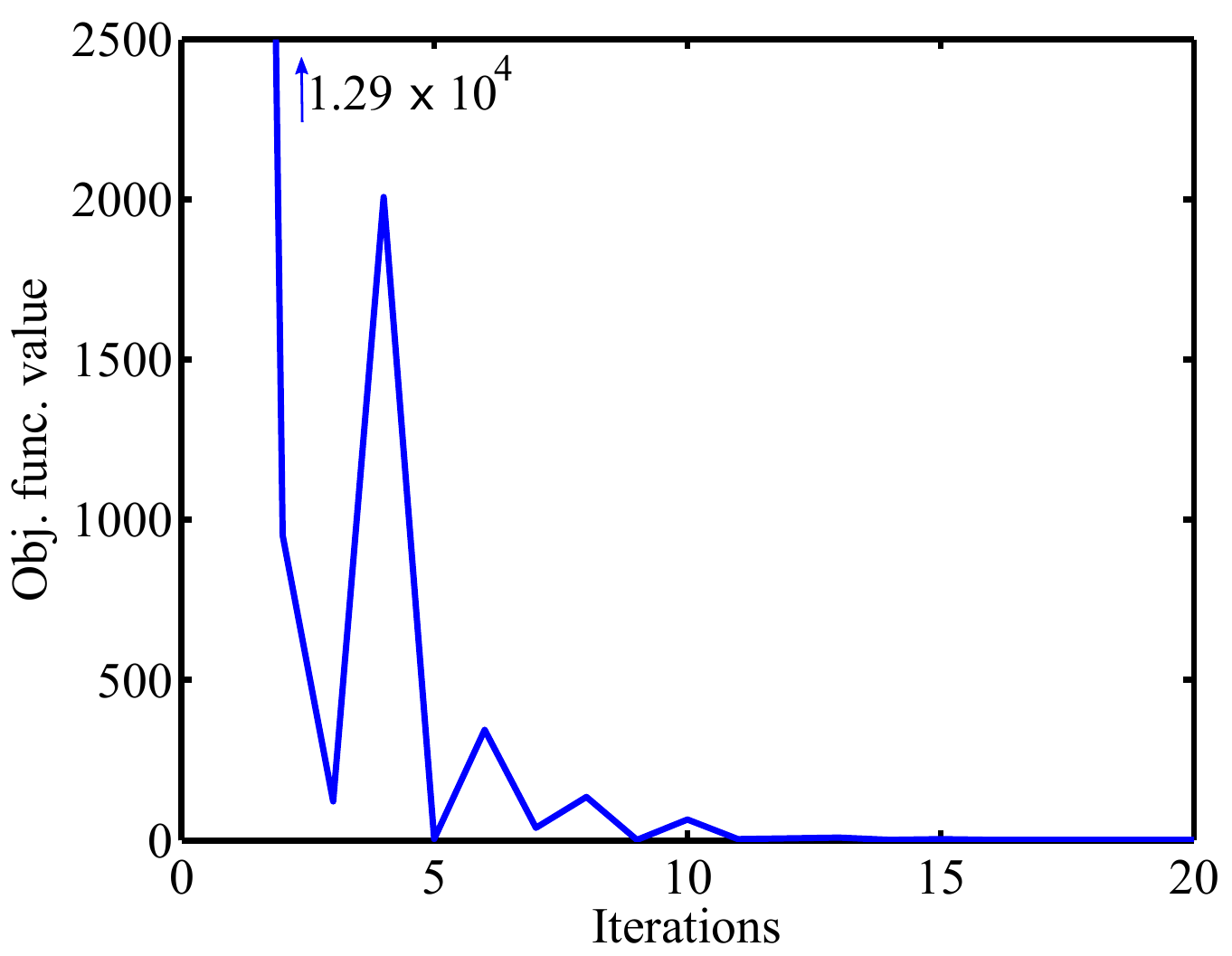}}
	\subfigure[]{\includegraphics[width = 0.425\textwidth]{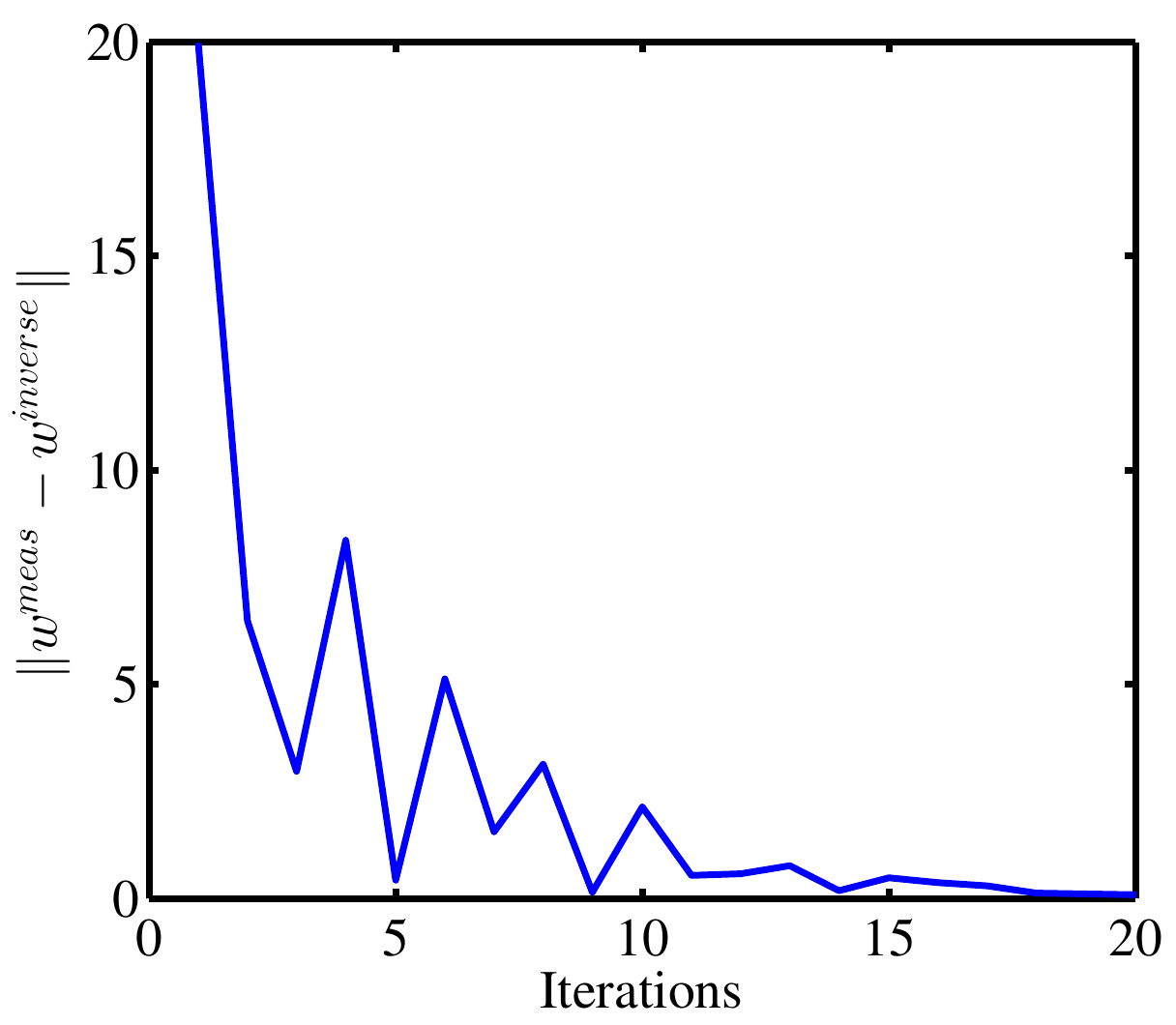}}
	\caption{Hinged cylindrical shell subjected to the displacement $w_{cen}$: (a) the objective function versus the number of iterations, (b) convergence of $L^2$ error norm in parameter space $X$ over the number of iterations.}
	\label{fig:Scoderlis_Lo_inverse_analsen}
\end{center}
\end{figure}

\begin{figure}[htbp]
\begin{center}
	\subfigure[]{\includegraphics[width = 0.475\textwidth]{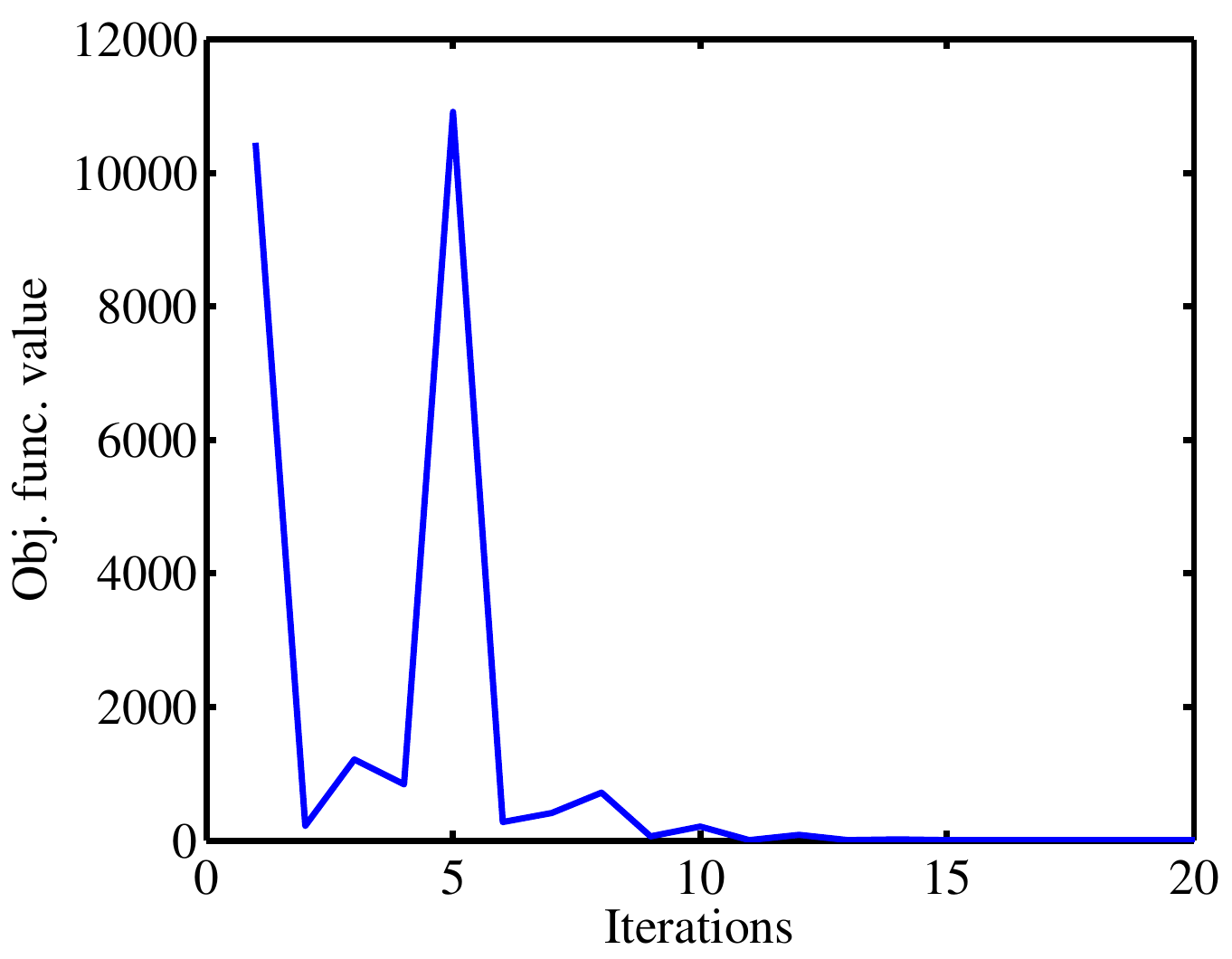}}
	\subfigure[]{\includegraphics[width = 0.425\textwidth]{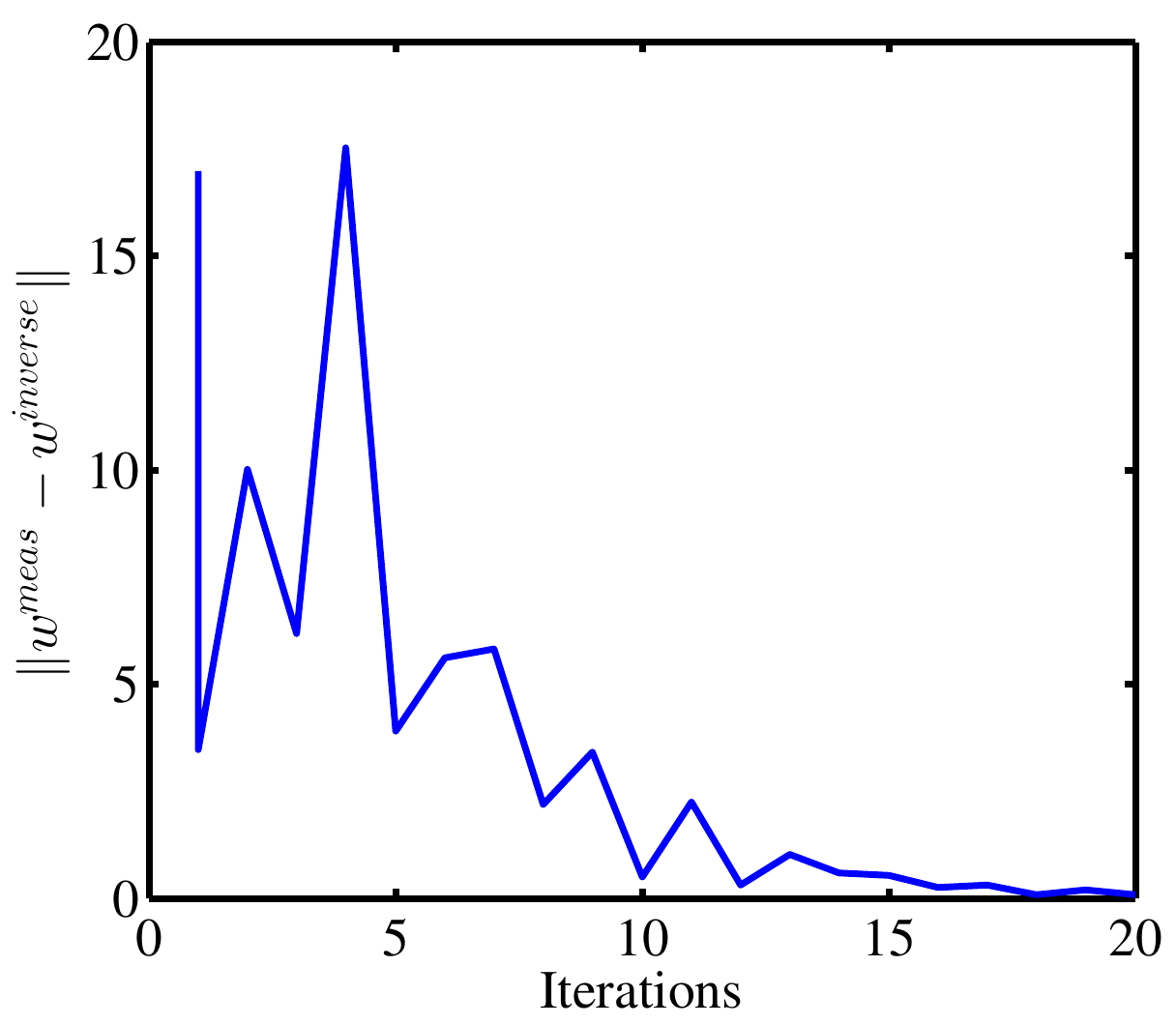}}
	\subfigure[]{\includegraphics[width = 0.45\textwidth]{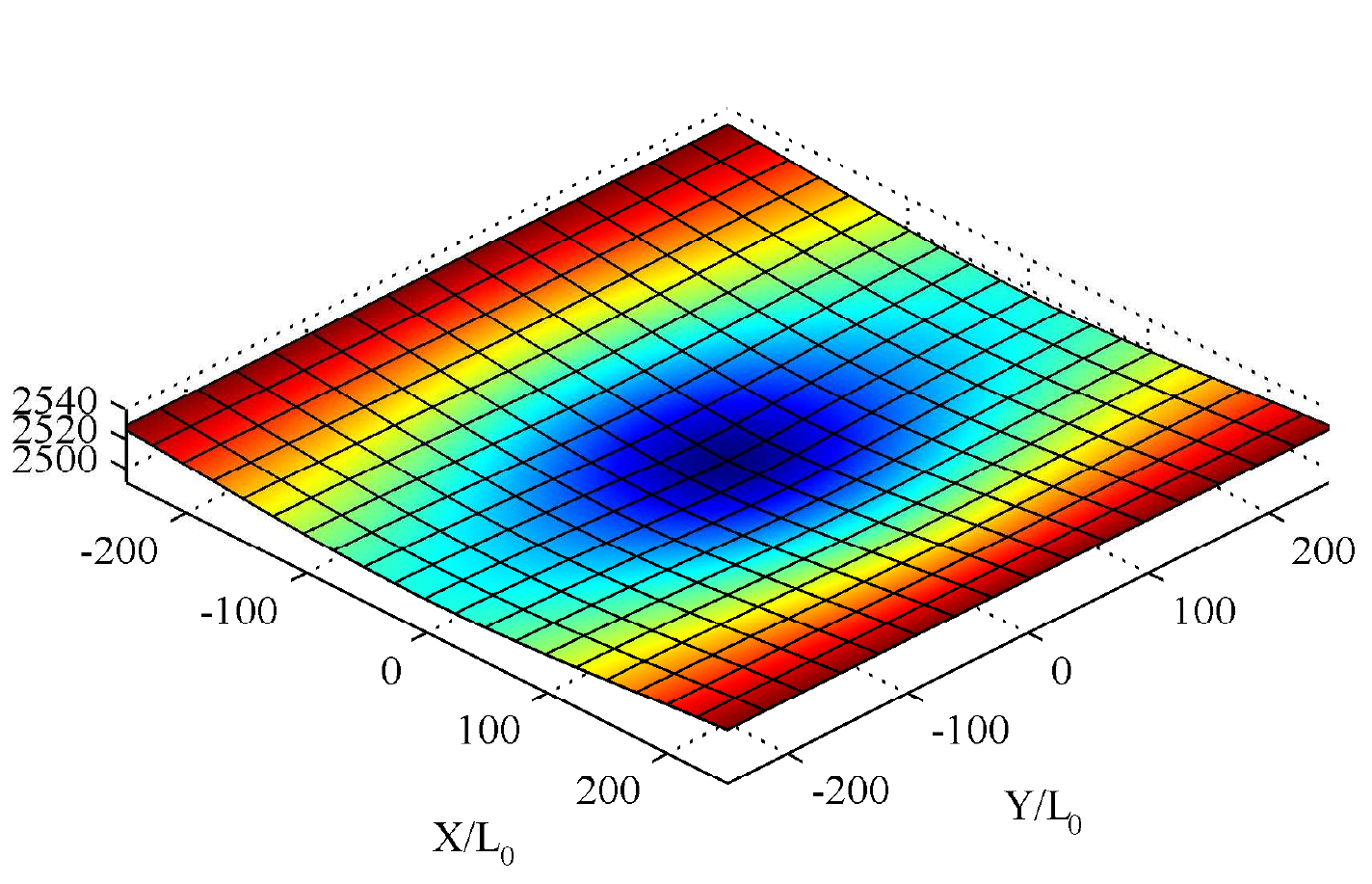}}
	\subfigure[]{\includegraphics[width = 0.45\textwidth]{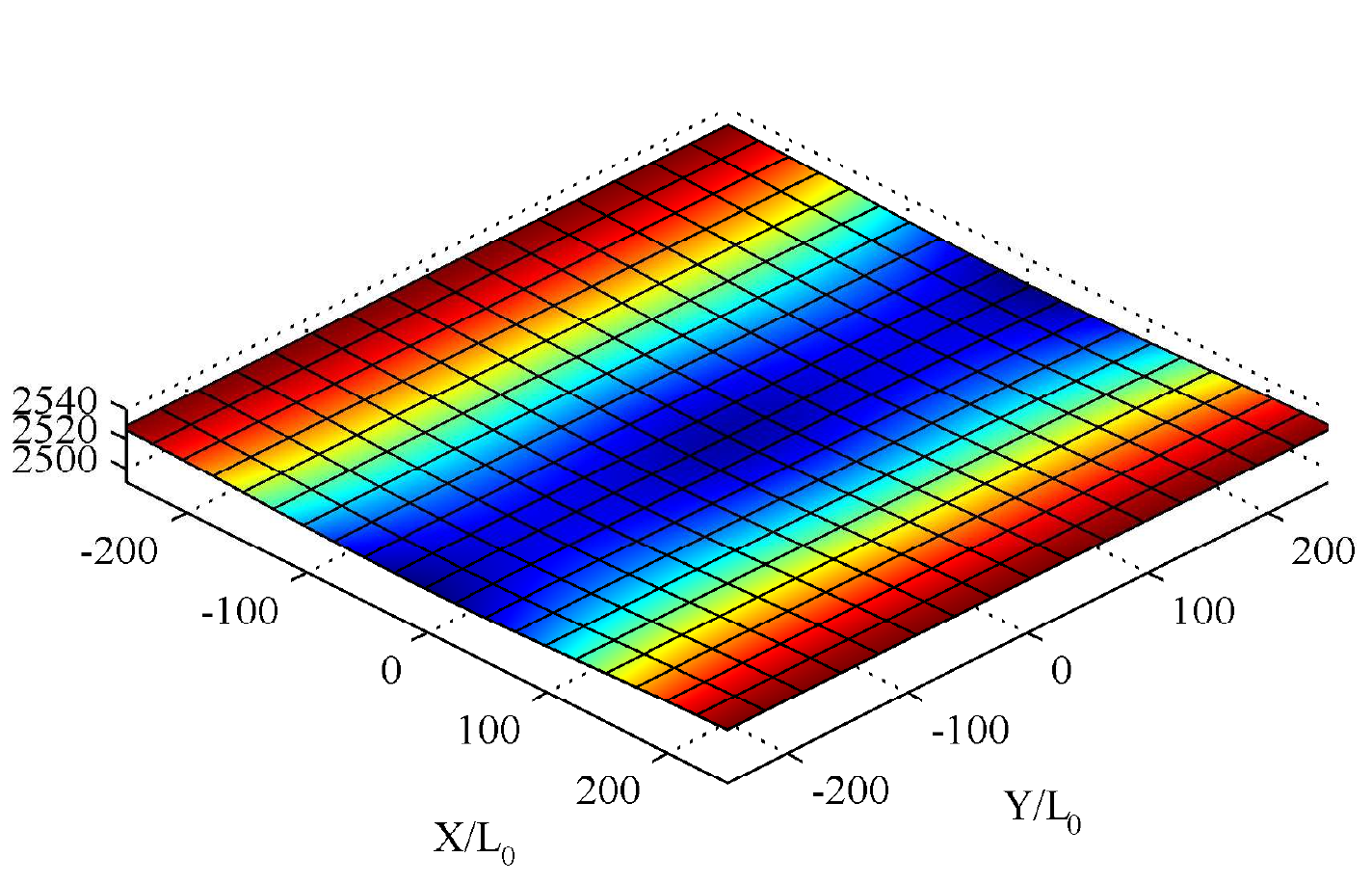}}
	\subfigure[]{\includegraphics[width = 0.45\textwidth]{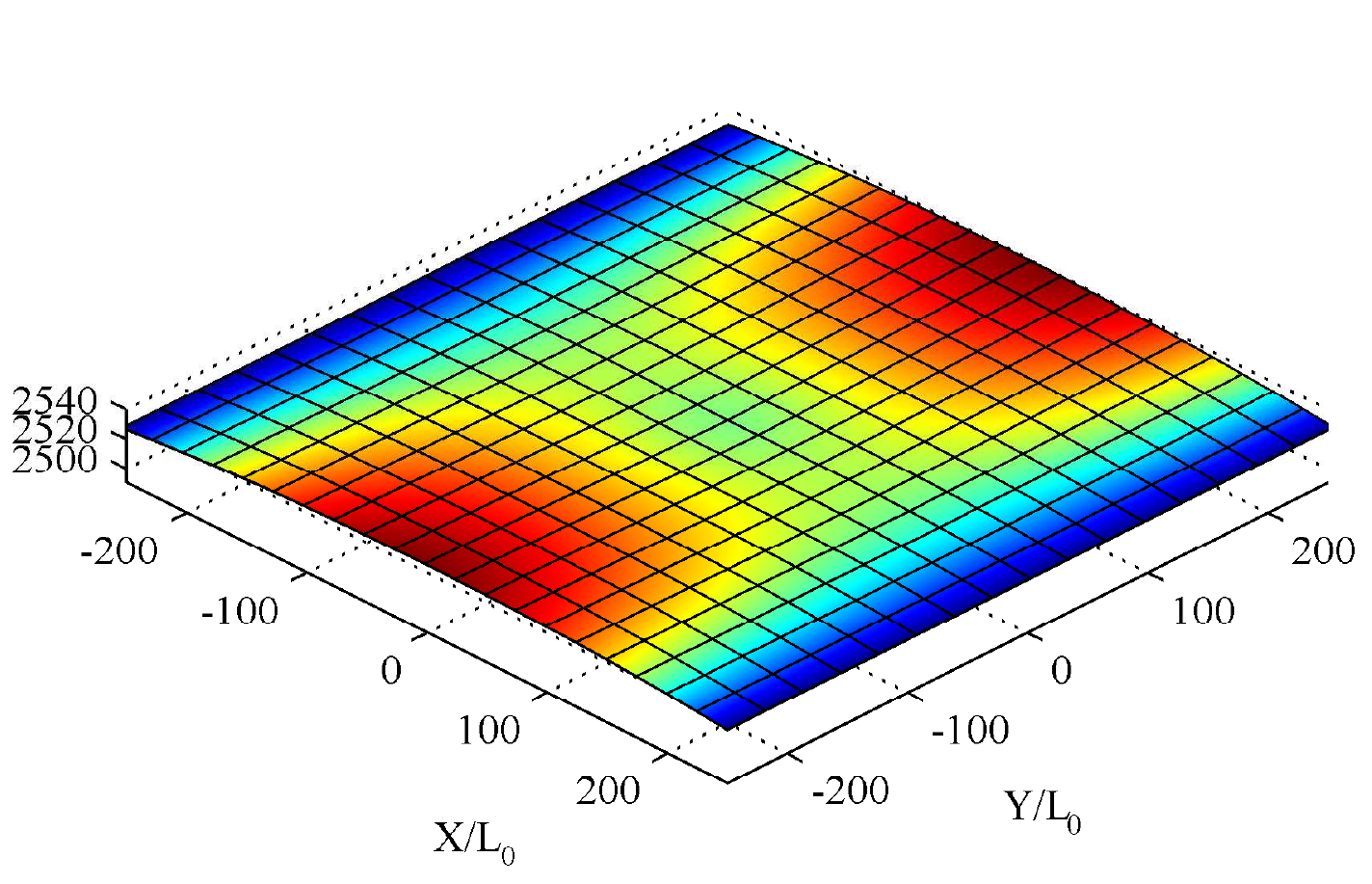}}
	\subfigure[]{\includegraphics[width = 0.45\textwidth]{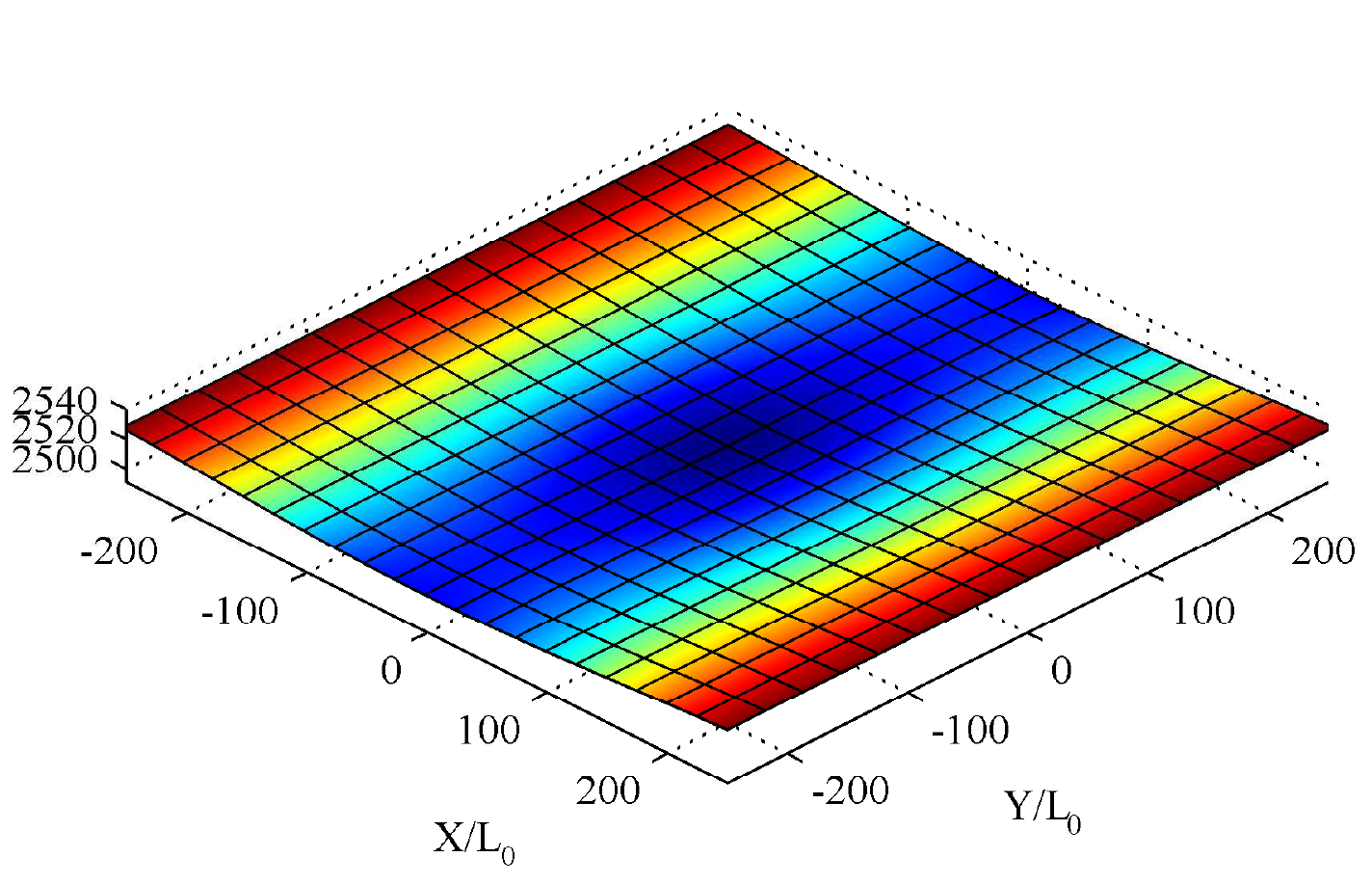}}
	\caption{Hinged cylindrical shell subjected to the displacement $w_{cen}$: (a) the objective function versus the number of iterations, (b) convergence of $L^2$ error norm in parameter space $X$ over the number of iterations, (c) the reconstructed deformation after $3$ iterations, (d) the reconstructed deformation after $4$ iterations, (e) the reconstructed deformation after $5$ iterations, (f) the reconstructed deformation after $20$ iterations.}
	\label{fig:Scoderlis_Lo_roof_inverse}
\end{center}
\end{figure}

Next, the determination of the prescribed center displacement $w_{cen}$ using the target shape shown in Figure \ref{fig:Scoderlis_Lo_roof_model}(b) is performed in the framework of compressible Neo-Hookean hyperelasticity to take nonlinear bending and stretching behaviors into account. Analytical and semi-analytical sensitivities are utilized to compute the derivatives in Equation(\ref{eq:derv_obj_func_adj}). Computation of the analytical sensitivities for prescribed displacement is provided in section \ref{appendix:sensitivities_displacement}. As shown in Figure \ref{fig:Scoderlis_Lo_inverse_analsen}, the objective function and the $L^2$ norm converge after $20$ iterations. The semi-analytical sensitivities are also used to provide the derivatives in Equation (\ref{eq:semianal_sens_anal}) based on how the inverse problem is carried out. The accuracy of the inverse method is highlighted in Figures \ref{fig:Scoderlis_Lo_roof_inverse}(a+b) where convergence is gained after $20$ iterations. The solution of the inverse problem over iterations is shown in Figures \ref{fig:Scoderlis_Lo_roof_inverse}(c+d+e+f). The reconstructed deformation after $20$ iterations shows good agreement with the target shape: $w^{meas}_{cen} = -24.98$ compares well to the maximal displacement $w^{inverse}_{cen} = -24.90$ at the central point. It should be noted that the reconstruction of the target shape is performed in the presence of a snap-through instability.

\begin{figure}[htbp]
\begin{center}
	\subfigure[]{\includegraphics[width = 0.5\textwidth]{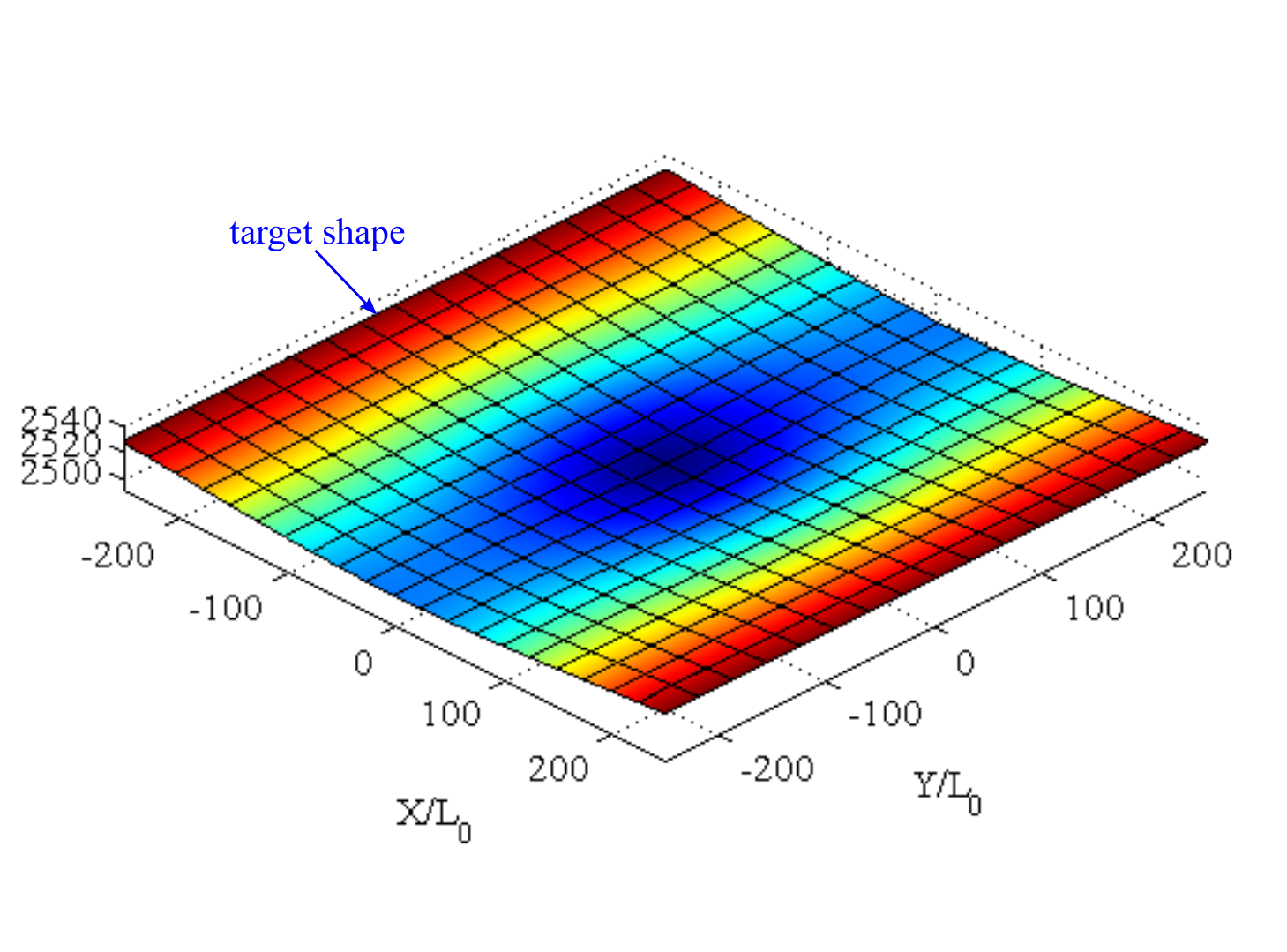}}
	\subfigure[]{\includegraphics[width = 0.45\textwidth]{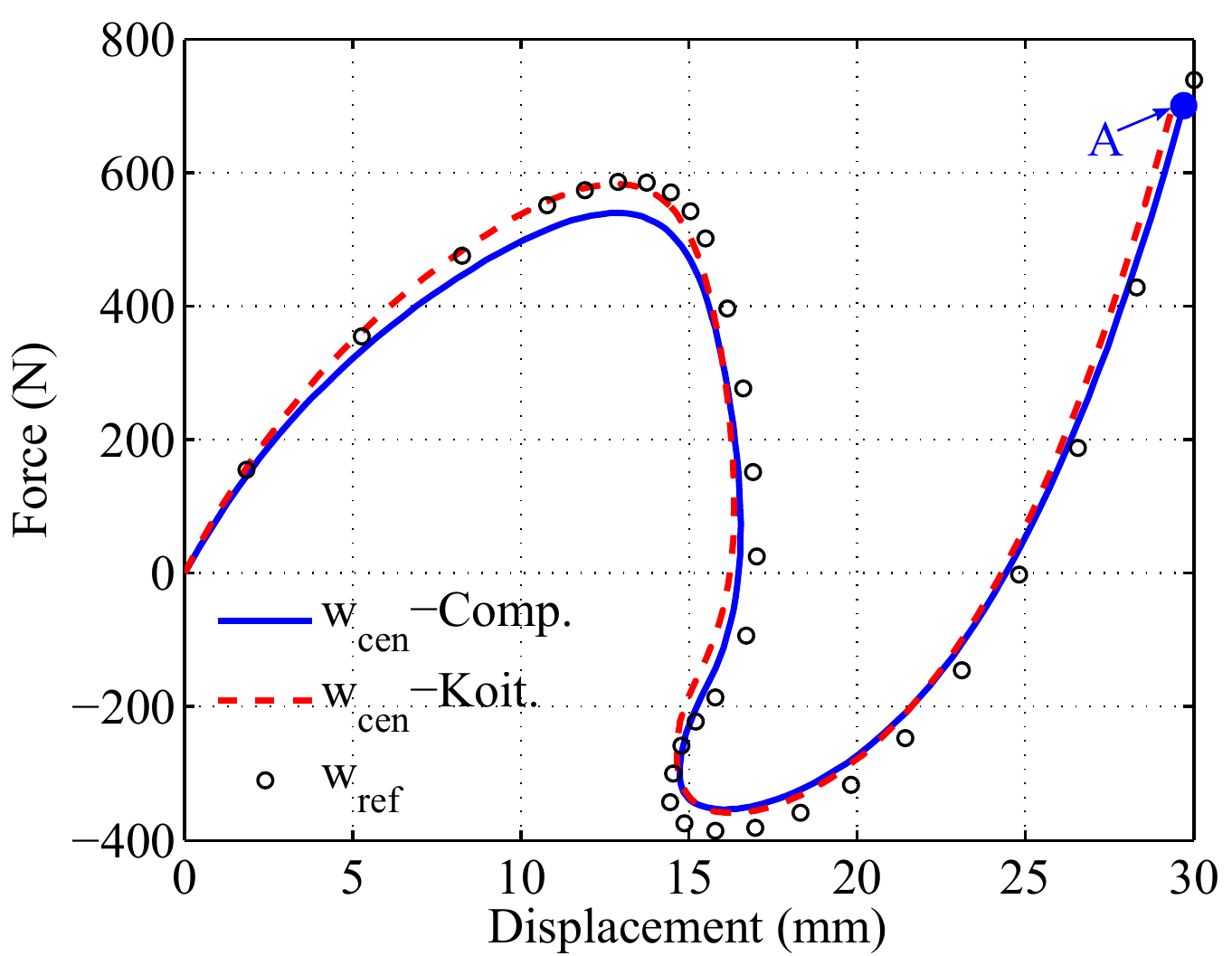}}
	\caption{Hinged cylindrical shell subject to central point load $F$: (a) deformed shape at external load $F$ using the projected shell model. This shape corresponds to point \textcolor{blue}{A} in (b), (b) reaction forces versus vertical displacement in $z$ direction for the Koiter shell model (red dashed line) and the projected shell model (blue solid line) using the compressible Neo-Hookean material. The force-displacement curves are in good agreement with the counterpart (black $\circ$) shown in \cite{Sze:2004}.}
	\label{fig:Scoderlis_Lo_snapback}
\end{center}
\end{figure}

In order to take the uniqueness of the solution into consideration, the hinged shell problem with smaller thickness ($T = 6.35 ~mm$) and snap-back behavior is also simulated. For this model, the prescribed center displacement $w_{cen}$ in Figure \ref{fig:Scoderlis_Lo_roof_model} is replaced by a central point load $F = 700 ~N$. The forward problem is implemented using the arc length method. The force-displacement responses for the Koiter model and the projected model with compressible Neo-Hookean material are shown in Figure \ref{fig:Scoderlis_Lo_snapback}. The target configuration corresponding to point \textcolor{blue}{A} on the blue force-displacement curve is obtained and used for the identification of the applied external load.

\begin{figure}[htbp]
\begin{center}
	\subfigure[]{\includegraphics[width = 0.475\textwidth]{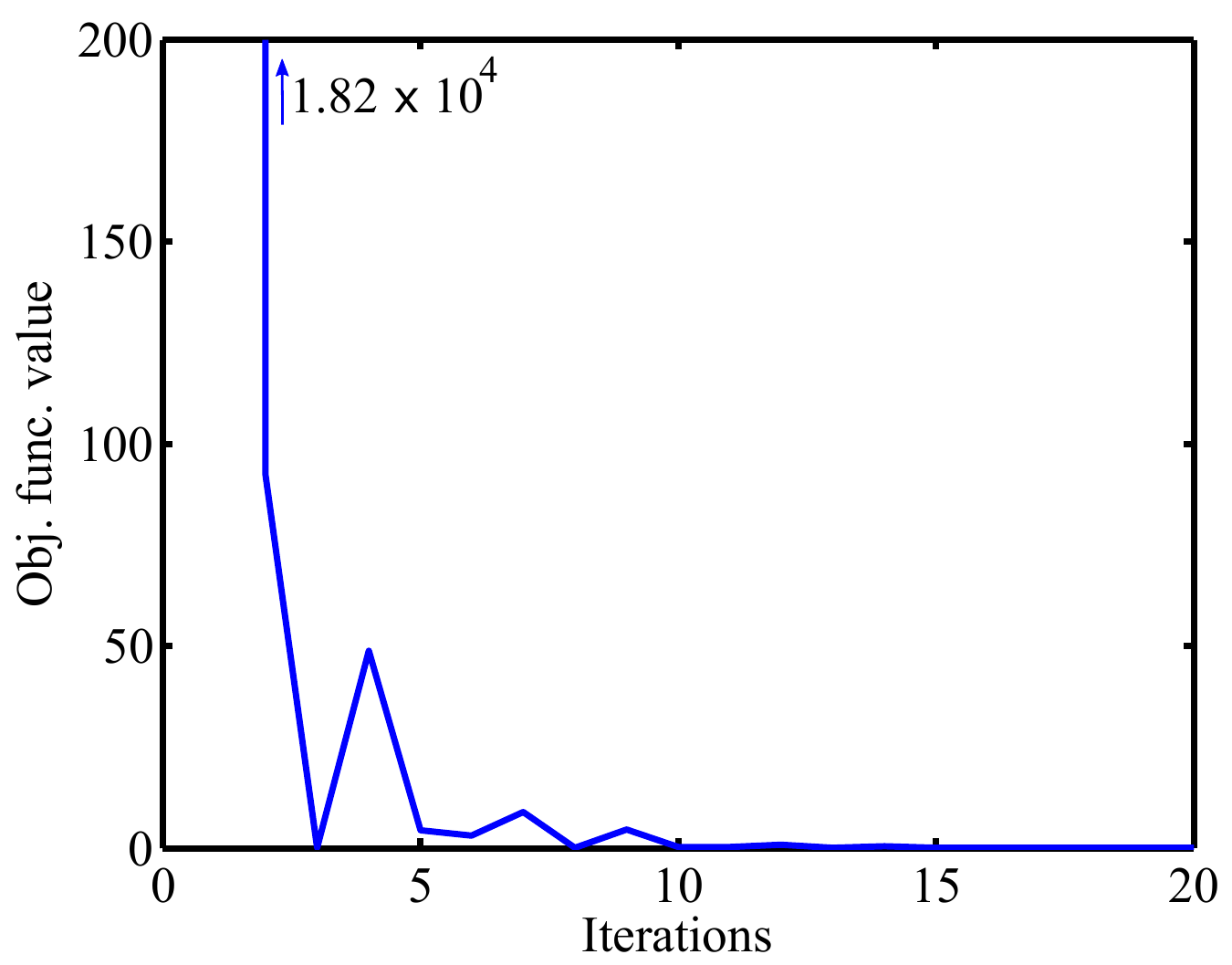}}
	\subfigure[]{\includegraphics[width = 0.435\textwidth]{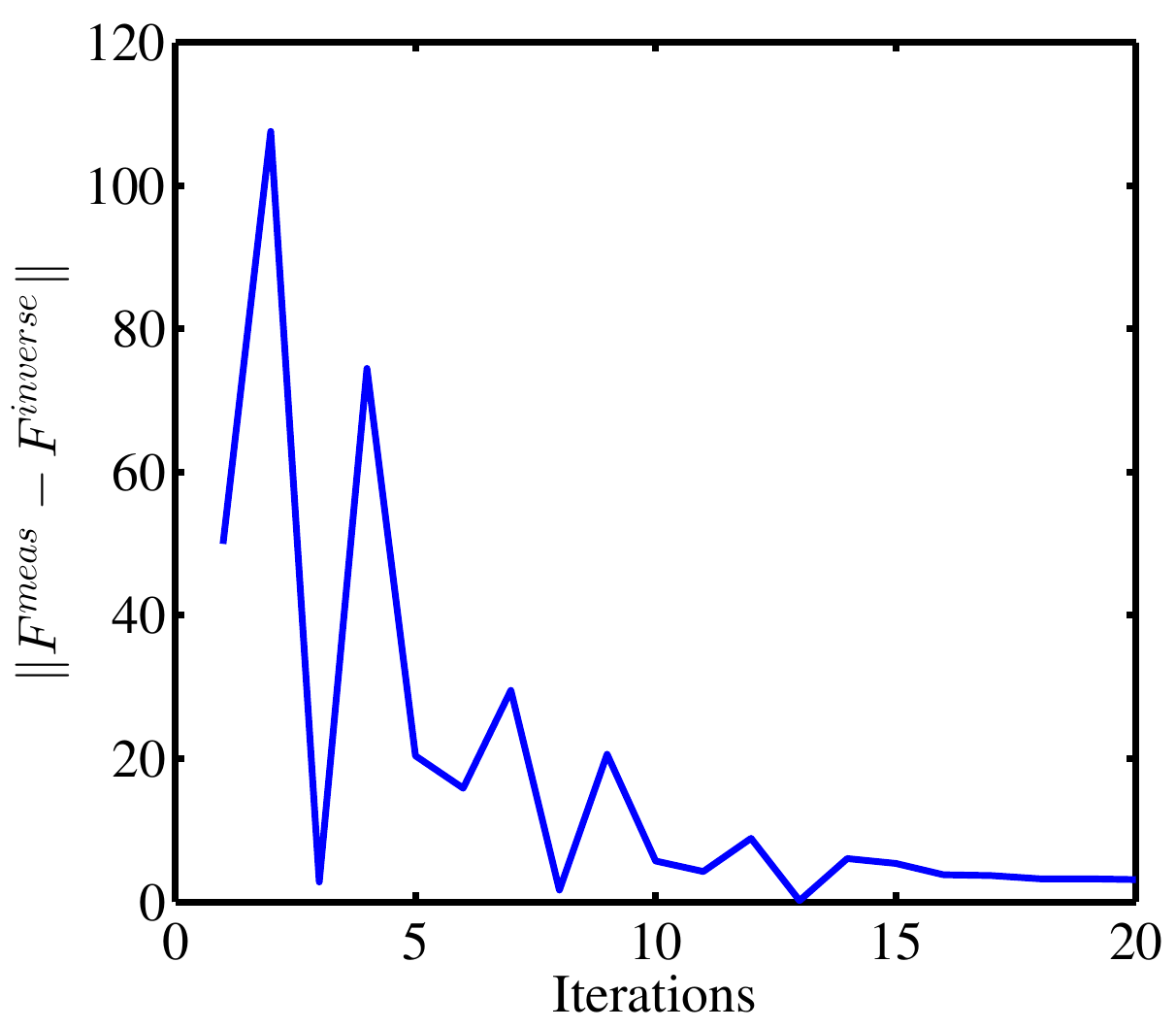}}
	\subfigure[]{\includegraphics[width = 0.45\textwidth]{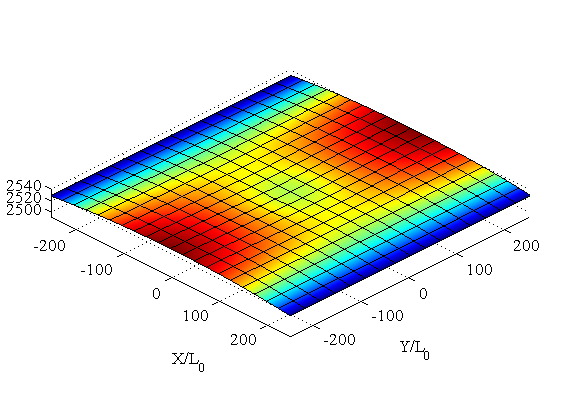}}
	\subfigure[]{\includegraphics[width = 0.435\textwidth]{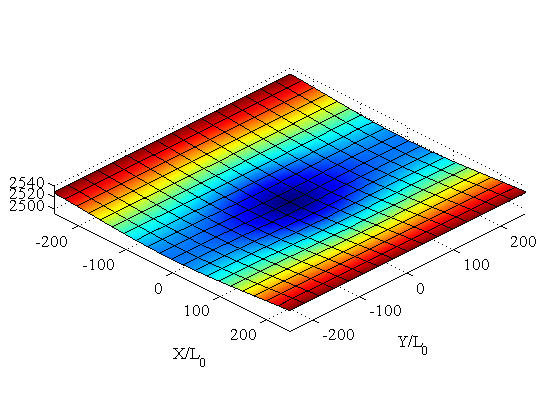}}
	\caption{Hinged cylindrical shell subjected to the central point load $F$: (a) the objective function versus the number of iterations, (b) convergence of $L^2$ error norm in parameter space $X$ over the number of iterations, (c) the reconstructed deformation after $1$ iterations, (d) the reconstructed deformation after $20$ iterations.}
	\label{fig:Scoderlis_Lo_inverse_snapback}
\end{center}
\end{figure}

Inverse analysis using the analytical sensitivity presented in Equation (\ref{eq:fstartnln_pointload}) is then performed based on the experiment-like data obtained from the forward problem. The projected model using the compressible Neo-Hookean material is employed. The solution to the inverse problem is obtained after $20$ iterations as depicted in Figure \ref{fig:Scoderlis_Lo_inverse_snapback}. The reconstructed configuration shown in Figure \ref{fig:Scoderlis_Lo_inverse_snapback}(d) under the inverse solution $F^{inverse} = 678.3 ~N$ shows good agreement with the target configuration. It should be noted that this configuration is reconstructed under a snap-back instability.

\subsection{A flat strip subjected to compressive displacement and bending moment} \label{subsec:ex3}

In reality, a perfect structure with perfect loading does not exist. Inevitable imperfections may be related to factors such as initial curvature, load eccentricity or small disturbing loads. Small imperfections do not matter in linear analysis, however in nonlinear problems, they are quite important. Hence, instabilities of shell structures that are sensitive to geometrical and loading imperfections must be considered. The shape changes occurring due to nonlinear buckling will be examined in the following examples.

\begin{figure}[htbp]
\begin{center}
	\subfigure[]{\includegraphics[width = 0.6\textwidth]{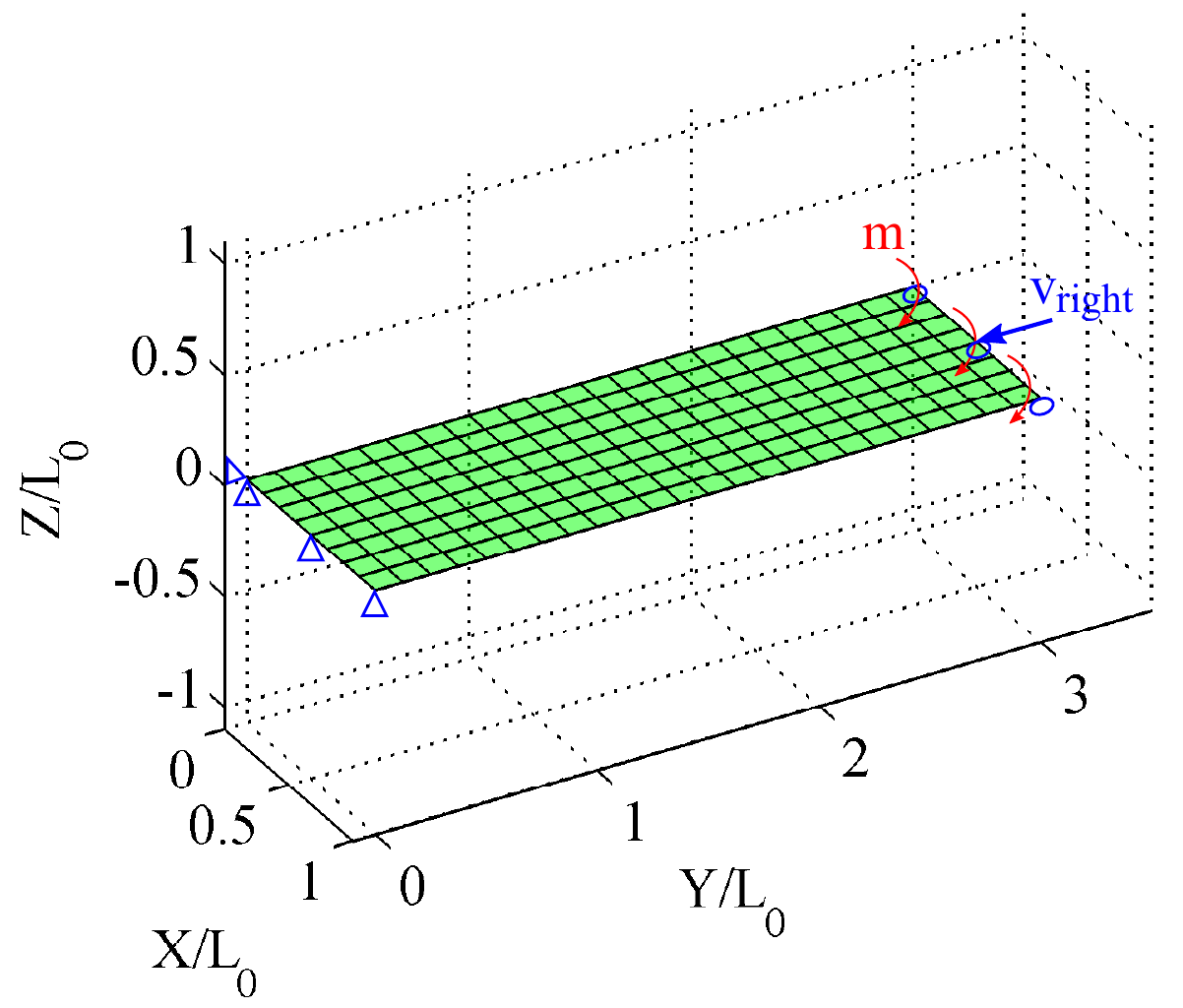}} \\
	\subfigure[]{\includegraphics[width = 0.4625\textwidth]{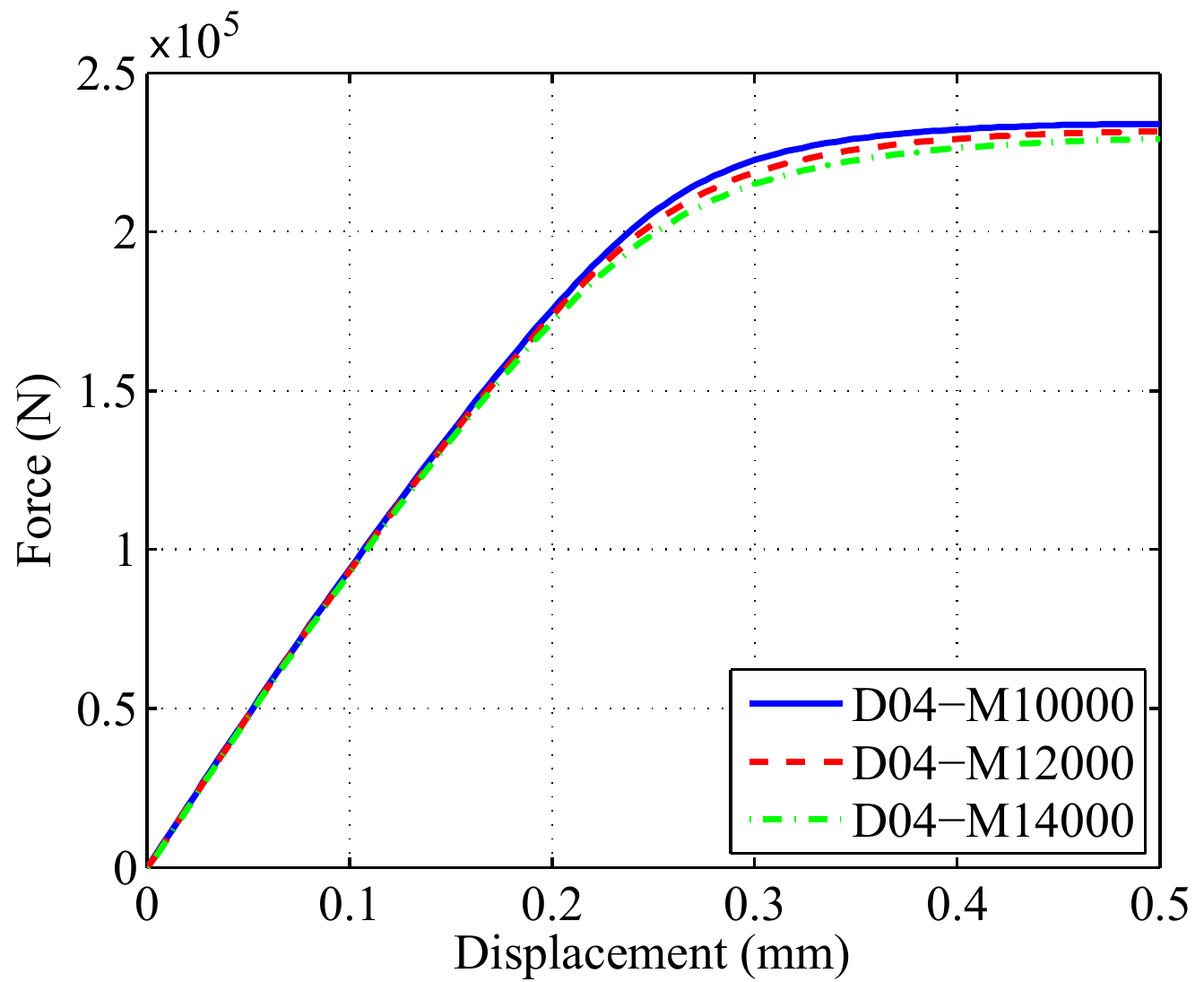}}
	\subfigure[]{\includegraphics[width = 0.45\textwidth]{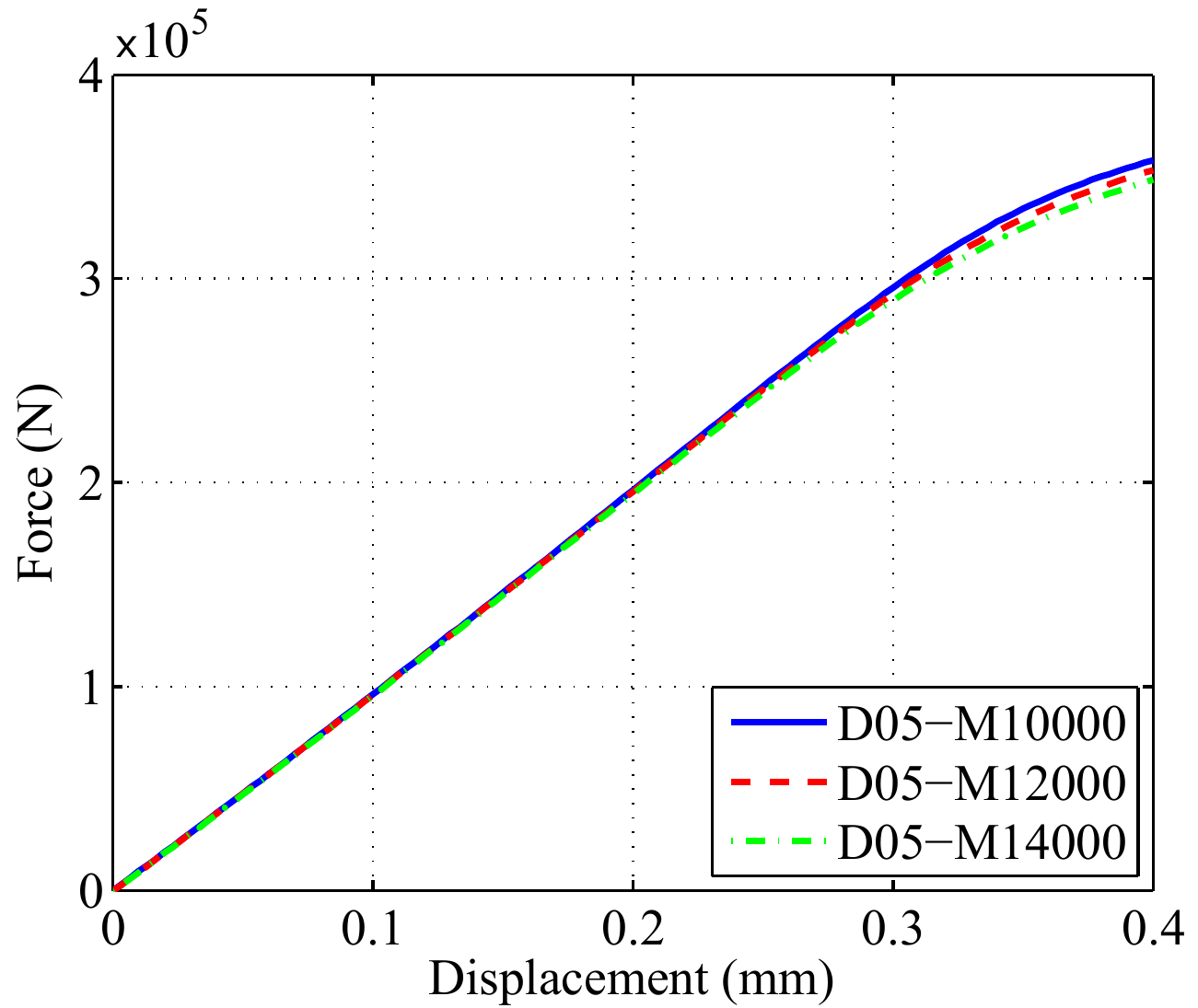}}
	\caption{Flat strip subjected to compressive displacement $v_{right}$ and bending moment $m$: (a) undeformed shape, (b) axial force-displacement responses for different moments $m$ modeled by the Koiter model, (c) axial force-displacement responses for different moments $m$ modeled by the projected shell model.}
	\label{fig:column_eccentric_load_model}
\end{center}
\end{figure}

\begin{figure}[htbp]
\begin{center}
	\subfigure[]{\includegraphics[width = 0.5\textwidth]{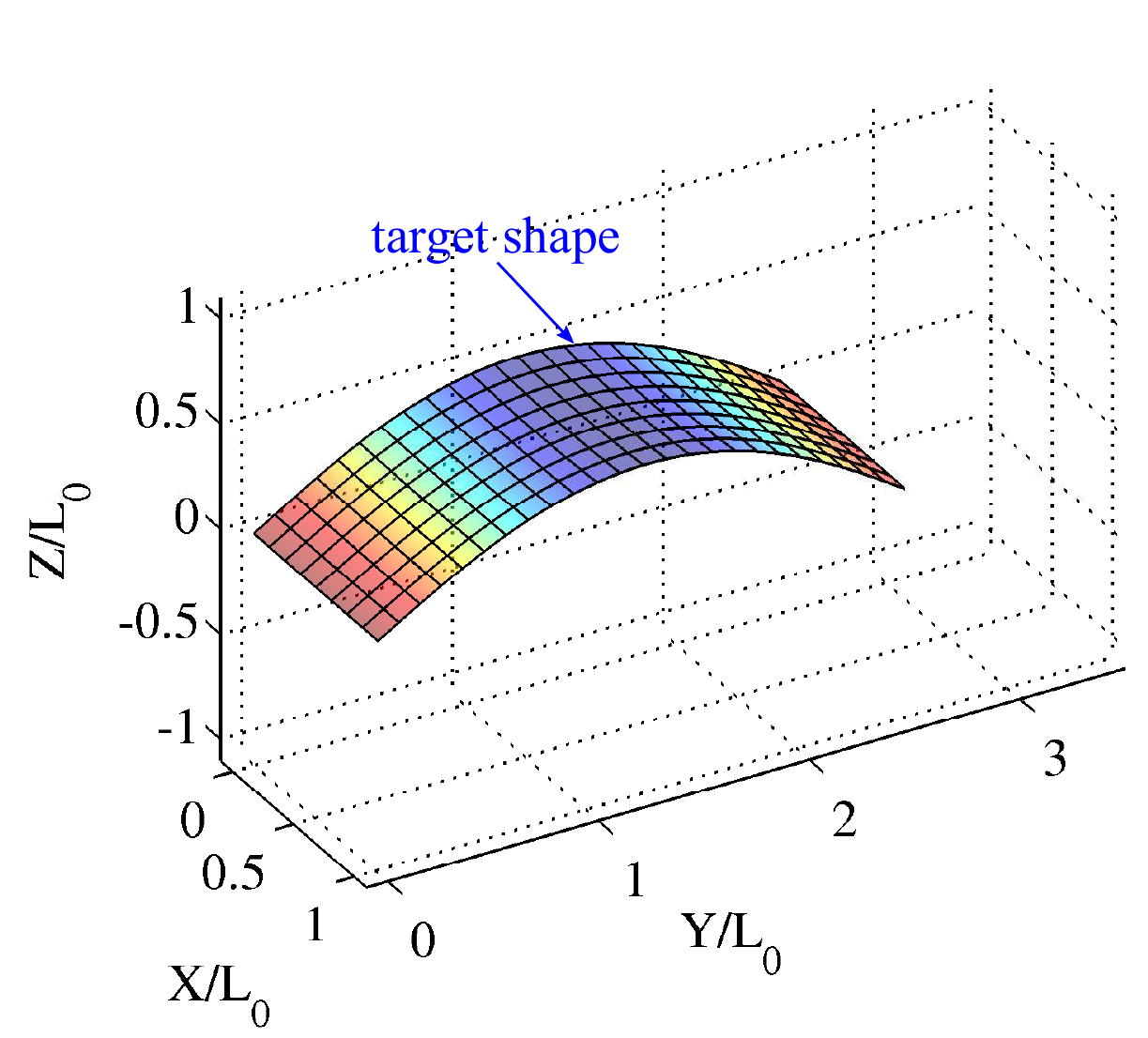}}
	\subfigure[]{\includegraphics[width = 0.45\textwidth]{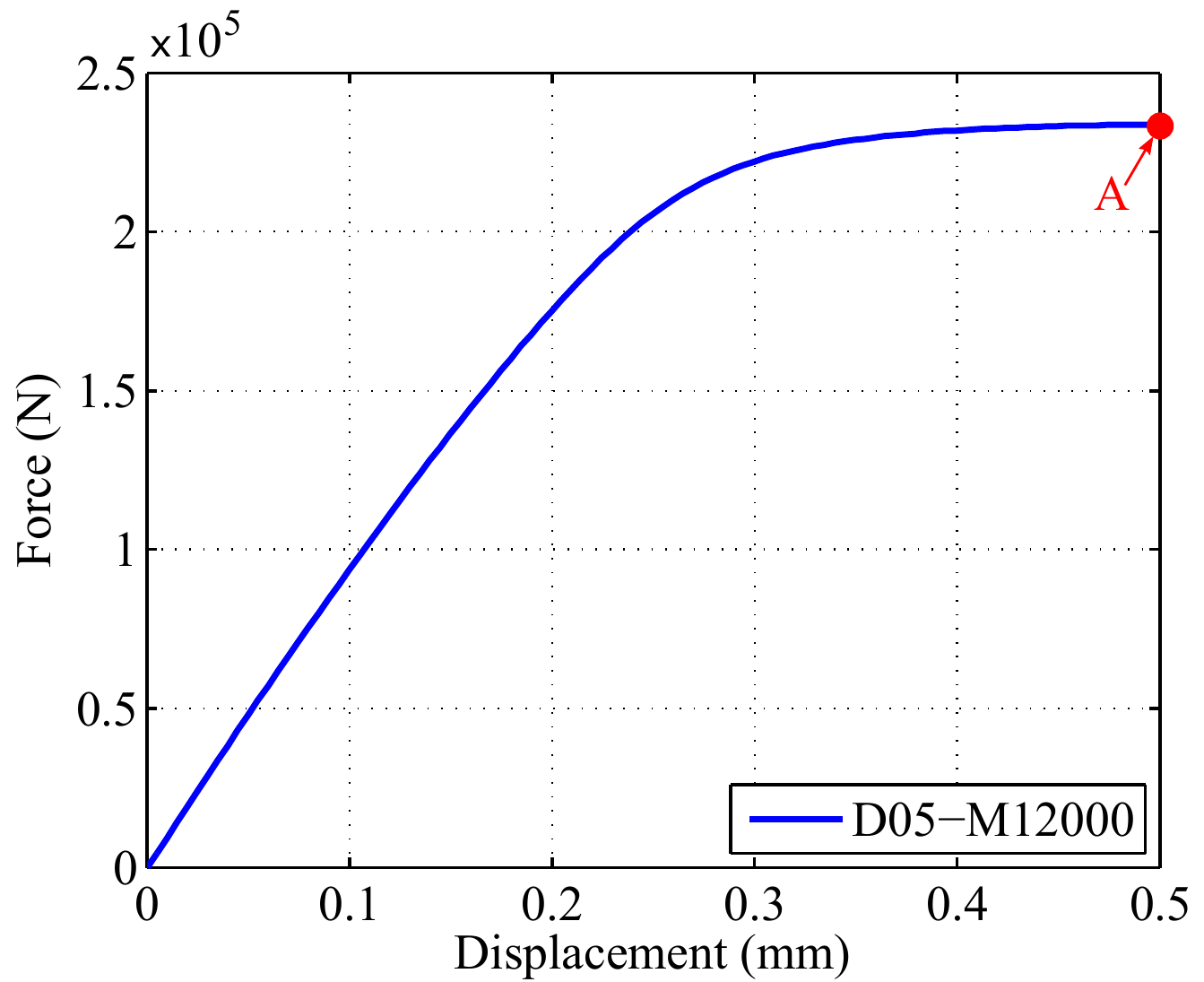}}
	\caption{Flat strip under the compressive displacement $v_{right}$ and the bending moment $m$ is modeled by the Koiter model: (a) deformed shape at displacement $v^{meas}_{Koit,right} = -0.5$, (b) reaction force versus displacement in $y$ direction.}
	\label{fig:column_eccetric_load_FEM_Koiter}
\end{center}
\end{figure}

\begin{figure}[htbp]
\begin{center}
	\subfigure[]{\includegraphics[width = 0.5\textwidth]{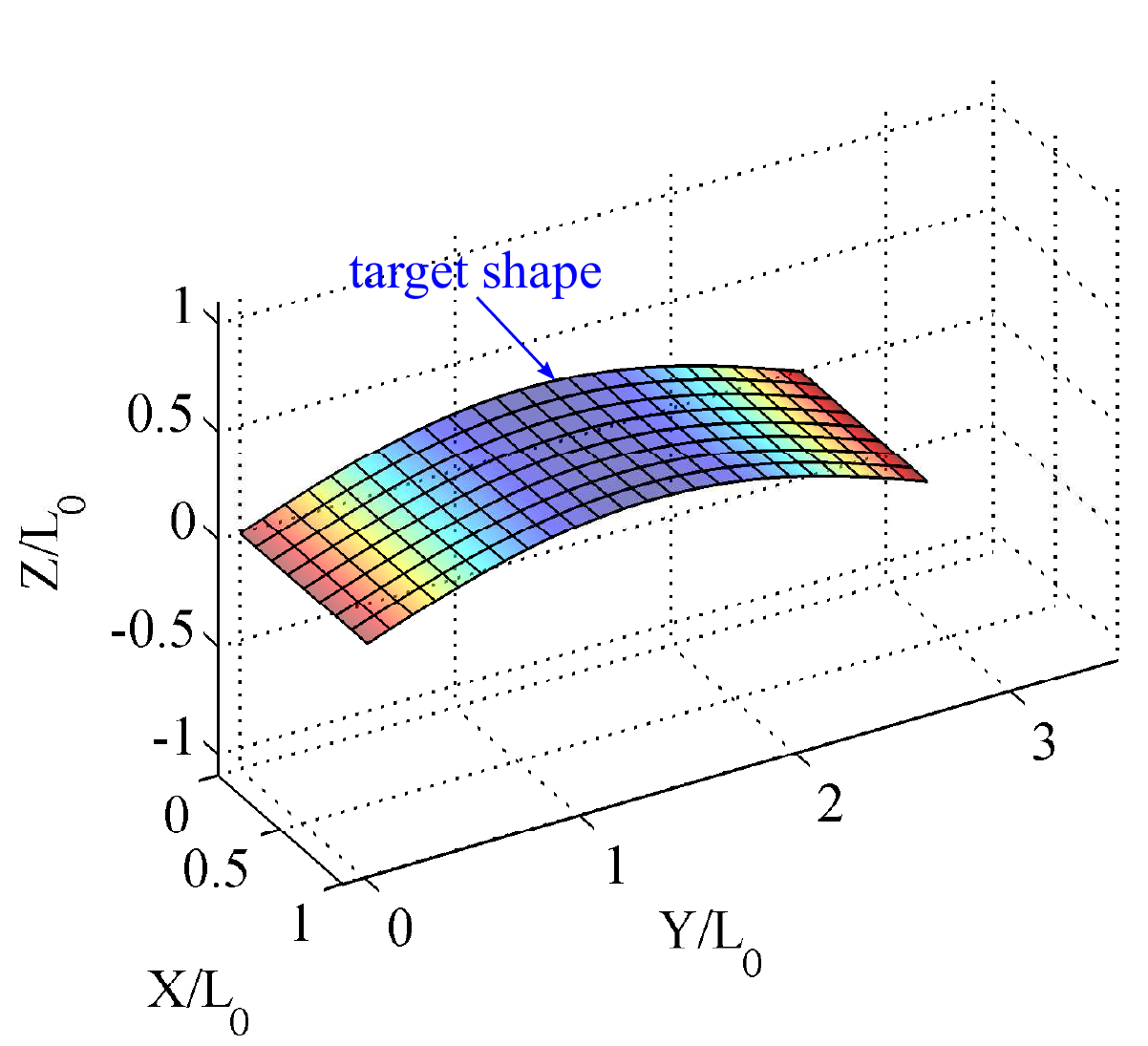}}
	\subfigure[]{\includegraphics[width = 0.45\textwidth]{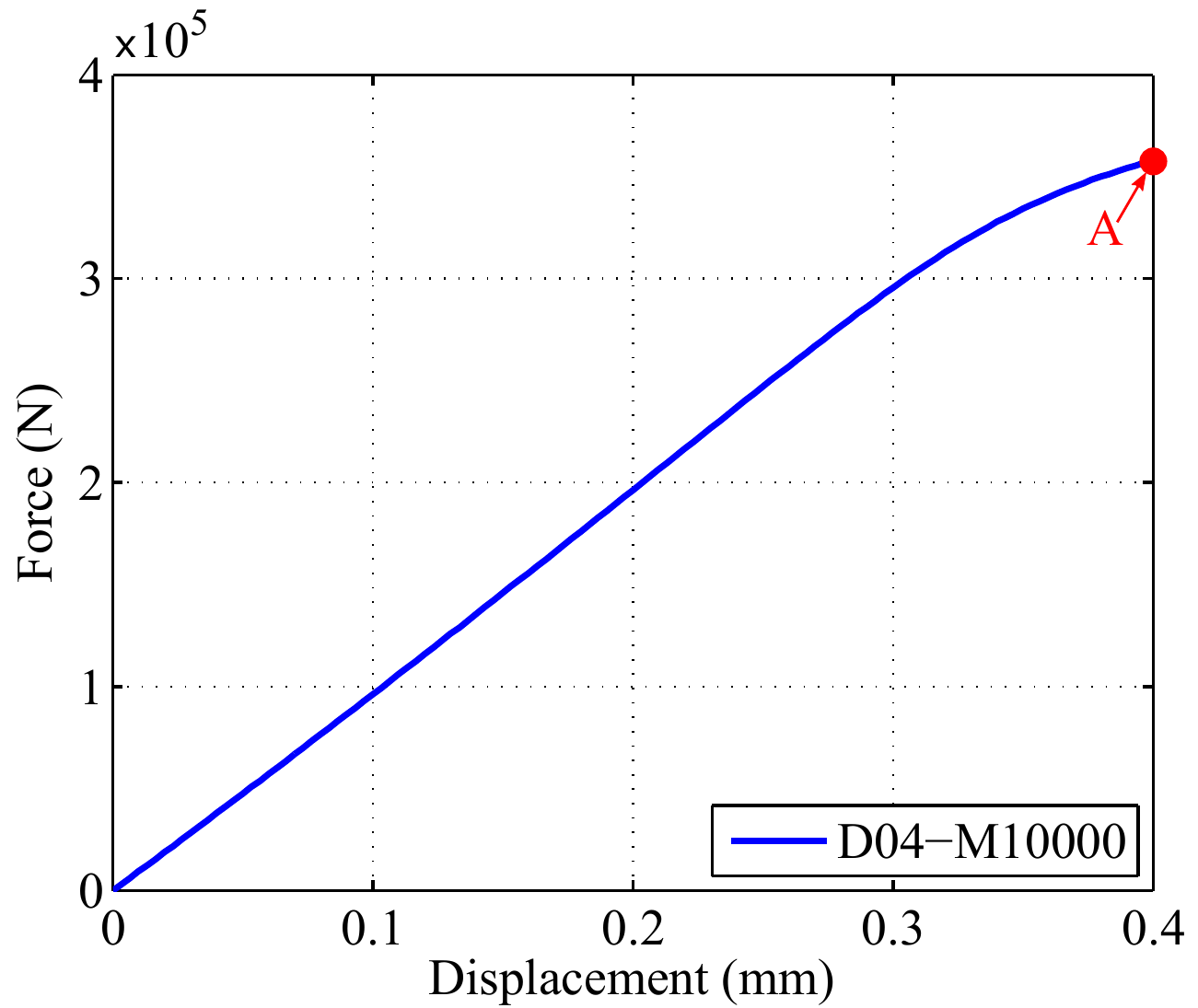}}
	\caption{Flat strip under the compressive displacement $v_{right}$ and the bending moment $m$ is modeled by the projected model: (a) deformed shape at displacement $v^{meas}_{proj,right} = -0.4$, (b) reaction force versus displacement in $y$ direction}
	\label{fig:column_eccentric_load_FEM_projected}
\end{center}
\end{figure}

We consider a pinned-end flat strip that it is perfectly straight before the displacement is prescribed. The strip is prescribed by compressive displacement and end distributed moment as described in Figure \ref{fig:column_eccentric_load_model}. The moment $m$ is applied first and we then apply the compressive displacement $v_{right}$. A FE solution for of $8 \times 24$ quadratic NURBS elements is first determined for the above structure with both the Koiter and projected shell models using the compressible Neo-Hookean formulation. The 3D Young's modulus $E = 3 \times 10^6 ~N/mm^2$ and Poisson's ratio $\nu = 0.3$ are adopted. The FE force vector associated with the distributed moment is given by Equation (\ref{eq:ext_force}) and the corresponding tangent matrix is given by Equation (\ref{eq:ext_moment_tangent_matrix}.1). The effect of the distributed moment on the force-displacement response for both models is indicated in Figures \ref{fig:column_eccentric_load_model}(b+c), respectively. The unstably buckled shape under $v^{meas}_{Koit,right} = -0.5$ and $m^{meas}_{Koit} = 12 \times 10^3 ~N.mm/mm$ is illustrated in Figure \ref{fig:column_eccetric_load_FEM_Koiter} for the Koiter model. Furthermore, Figure \ref{fig:column_eccentric_load_FEM_projected} shows the unstably buckled shape under $v^{meas}_{proj,right} = -0.4$ and $m^{meas}_{proj} = 10 \times 10^3 ~N.mm/mm$ for the projected model. The nodal displacements $\bm{u}^{meas}$ corresponding to point A on the curves in Figures \ref{fig:column_eccetric_load_FEM_Koiter}(b) and \ref{fig:column_eccentric_load_FEM_projected}(b) are then computed which are used as the experiment-like data.

\begin{figure}[htbp]
\begin{center}
\vspace*{-2cm}
	\subfigure[]{\includegraphics[width = 0.45\textwidth]{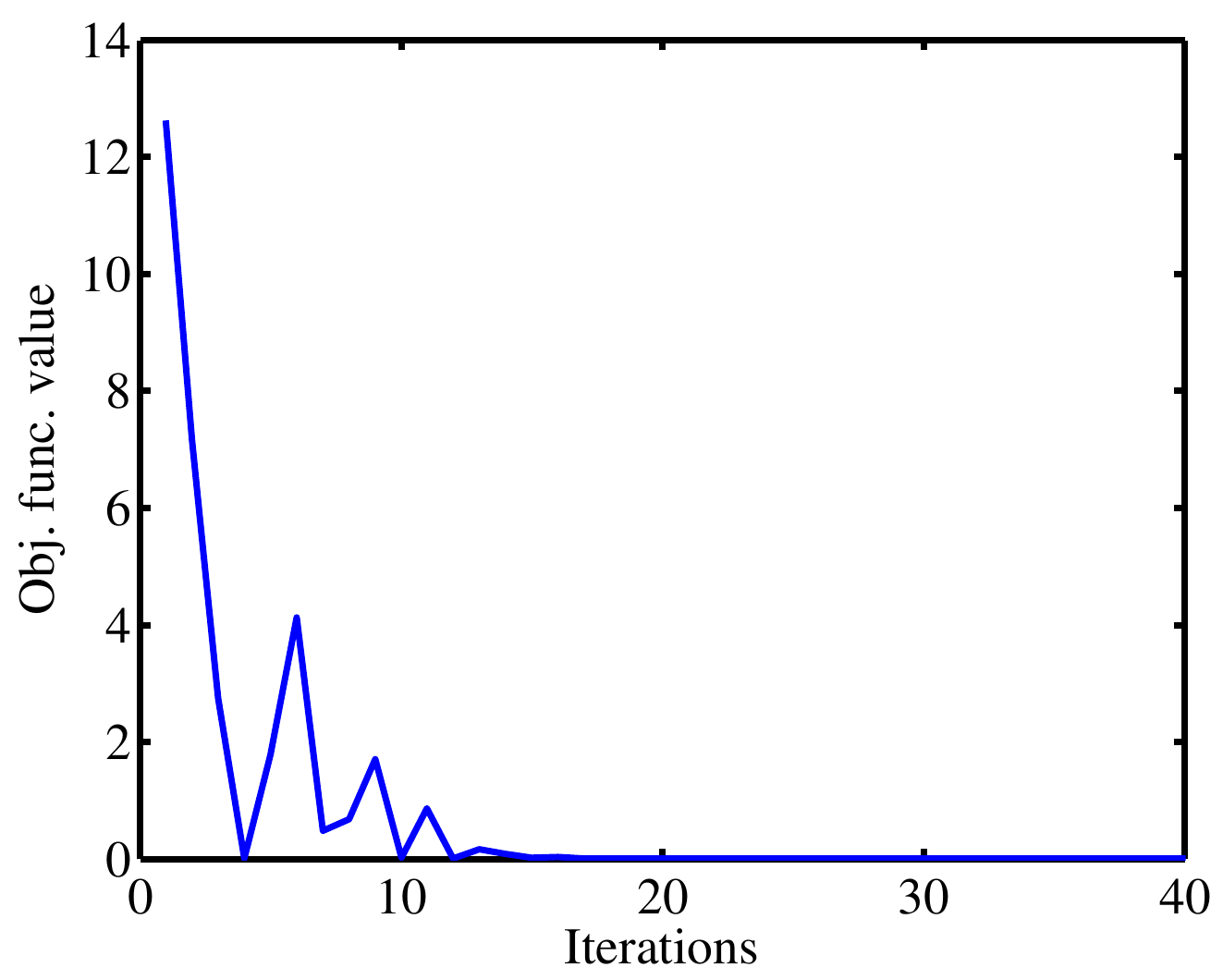}}
	\subfigure[]{\includegraphics[width = 0.45\textwidth]{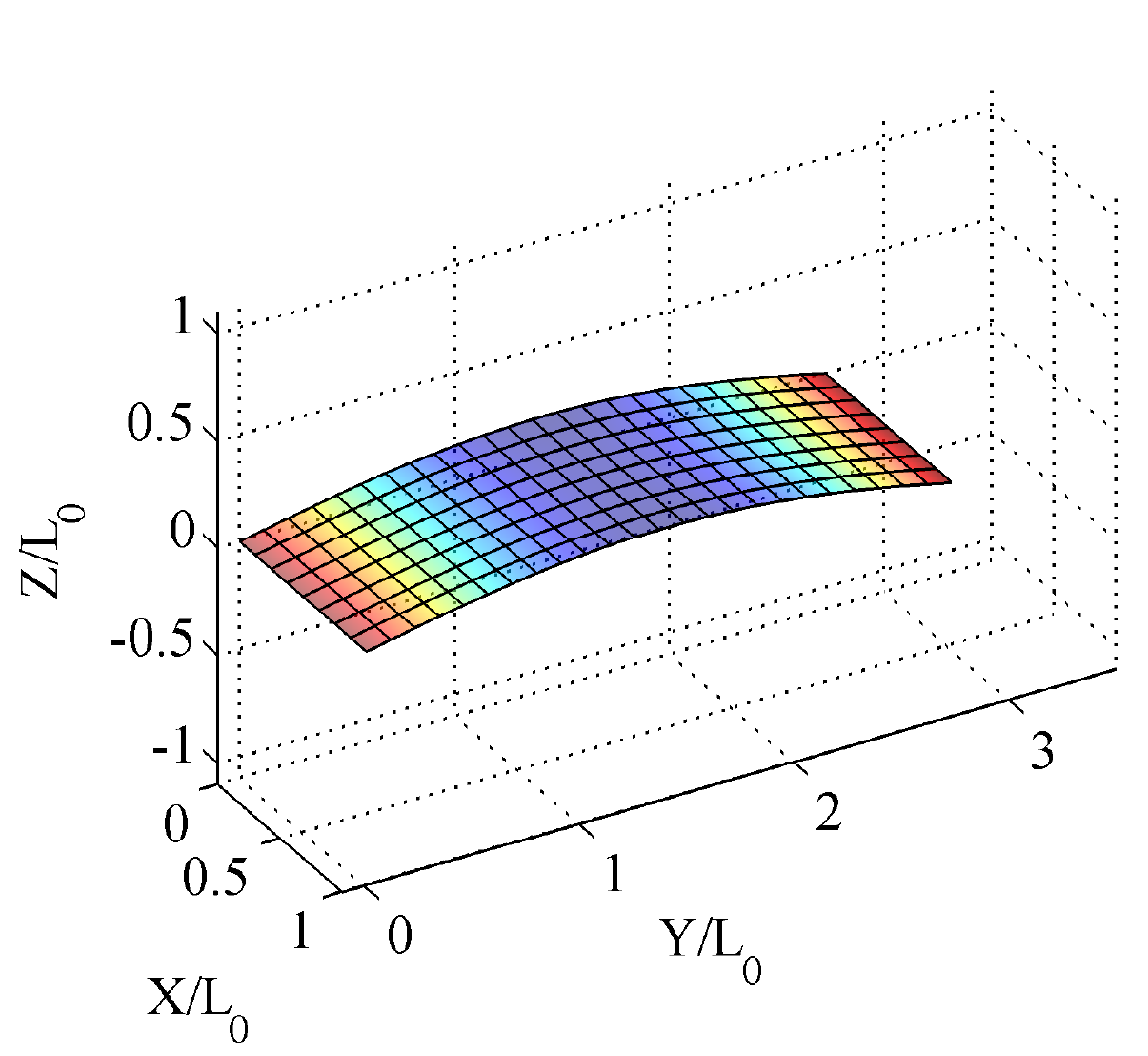}} \\
\vspace*{-0.5cm}
	\subfigure[]{\includegraphics[width = 0.45\textwidth]{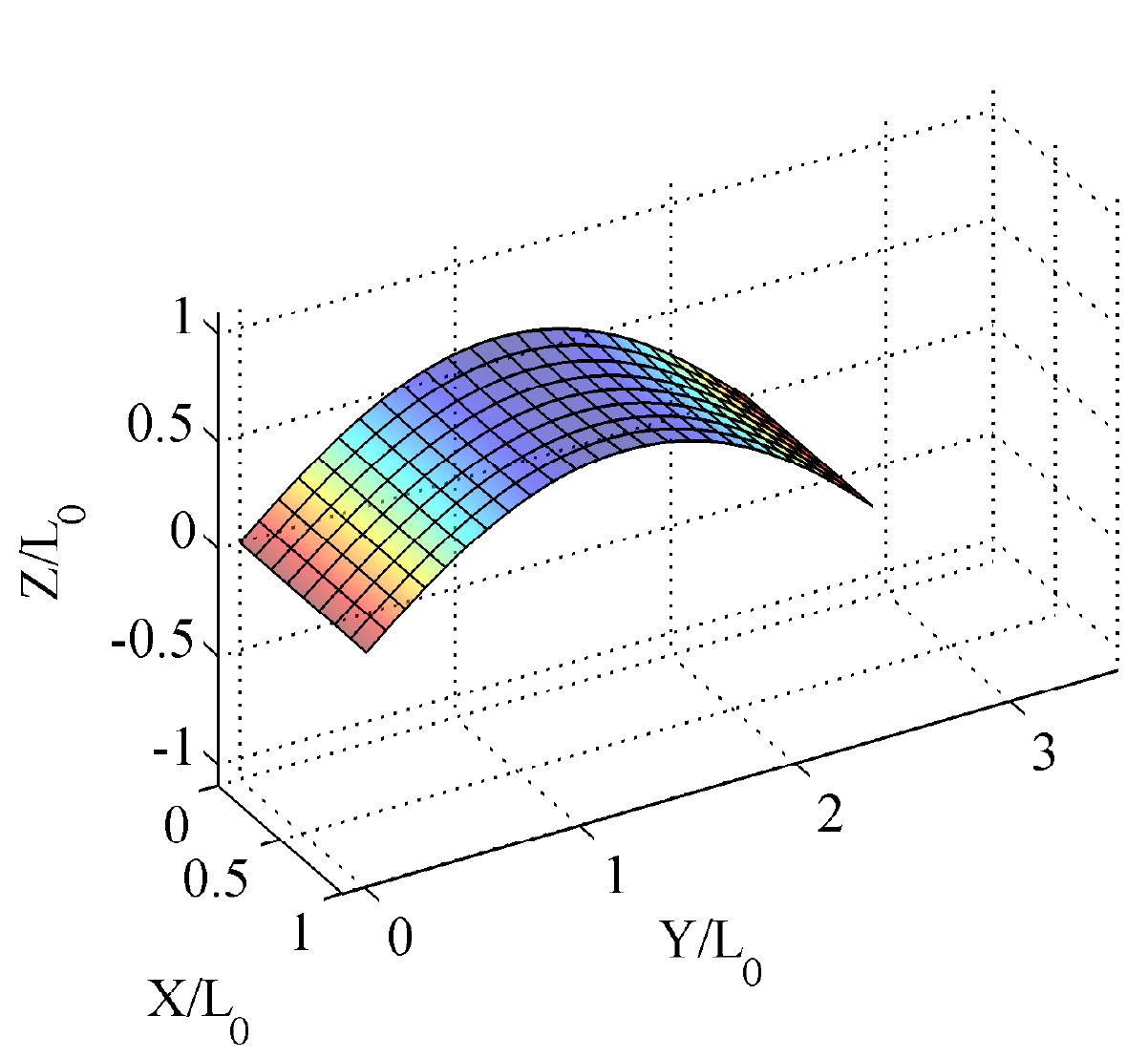}}
	\subfigure[]{\includegraphics[width = 0.45\textwidth]{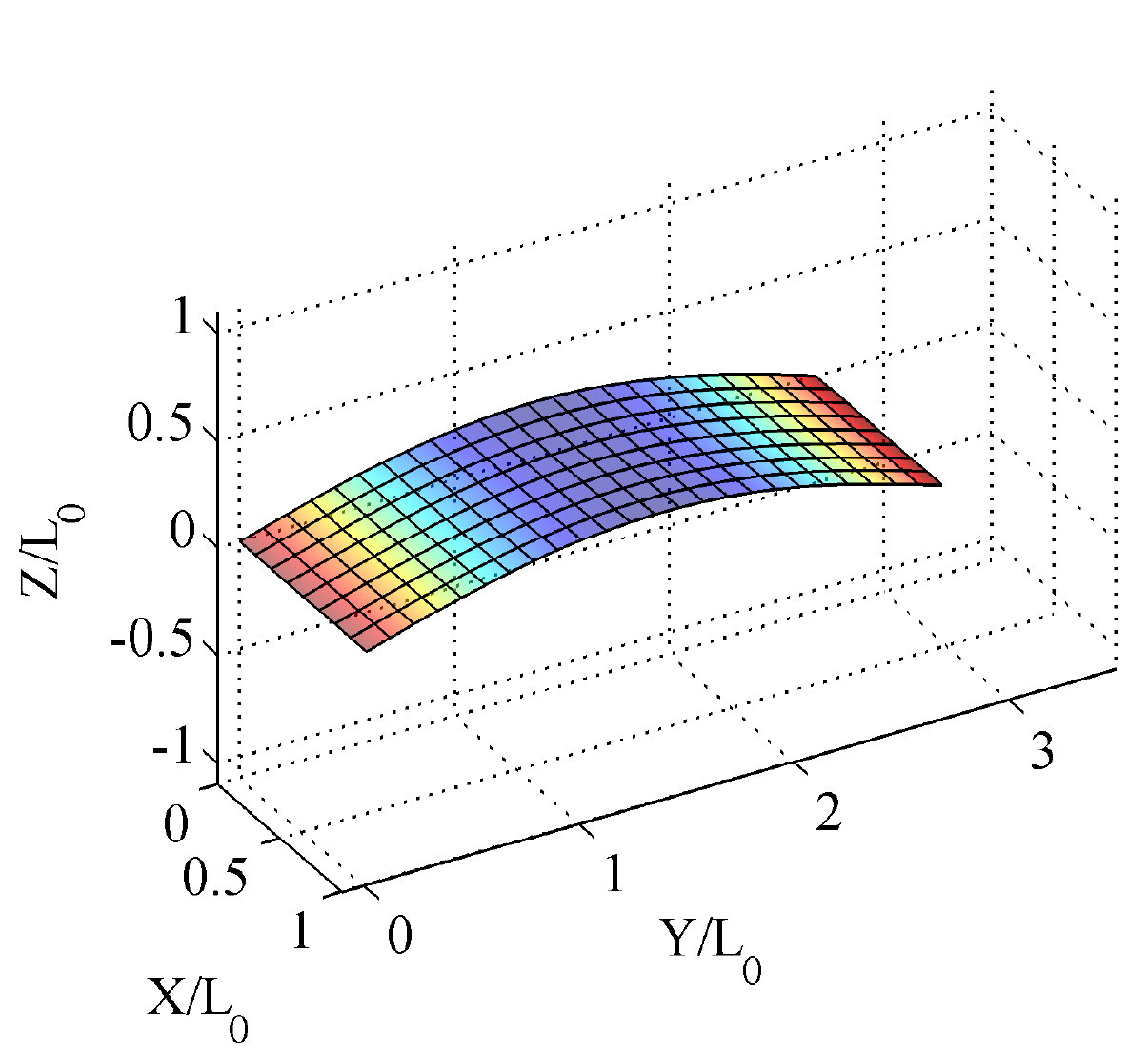}} \\
\vspace*{-0.5cm}
	\subfigure[]{\includegraphics[width = 0.45\textwidth]{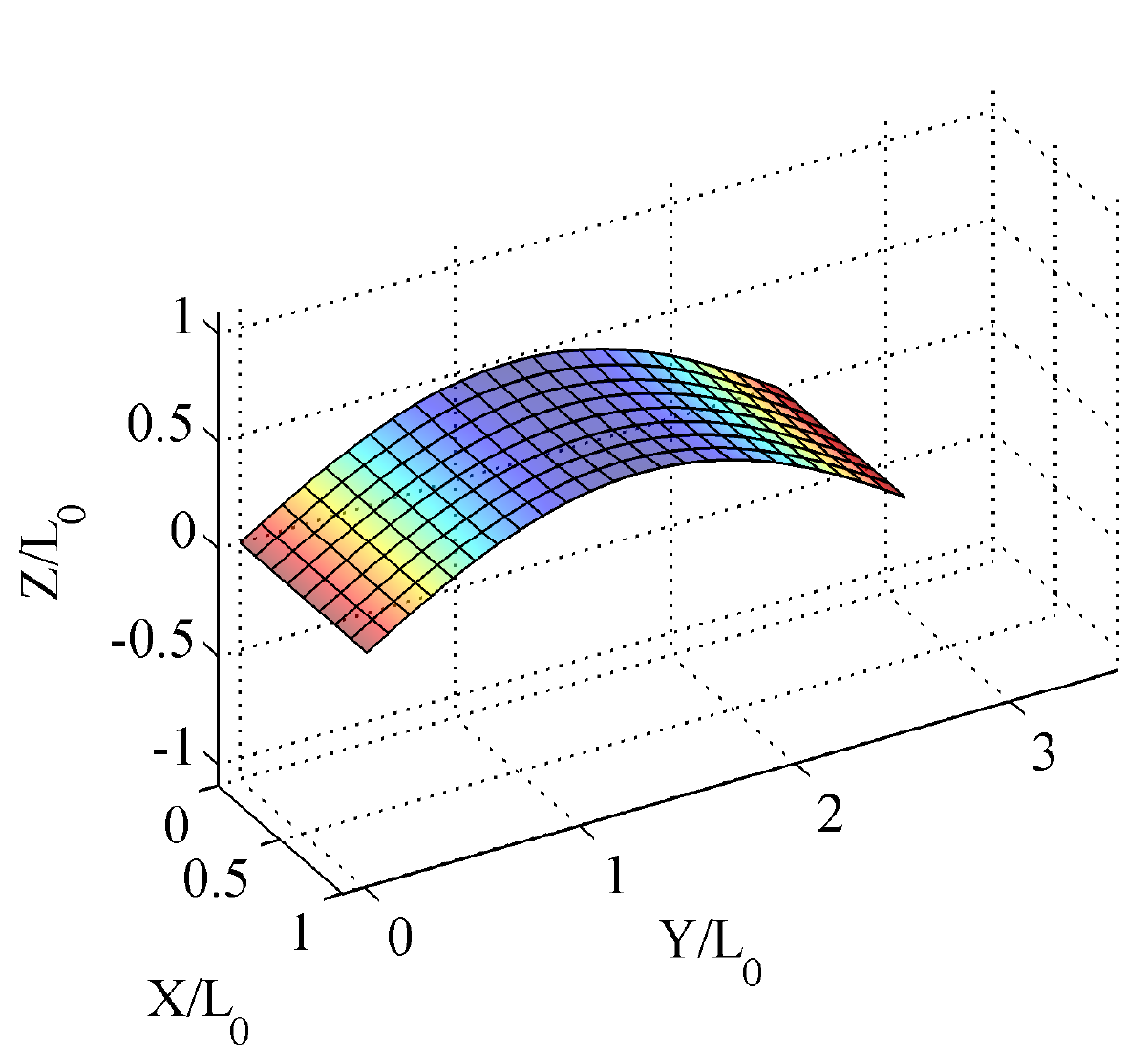}}
	\caption{Flat strip under the compressive displacement $v_{right}$ and the bending moment $m$ is modeled by the Koiter model: (a) objective function versus the number of iterations, (b) the reconstructed deformation at the second iteration, (c) the reconstructed deformation after $5$ iterations, (d) the reconstructed deformation after $6$ iterations, (e) the reconstructed deformation after $40$ iterations.}
	\label{fig:column_eccentric_load_inverse_Koiter}
\end{center}
\end{figure}

In the second step, an inverse approach based on the nodal displacement of target configuration (with additional noise) is employed to identify the applied displacement $v_{right}$ and moment $m$ for the two shell models: (1) the Koiter model and (2) the projected model with the compressible Neo-Hookean formulation. For the shell structure modeled by the Koiter model as shown in Figure \ref{fig:column_eccentric_load_inverse_Koiter}(a), convergence is gained after $40$ iterations and the optimal shape in Figure \ref{fig:column_eccentric_load_inverse_Koiter}(e) recovered after $40$ iterations, corresponding to the inverse solution $v^{inverse}_{Koit,right} = -0.5055$ and $m^{inverse}_{Koit} = 9.06 \times 10^3 ~N.mm/mm$, which is in excellent agreement with the target shape in Figure \ref{fig:column_eccetric_load_FEM_Koiter}(a).

\begin{figure}[htbp]
\begin{center}
\vspace*{-2cm}
	\subfigure[]{\includegraphics[width = 0.45\textwidth]{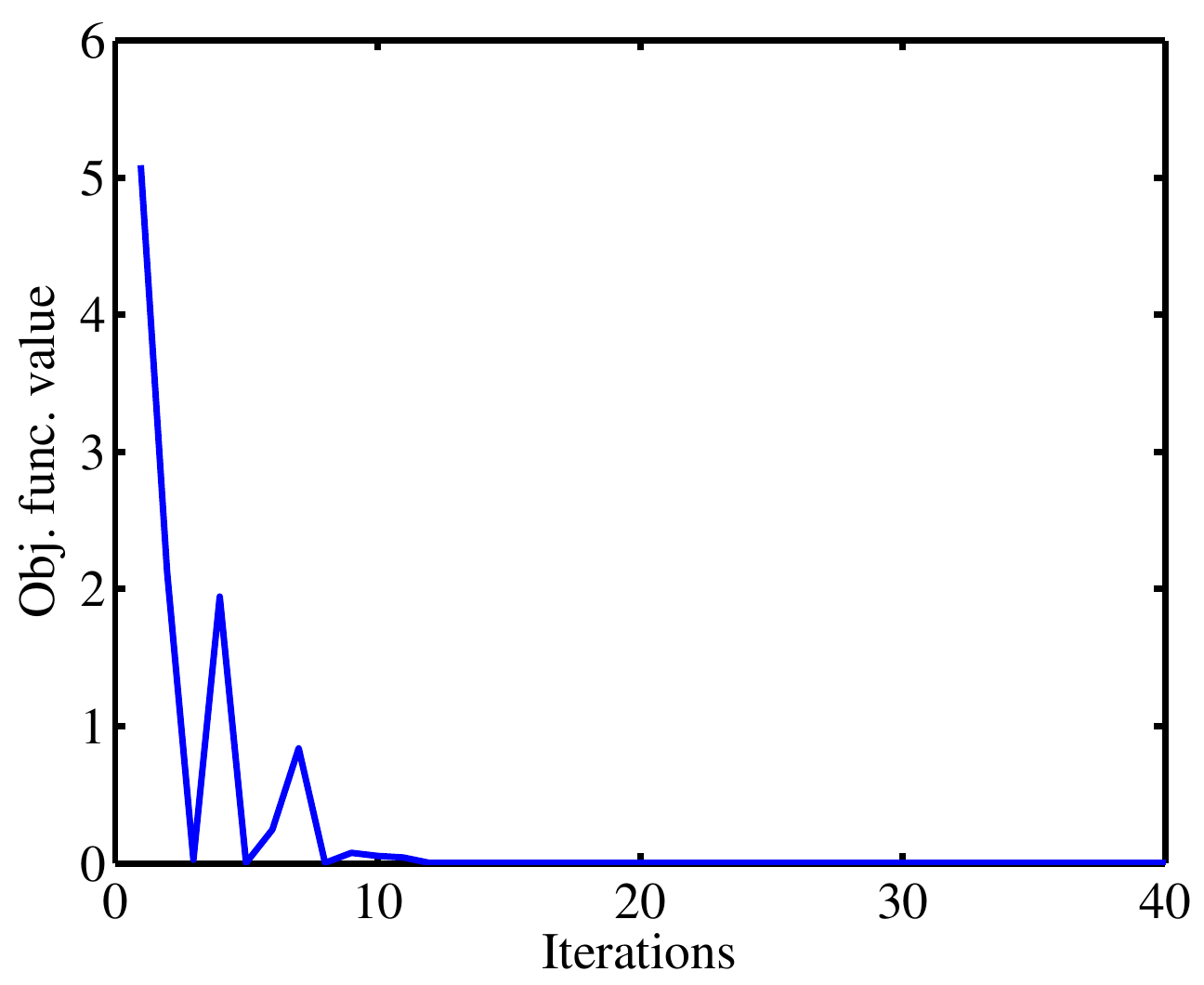}}
	\subfigure[]{\includegraphics[width = 0.45\textwidth]{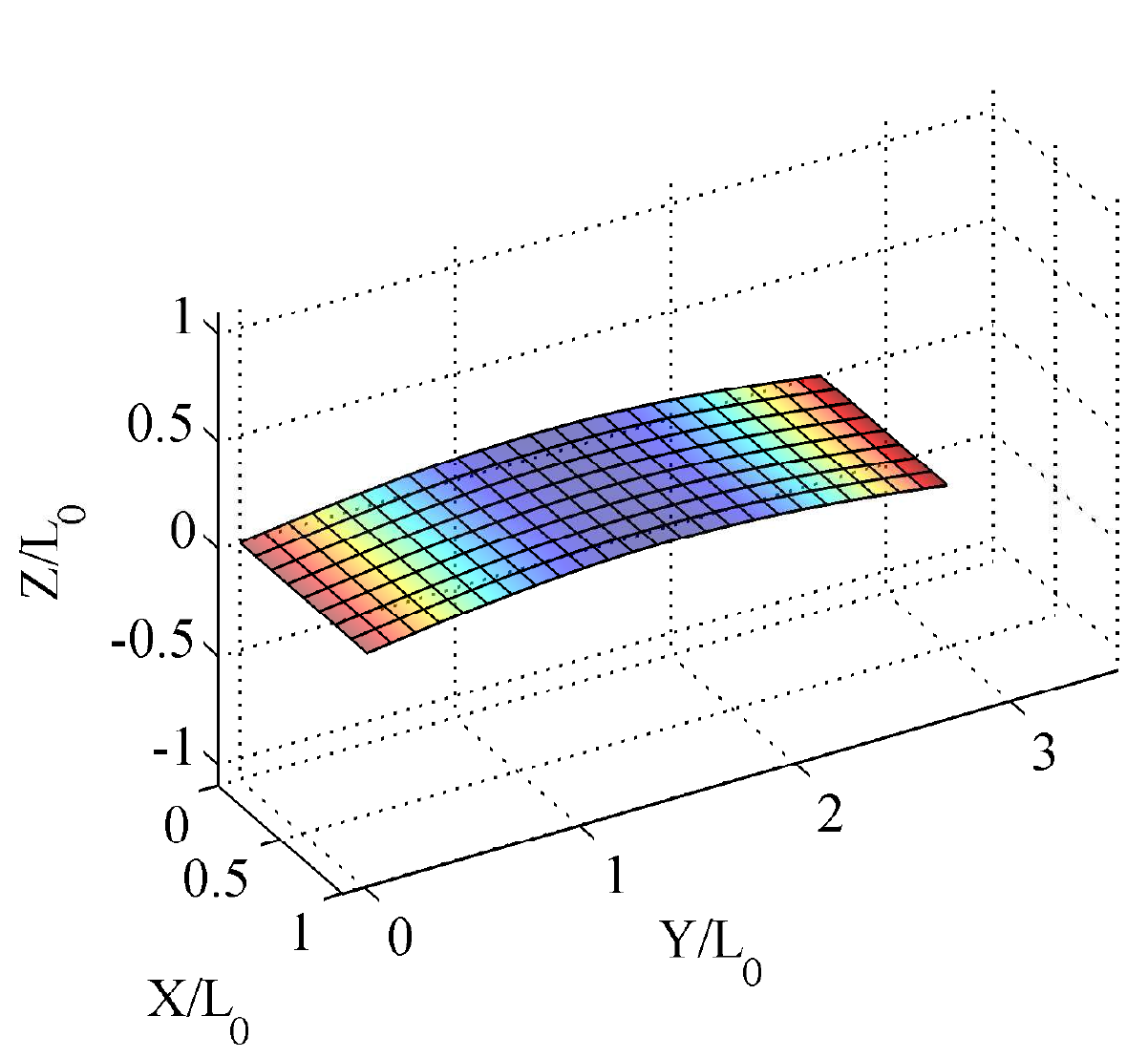}} \\
\vspace*{-0.5cm}
	\subfigure[]{\includegraphics[width = 0.45\textwidth]{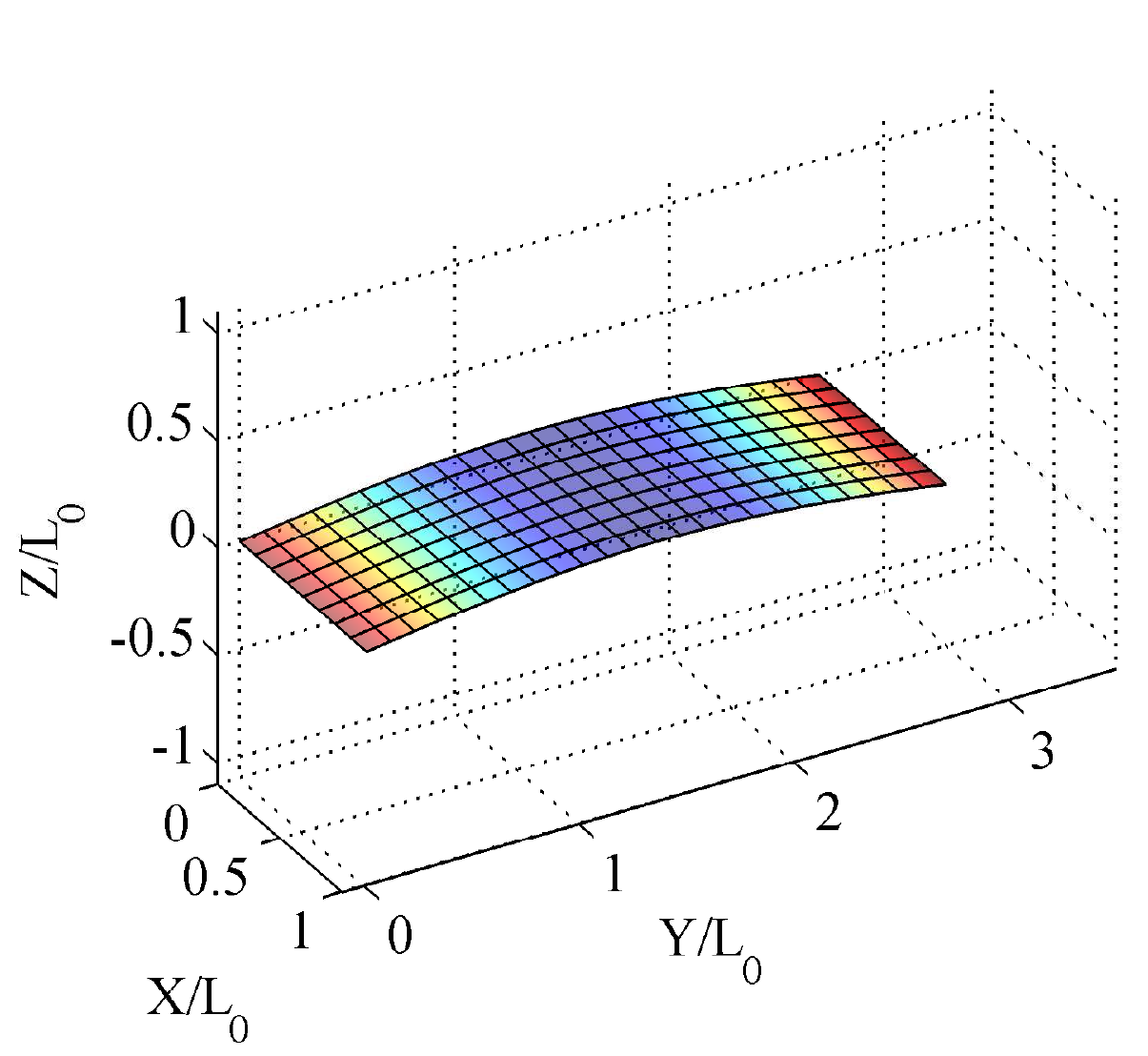}}
	\subfigure[]{\includegraphics[width = 0.45\textwidth]{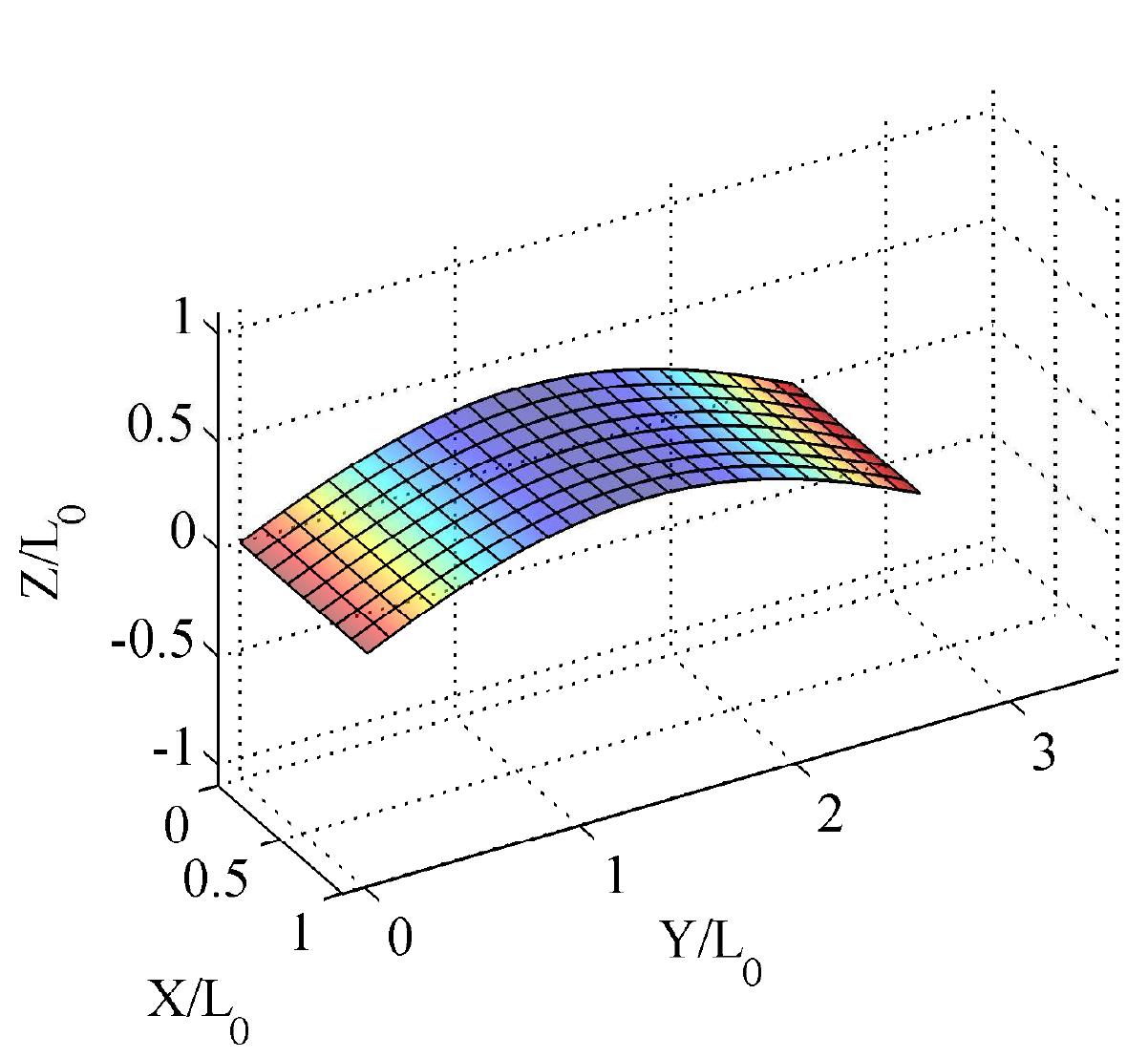}} \\
\vspace*{-0.5cm}
	\subfigure[]{\includegraphics[width = 0.45\textwidth]{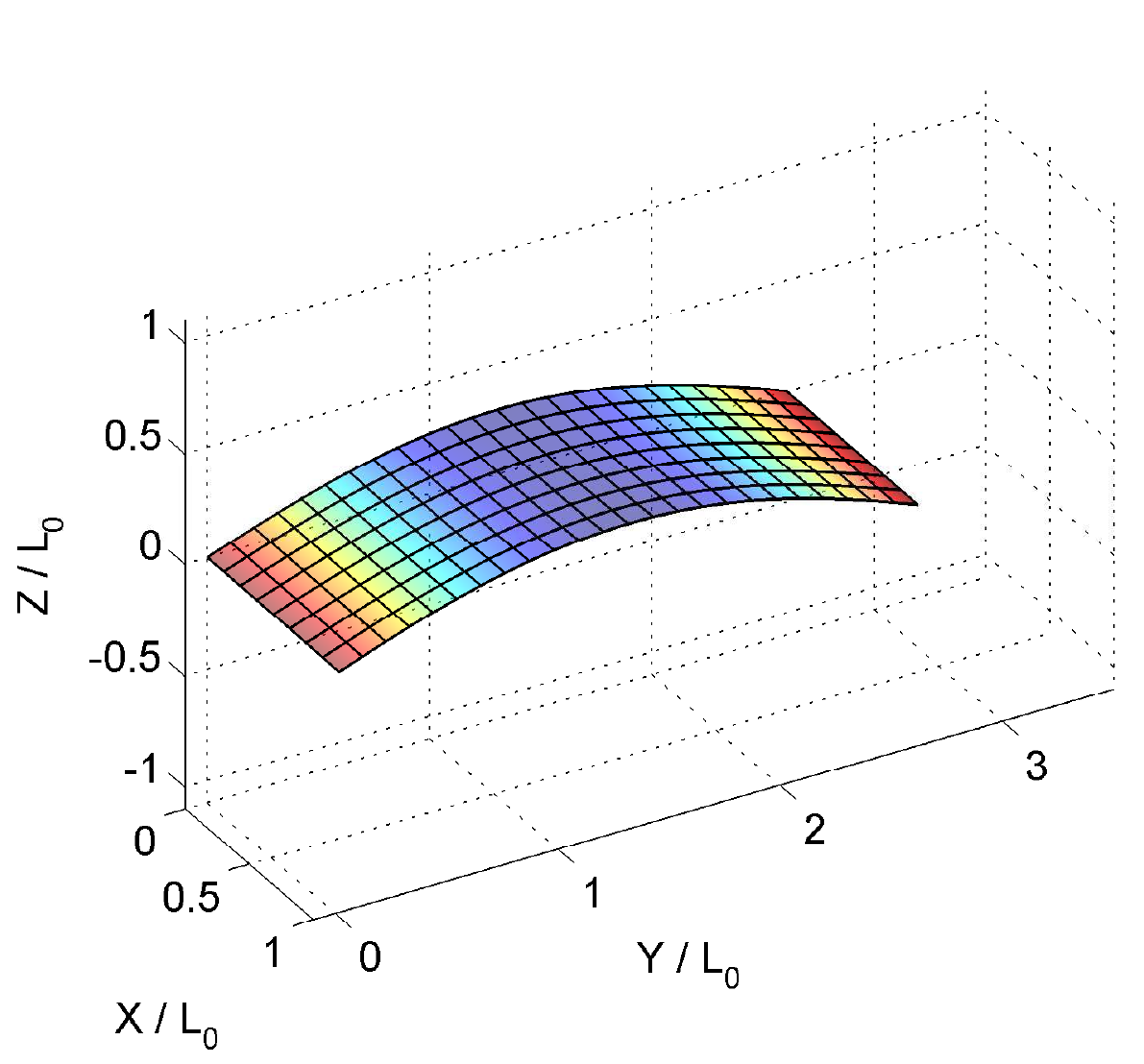}}
	\caption{Flat strip under the compressive displacement $v_{right}$ and the bending moment $m$ is modeled by the projected model: (a) objective function versus the number of iterations, (b) the reconstructed deformation at the second iteration, (c) the reconstructed deformation after $4$ iterations, (d) the reconstructed deformation after $6$ iterations, (e) the reconstructed deformation after $40$ iterations.}
	\label{fig:column_eccentric_load_inverse_projected}
\end{center}
\end{figure}

Similar observation can be made for the results obtained from the projected model. The convergence of the objective function over iterations and the reconstructed configurations with iterations are depicted after $40$ iterations Figures \ref{fig:column_eccentric_load_inverse_projected}. The recovered shape in Figure \ref{fig:column_eccentric_load_inverse_projected}(e), corresponding to the inverse solution $v^{inverse}_{proj,right} = -0.3989$ and $m^{inverse}_{proj} = 11.13 \times 10^3 ~N.mm/mm$, is almost identical with the measured shape in Figure \ref{fig:column_eccentric_load_FEM_projected}(a). In summary, the proposed inverse method is able to accurately determine the prescribed displacement $v_{right}$ and moment $m$ and the specified final shape can be reconstructed where the unstable deformation due to buckling are captured. Note that nodal displacements in the $z$ direction for both models are restricted to be larger than $0$ to ensure the convex shape of the measured shells in Figures \ref{fig:column_eccetric_load_FEM_Koiter}(a) and \ref{fig:column_eccentric_load_FEM_projected}(a), respectively.


\subsection{A flat strip subjected to compressive displacement and disturbing pressure} \label{subsec:ex4}

\begin{figure}[htbp]
\begin{center}
	\subfigure[]{\includegraphics[width = 0.6\textwidth]{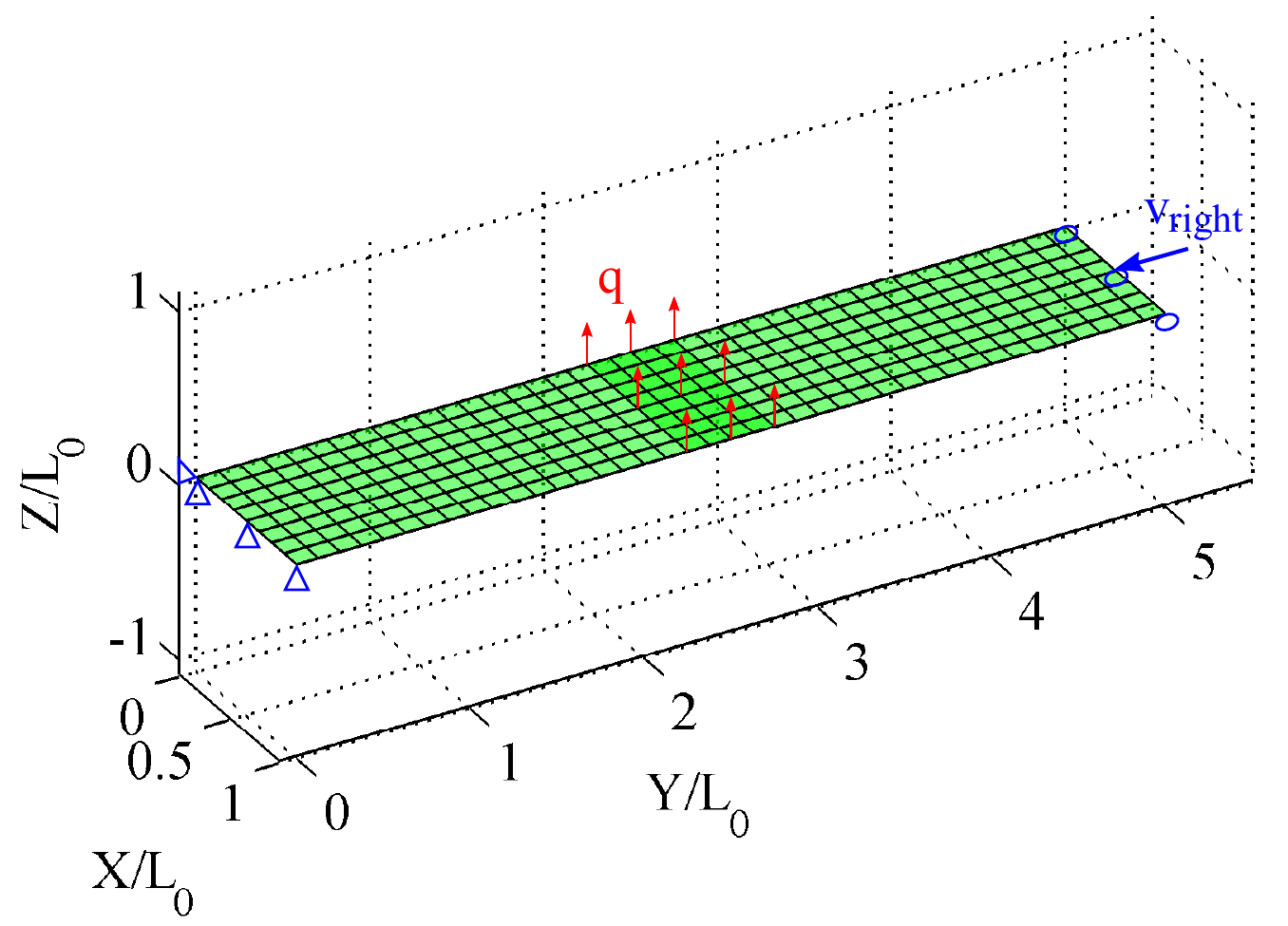}} \\
	\subfigure[]{\includegraphics[width = 0.45\textwidth]{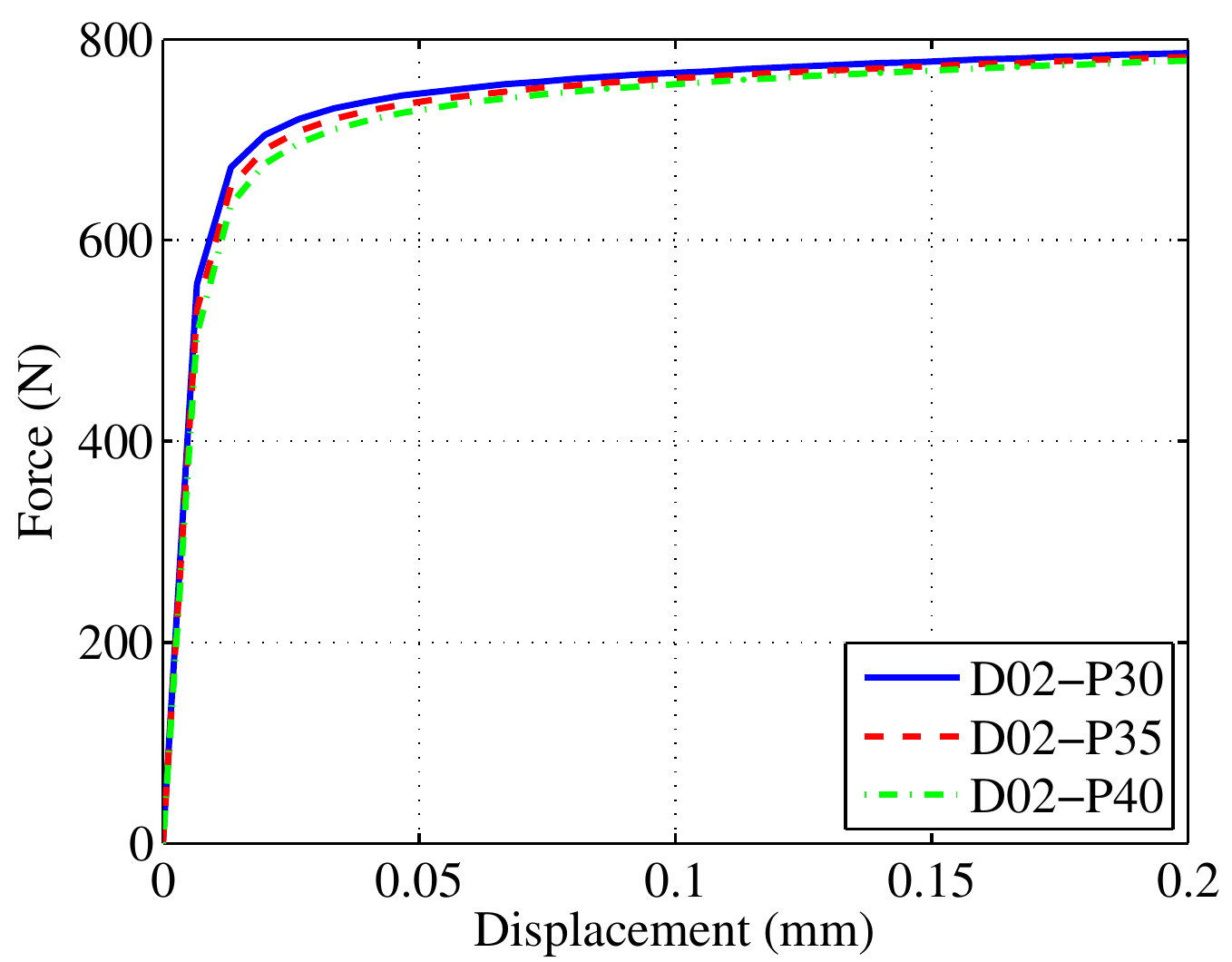}}
	\subfigure[]{\includegraphics[width = 0.45\textwidth]{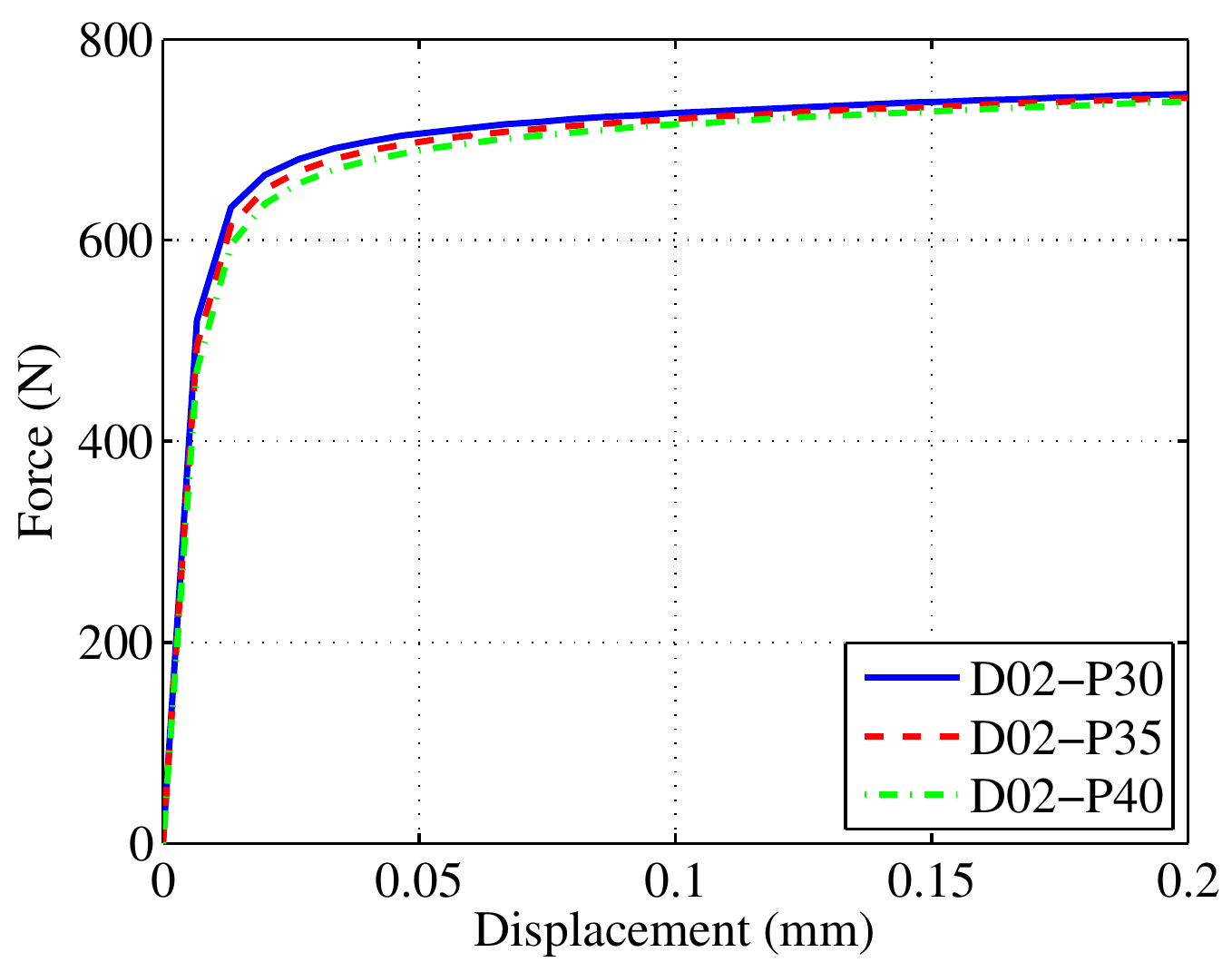}}
	\caption{Flat strip subjected to compressive displacement $v_{right}$ and lateral disturbing pressure $q$: (a) undeformed shape, (b) axial force-displacement responses for different pressures $q$ modeled by the Koiter model, (c) axial force-displacement response for different pressures $q$ modeled by the projected model.}
	\label{fig:flat_strip_pressure_model}
\end{center}
\end{figure}

\begin{figure}[htbp]
\begin{center}
	\subfigure[]{\includegraphics[width = 0.5\textwidth]{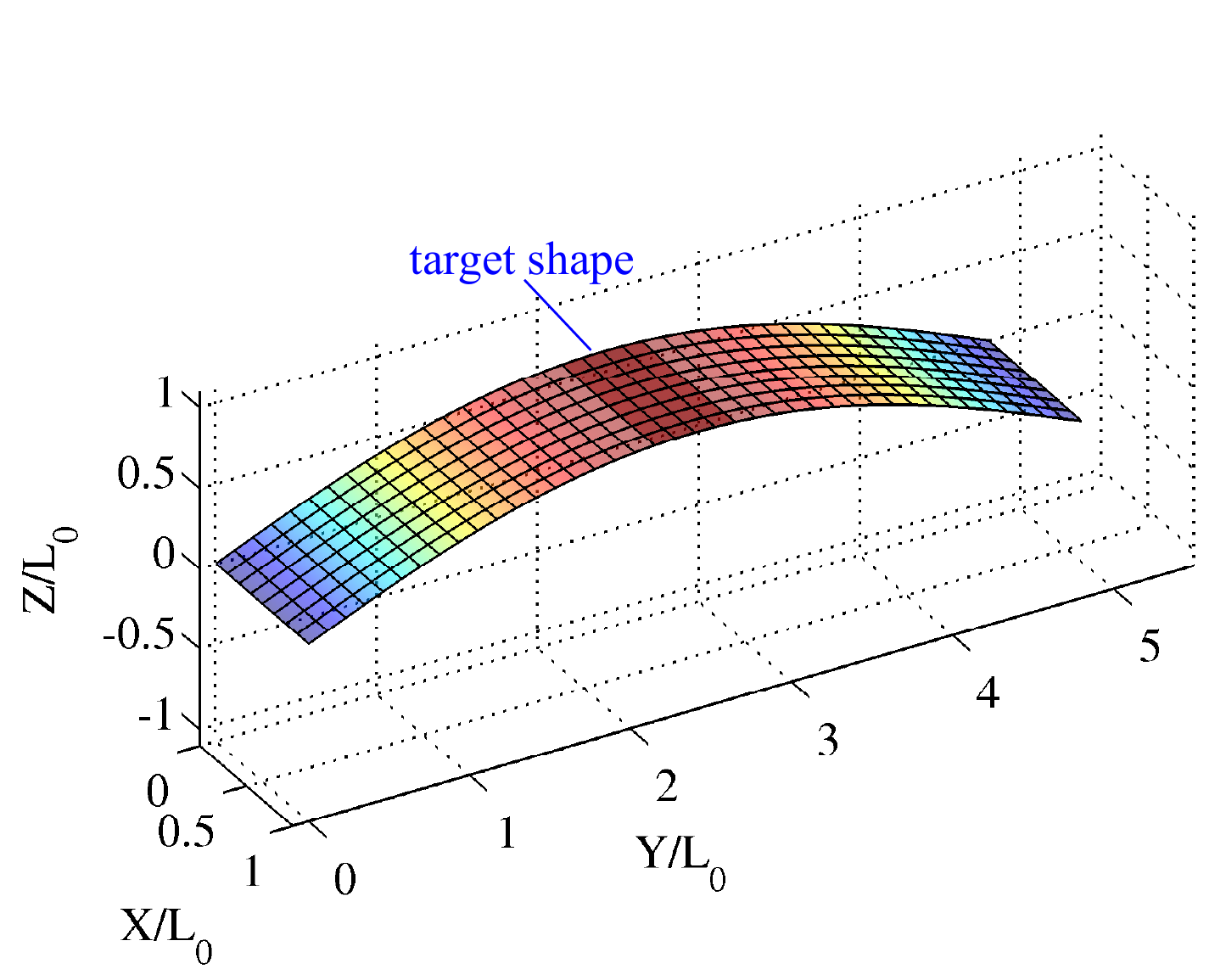}}
	\subfigure[]{\includegraphics[width = 0.45\textwidth]{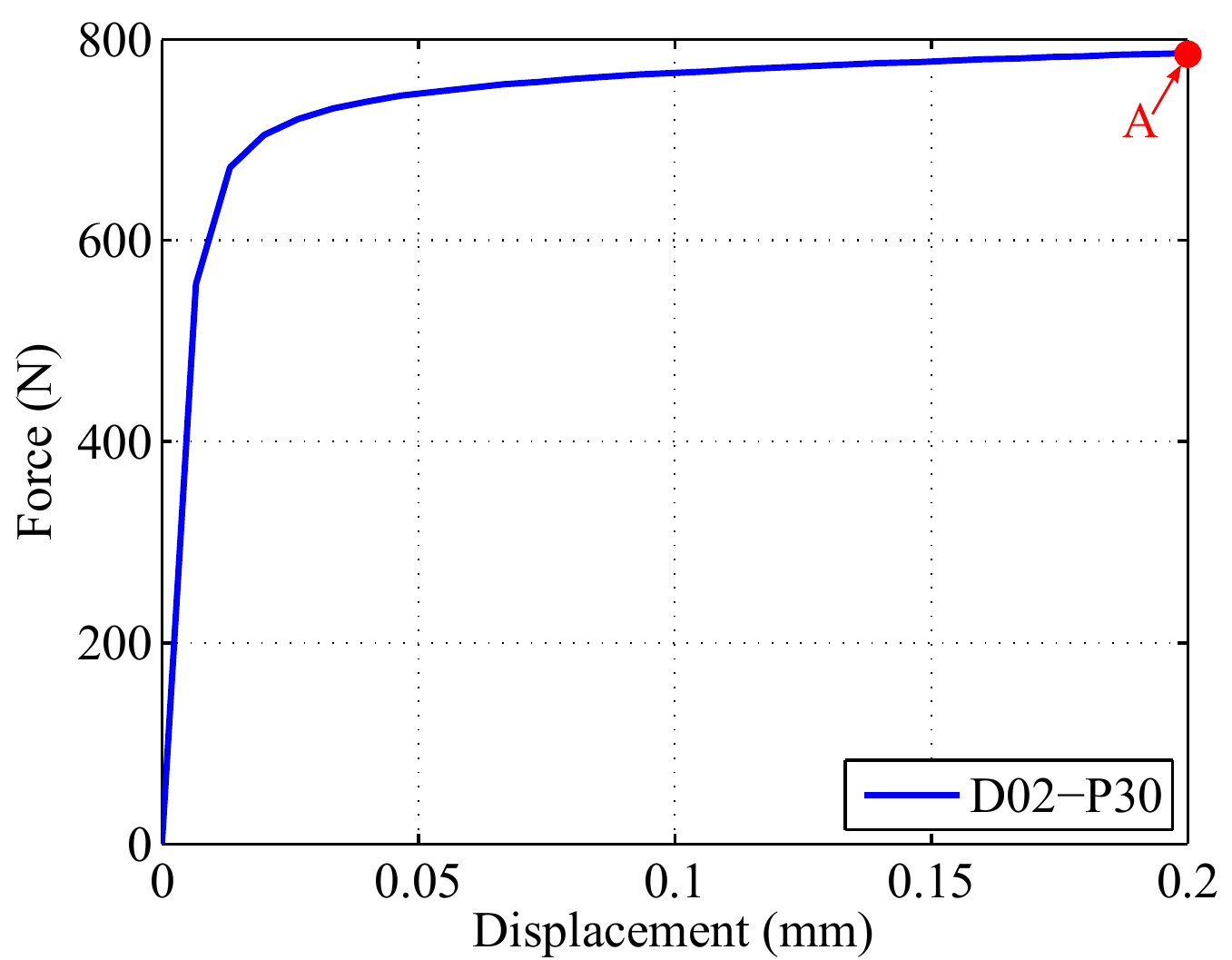}}
	\caption{Pin-ended flat strip under the compressive displacement $v_{right}$ and the lateral disturbing pressure $q$ modeled by the Koiter model: (a) deformed shape at displacement $v_{right} = -0.2$, (b) reaction force versus displacement in $y$ direction.}
	\label{fig:flat_strip_pressure_FEM_Koiter}
\end{center}
\end{figure}

\begin{figure}[htbp]
\begin{center}
	\subfigure[]{\includegraphics[width = 0.5\textwidth]{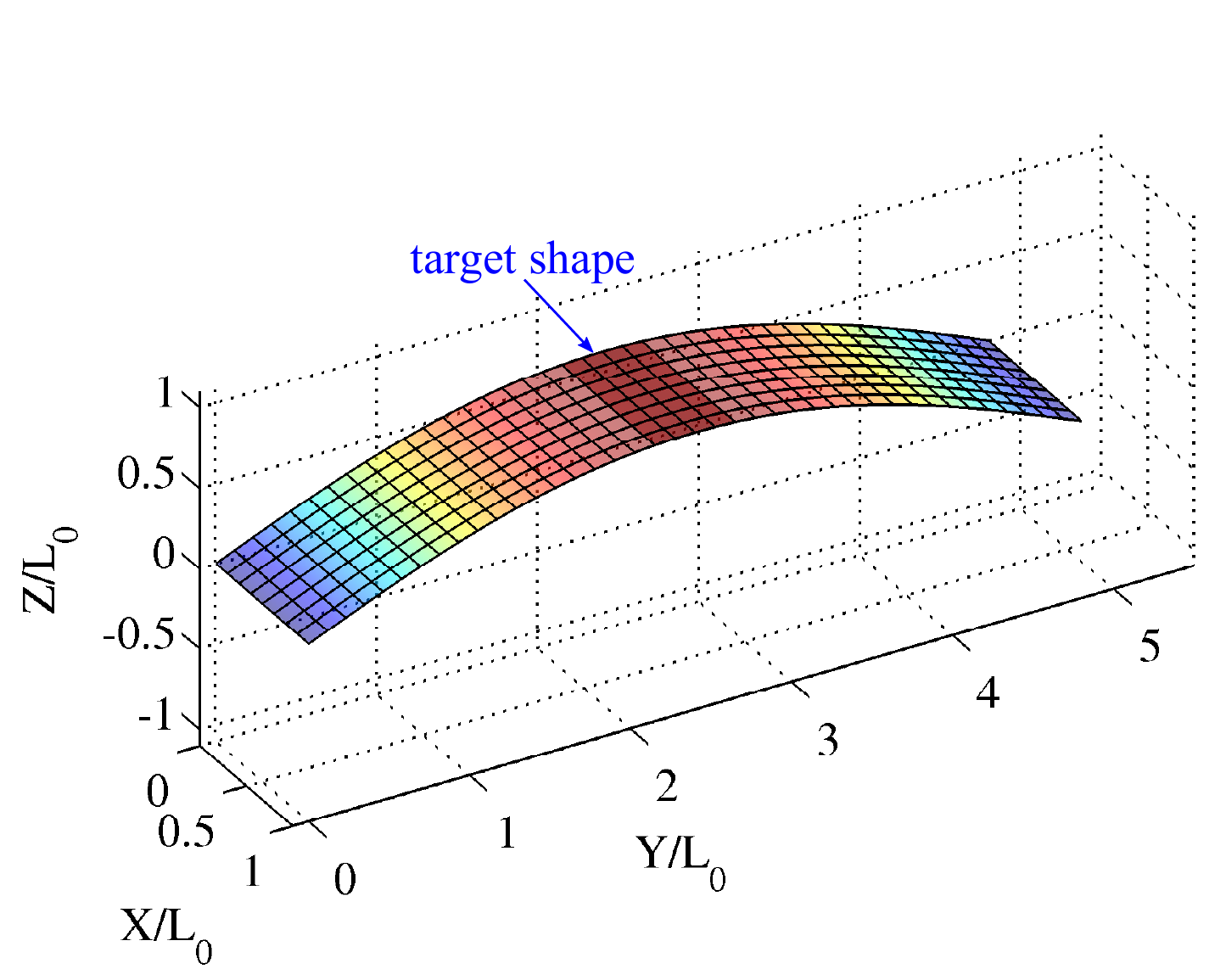}}
	\subfigure[]{\includegraphics[width = 0.45\textwidth]{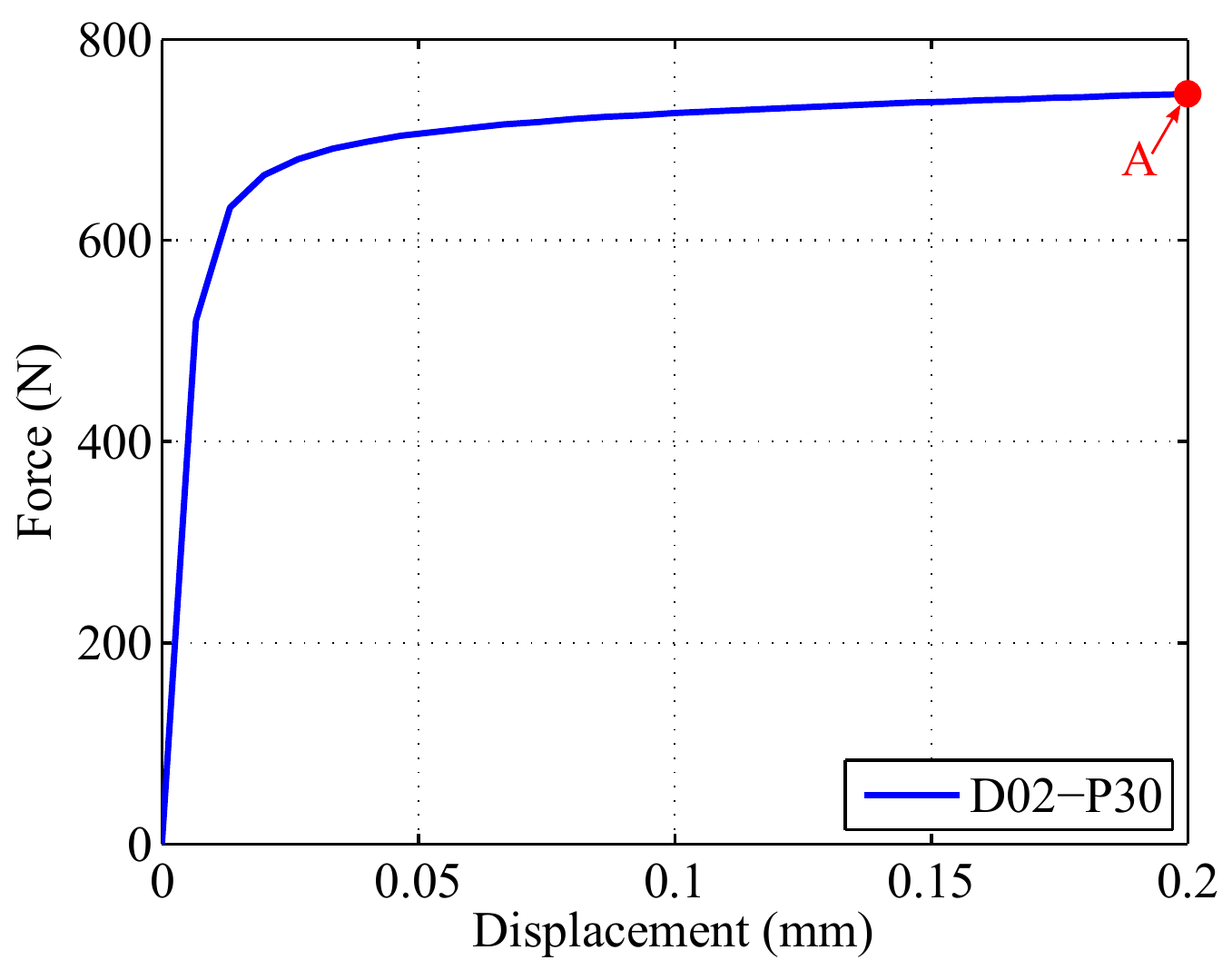}}
	\caption{Flat strip under the compressive displacement $v_{right}$ and the lateral disturbing pressure $q$ modeled by the projected model: (a) deformed shape at displacement $v_{right} = -0.2$, (b) reaction force versus displacement in $y$ direction.}
	\label{fig:flat_strip_pressure_FEM_projected}
\end{center}
\end{figure}

The pinned-end flat strip, with dimensions $L \times W \times T = 5 \times 1 \times 0.2 ~mm$ is considered again. We assume its initial shape is perfectly straight but subjected to axially prescribed displacement $v_{right}$ and lateral pressure $q$ as depicted in Figure \ref{fig:flat_strip_pressure_model}. First, NURBS-based finite elements are used to solve the forward problem. This example was analyzed using $8 \times 40$ NURBS elements with the material constants: $E = 3 \times 10^6 ~N/mm^2$ and $\nu = 0.3$. The FE force vector due to lateral pressure is given by Equation (\ref{eq:ext_force}.2) and the corresponding stiffness matrix $k^e_{\mathrm{ext} p}$ is determined by (\ref{eq:ext_moment_tangent_matrix}.2). The Koiter model and the projected model using the compressible Neo-Hookean formulation are considered. Figures \ref{fig:flat_strip_pressure_model}(b+c) respectively show the influence of the disturbing pressure on the force-displacement response for both models. The measured shapes and force-displacement response, corresponding to $v^{meas}_{Koit,right} = -0.2$, $q^{meas}_{Koit} = 30 ~N/mm^2$, are shown in Figure \ref{fig:flat_strip_pressure_FEM_Koiter} for the Koiter model and in Figure \ref{fig:flat_strip_pressure_FEM_projected} for the projected model with $v^{meas}_{proj,right} = -0.2$, $q^{meas}_{proj} = 30 ~N/mm^2$, respectively.

\begin{figure}[htbp]
\begin{center}
\vspace*{-1.5cm}
	\subfigure[]{\includegraphics[width = 0.45\textwidth]{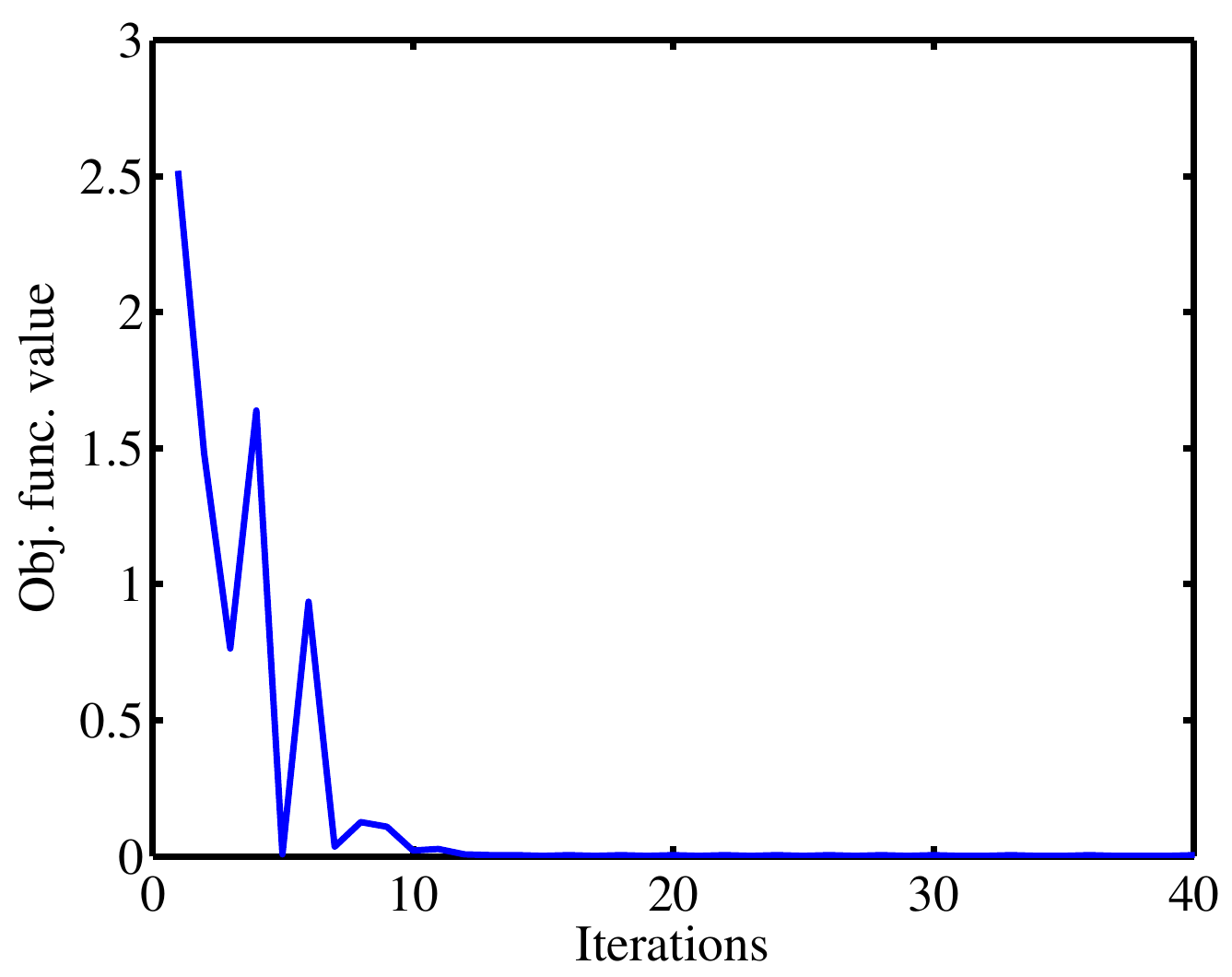}}
	\subfigure[]{\includegraphics[width = 0.45\textwidth]{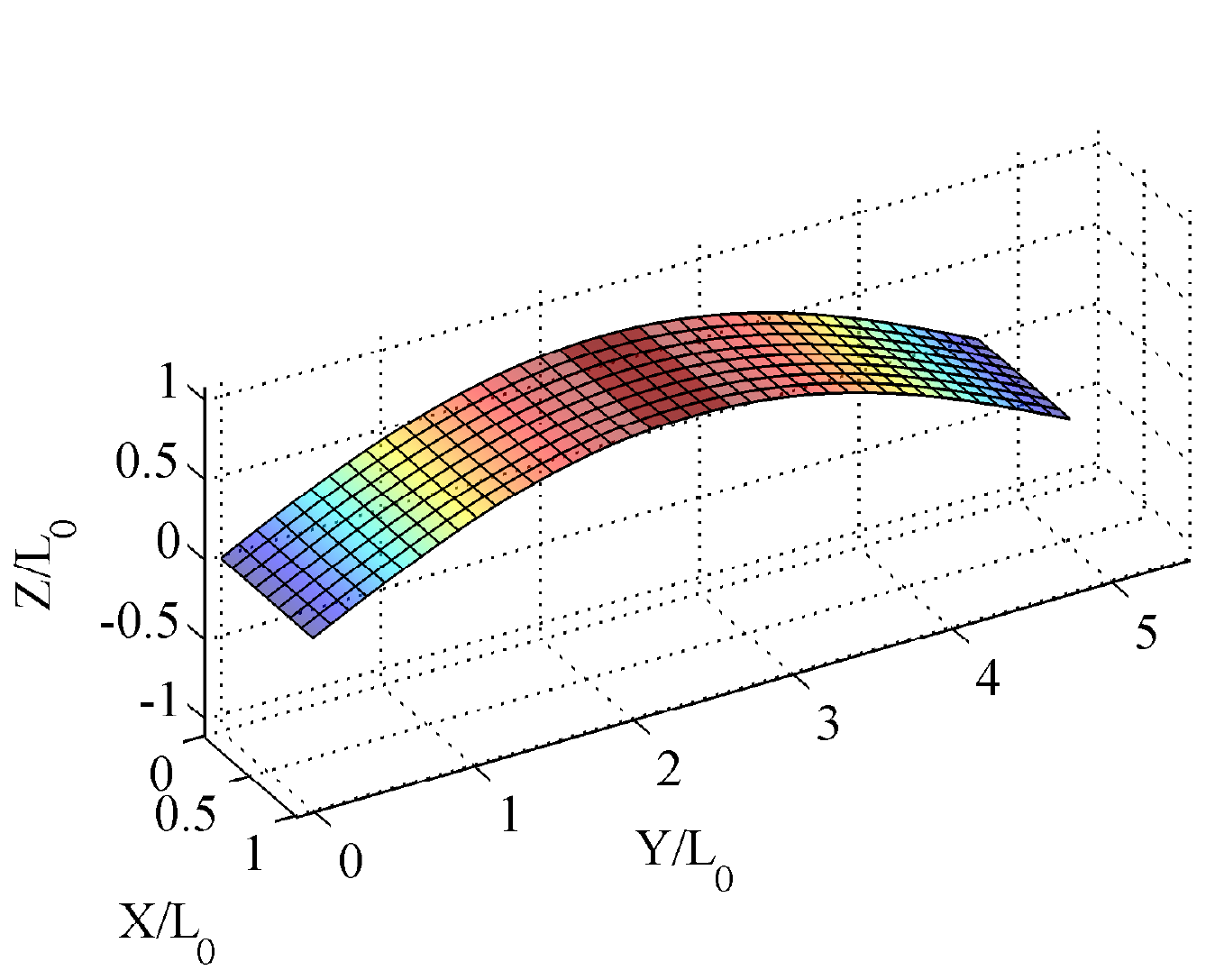}} \\
\vspace*{-0.25cm}
	\subfigure[]{\includegraphics[width = 0.45\textwidth]{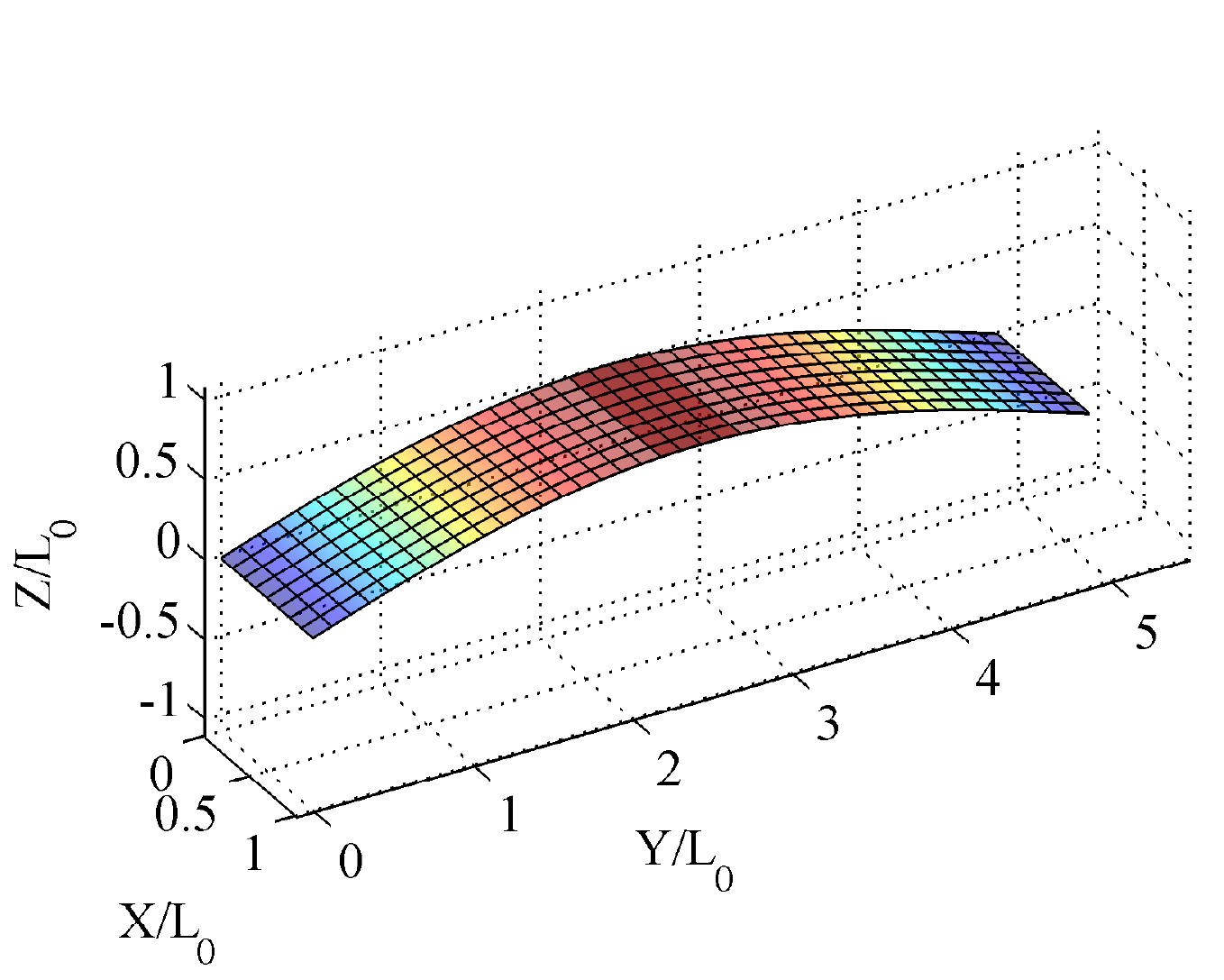}}
	\subfigure[]{\includegraphics[width = 0.45\textwidth]{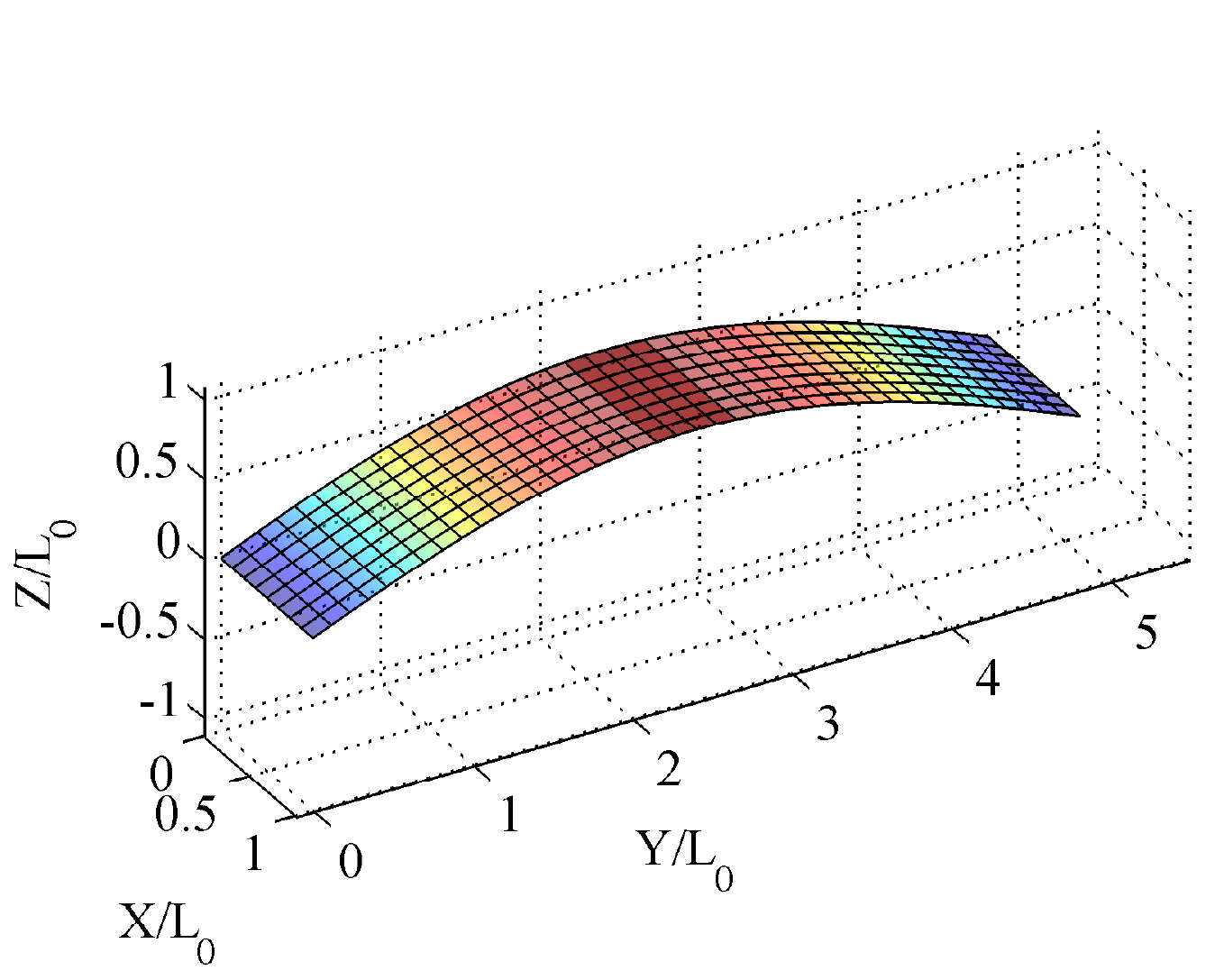}} \\
\vspace*{-0.25cm}
	\subfigure[]{\includegraphics[width = 0.45\textwidth]{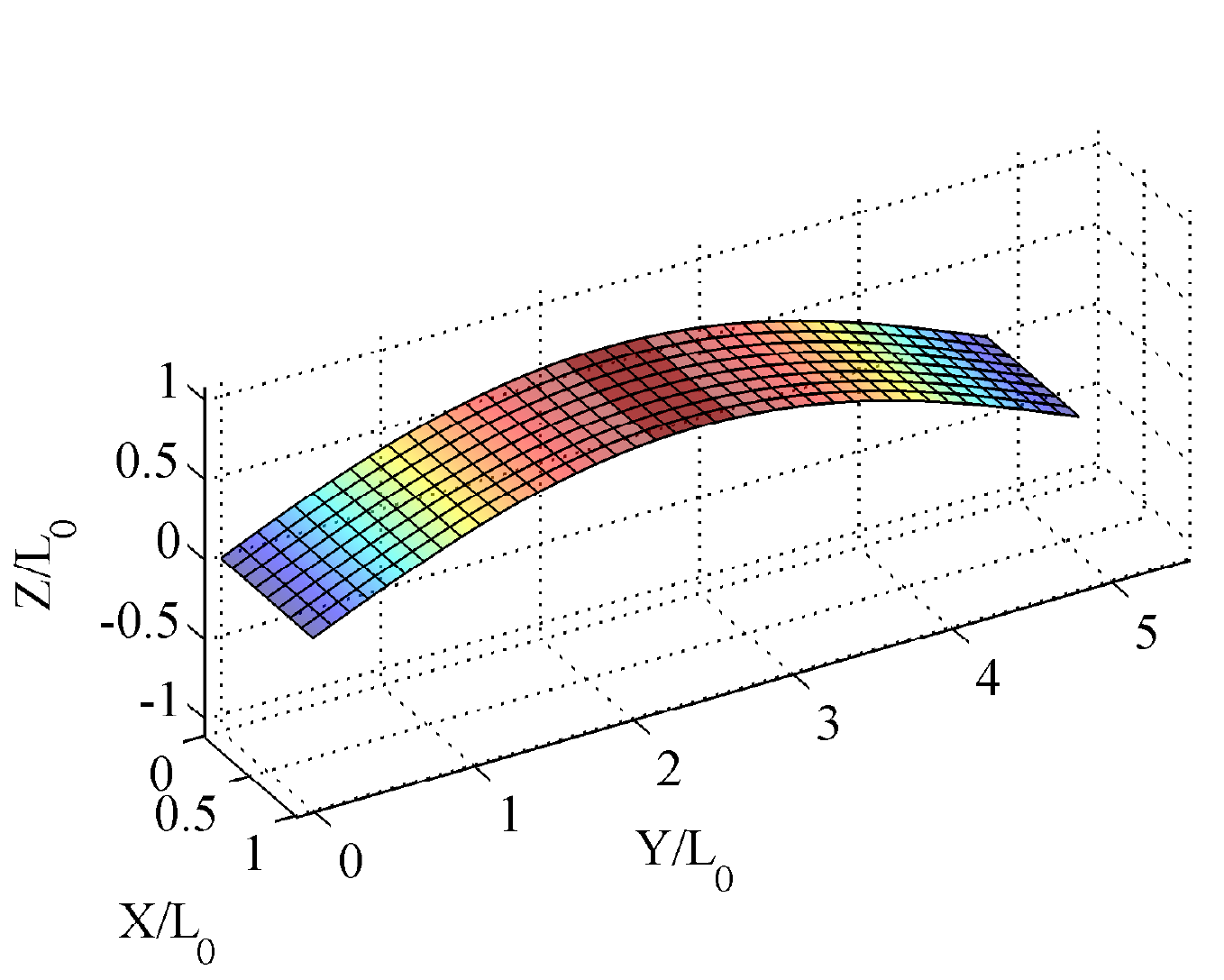}}
	\caption{Flat strip under the compressive displacement $v_{right}$ and the lateral disturbing pressure $q$ modeled by the Koiter model: (a) objective function versus the number of iterations, (b) the reconstructed deformation at the second iteration, (c) the reconstructed deformation after $5$ iterations, (d) the reconstructed deformation after $6$ iterations, (e) the reconstructed deformation after $40$ iterations.}
	\label{fig:flat_strip_pressure_inverse_Koiter}
\end{center}
\end{figure}

\begin{figure}[htbp]
\begin{center}
\vspace*{-1.5cm}
	\subfigure[]{\includegraphics[width = 0.45\textwidth]{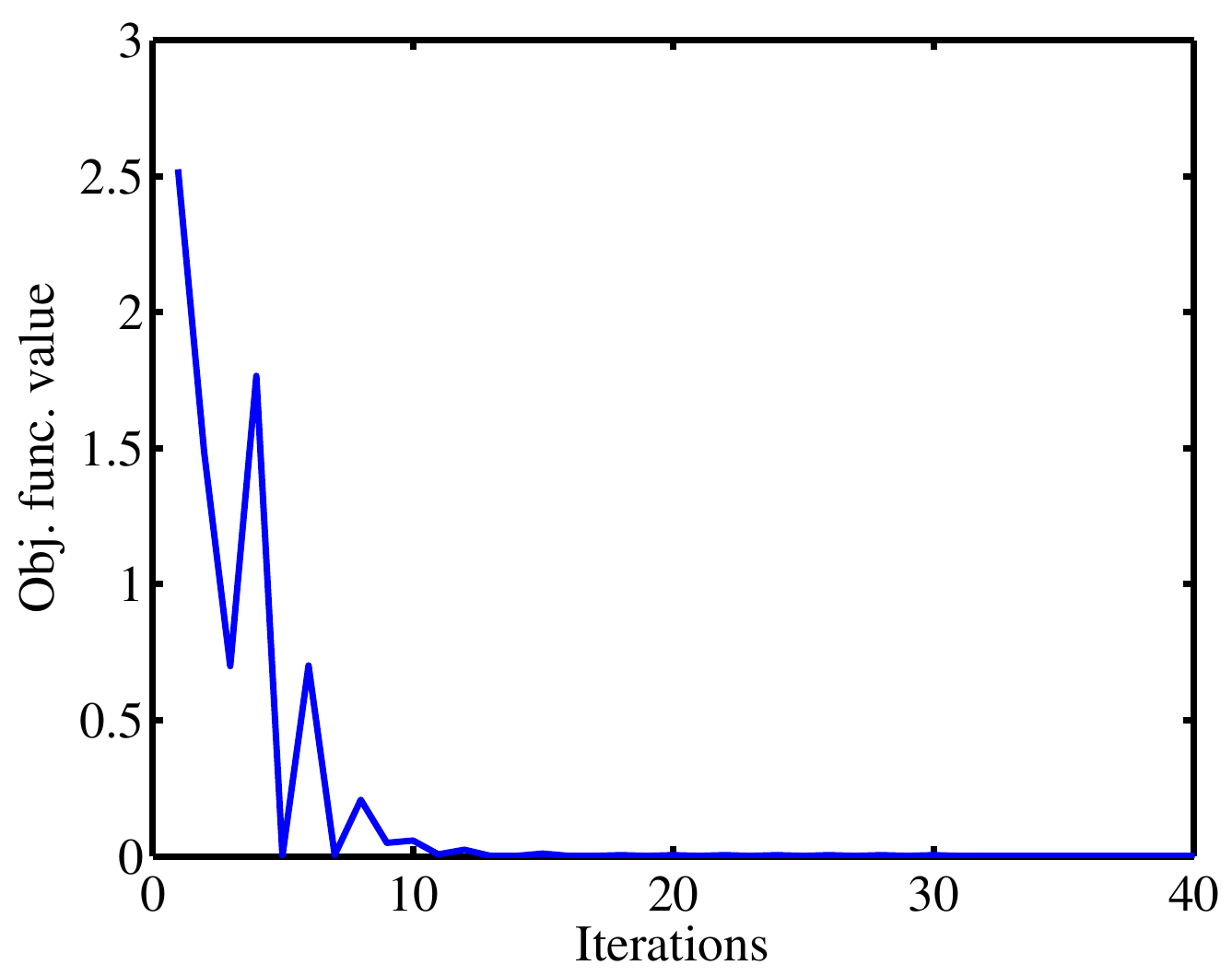}}
	\subfigure[]{\includegraphics[width = 0.45\textwidth]{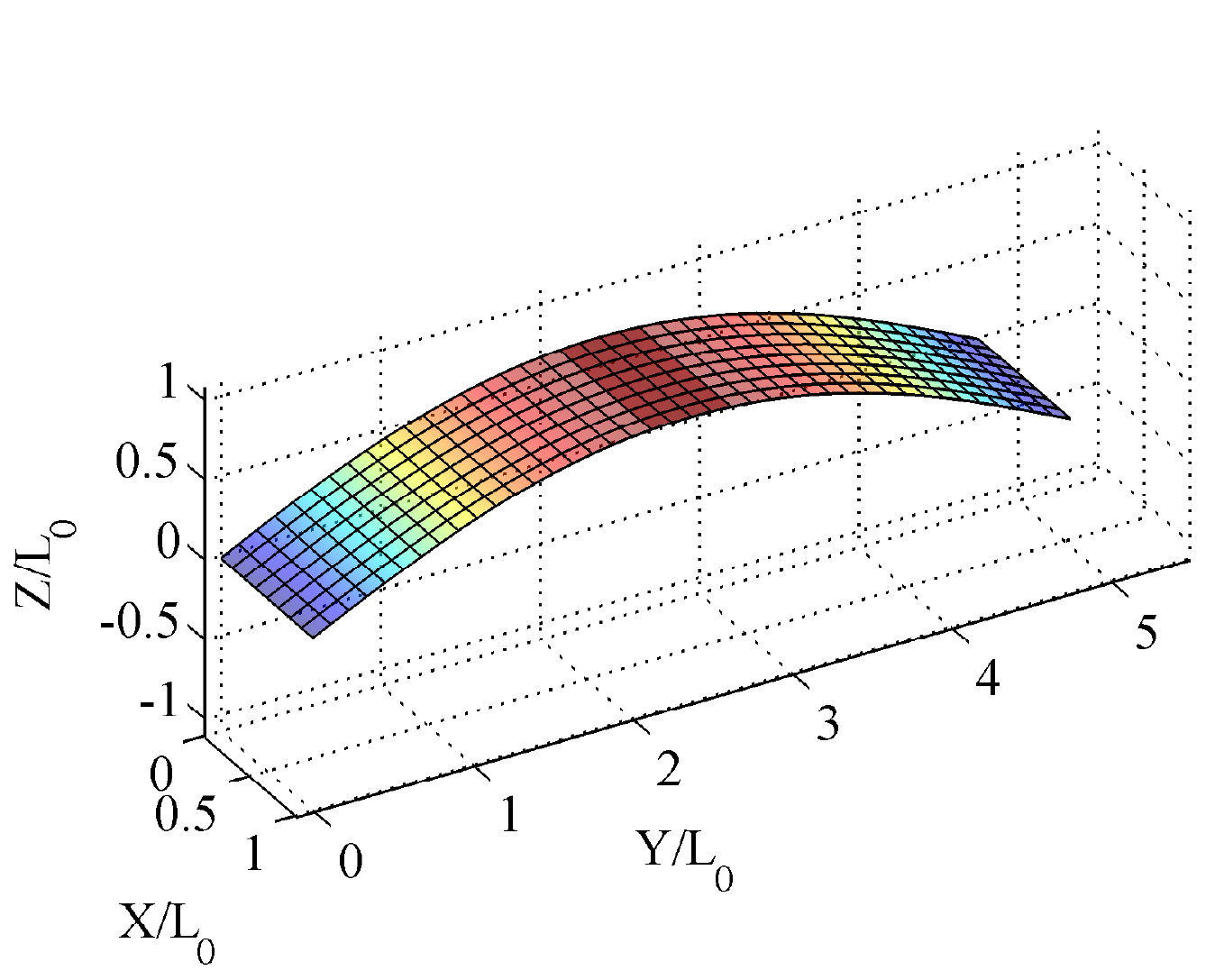}} \\
\vspace*{-0.25cm}
	\subfigure[]{\includegraphics[width = 0.45\textwidth]{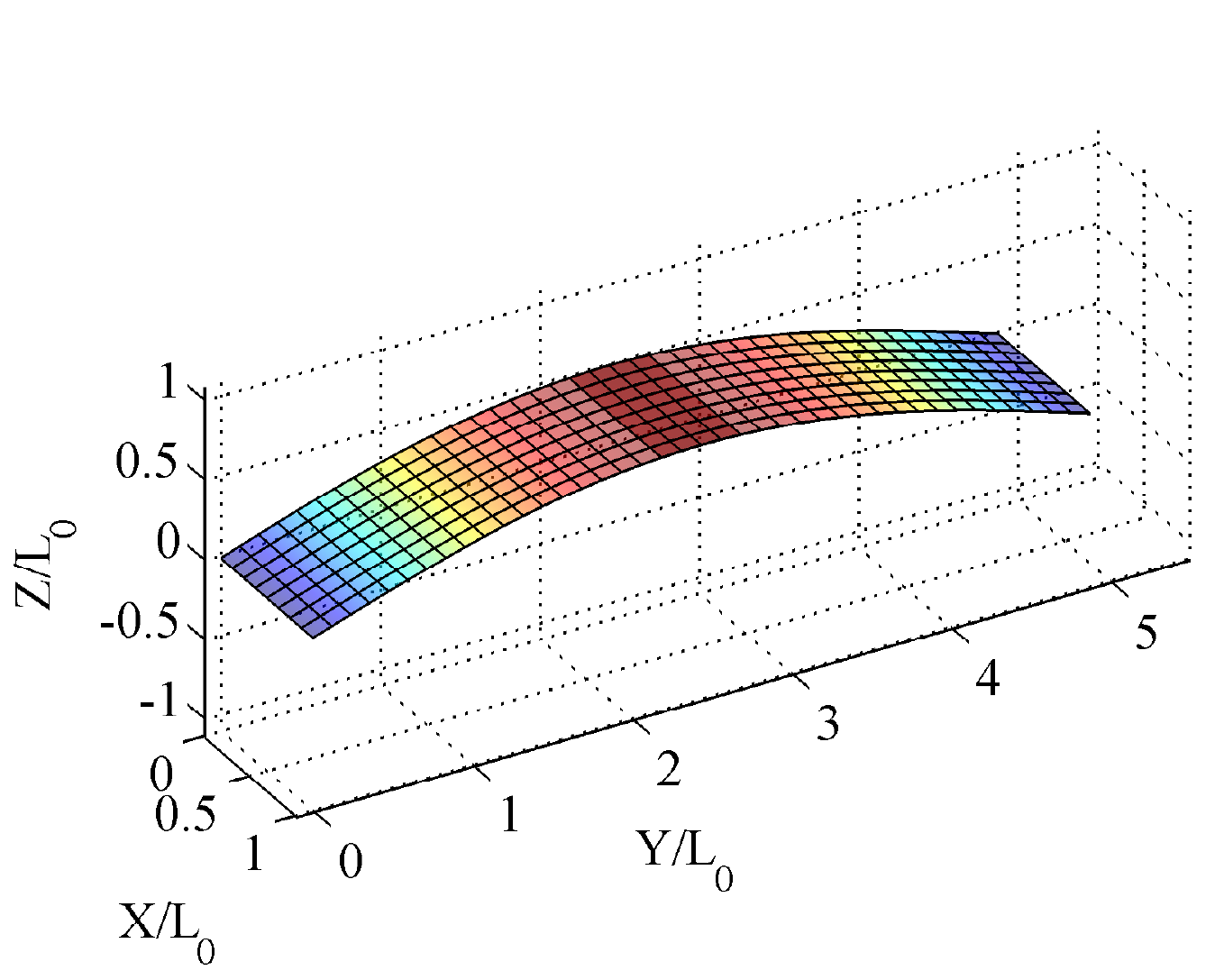}}
	\subfigure[]{\includegraphics[width = 0.45\textwidth]{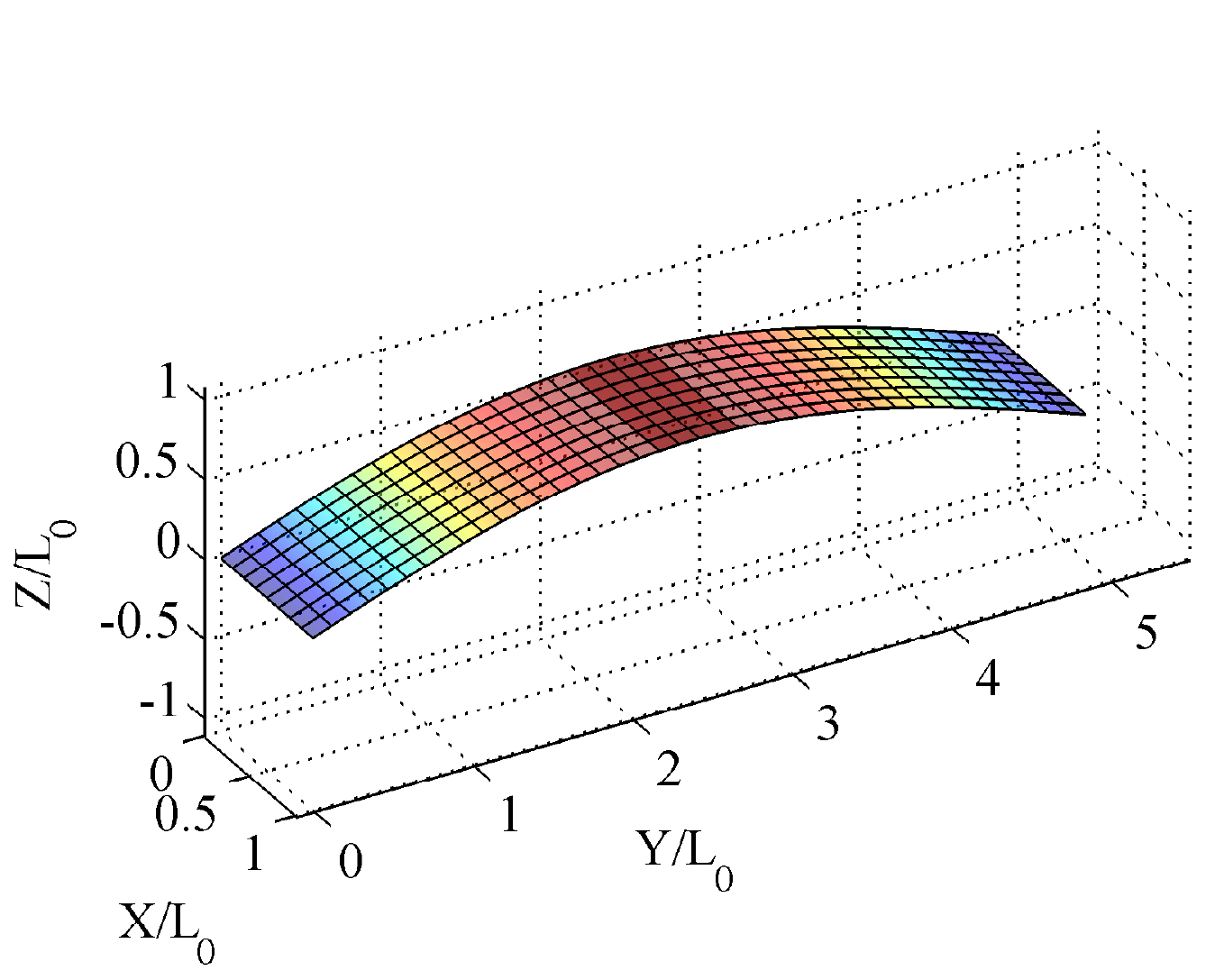}} \\
\vspace*{-0.25cm}
	\subfigure[]{\includegraphics[width = 0.45\textwidth]{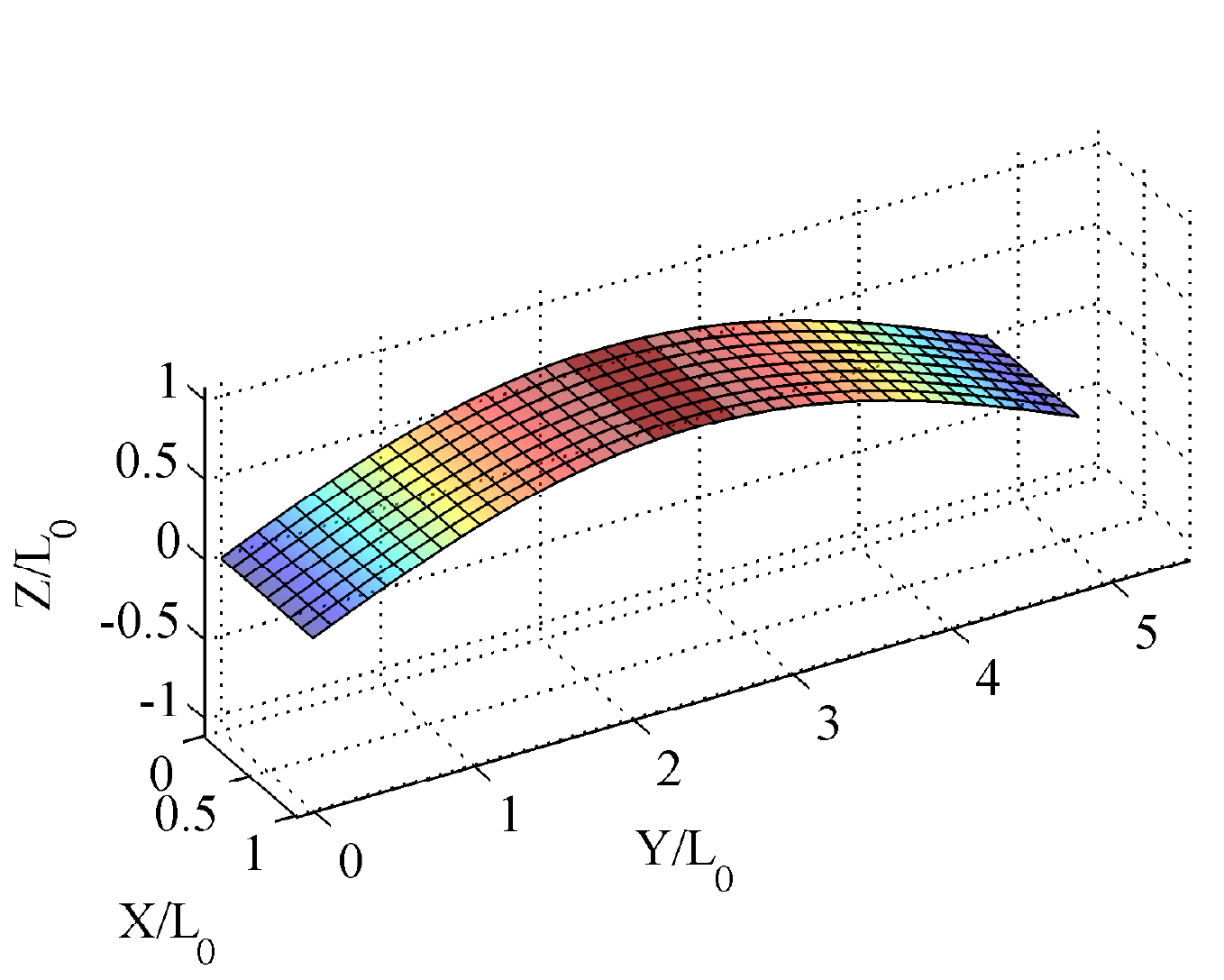}}
	\caption{Flat strip under the compressive displacement $v_{right}$ and the lateral disturbing pressure $q$ modeled by the projected model: (a) objective function versus the number of iterations, (b) the reconstructed deformation at the second iteration, (c) the reconstructed deformation after $5$ iterations, (d) the reconstructed deformation after $6$ iterations, (e) the reconstructed deformation after $40$ iterations.}
	\label{fig:flat_strip_pressure_inverse_projected}
\end{center}
\end{figure}

Based on the nodal displacements $\bm{u}^{meas}$ corresponding to point A on the curves obtained from the forward problem, we conduct an inverse analysis for the strip modeled by (1) the Koiter model and (2) the projected model with the compressible Neo-Hookean formulation. The convex shape of the reconstructed deformations is ensured by restricting nodal displacements in the $z$ direction to be larger than $0$. For the former model, convergence is reached after $40$ iterations, see Figure \ref{fig:flat_strip_pressure_inverse_Koiter}(a) and the resultant shape under the identified $v^{inverse}_{Koit,right} = -0.1992$, $q^{inverse}_{Koit} = 35.18 ~N/mm^2$ is accurately reconstructed after $40$ iterations in Figure \ref{fig:flat_strip_pressure_inverse_Koiter}(e) compared to the one in Figure \ref{fig:flat_strip_pressure_FEM_Koiter}(a). Similarly, the objective function versus iterations and the reconstructed configurations are illustrated in Figure \ref{fig:flat_strip_pressure_inverse_projected} for the latter. By comparing Figure \ref{fig:flat_strip_pressure_inverse_projected}(e) to Figure \ref{fig:flat_strip_pressure_FEM_projected}(a), it can be seen that the specified target shape under the identified $v^{inverse}_{proj} = -0.1999$, $q^{inverse}_{proj} = 35.74 ~N/mm^2$ can be recovered nearly exactly. Overall, the inverse formulation correctly recovers the prescribed displacement $v_{right}$, the lateral pressure $q$ and the measured deformation due to buckling.

\section{Conclusions} \label{sec:conclusions}

In this article, we have developed a new method to combine inverse analysis with material and geometric nonlinearities to identify the unknown applied loads from given displacements and reconstruct the deformations of thin shell structures. NURBS-based finite elements, which offer geometrically exact discretization, are used to ensure $C^1$-continuity of the 3D surface. Both a Koiter material model and a projected shell material model with compressible Neo-Hookean formulation are considered. Gradient-based optimization algorithms with analytical and semi-analytical sensitivities were utilized to tackle the inverse problems for given experiment-like displacements at specified locations. Various numerical simulations are performed to capture either the stable or unstable (i.e. snap-through, snap-back and buckling) shape changes. The solution of the inverse analysis demonstrates that the proposed method is able to recover target shapes with high accuracy. We believe this research will open a new path for computer-aided manufacturing of shell structures.

\section{Acknowledgments} \label{sec:acknowledgments}

We gratefully acknowledge the support by ERC COMBAT project (project number 615132). We would like to thank Prof. Krister Svanberg from Royal Institute of Technology for providing the MMA code.

\appendix

\section{FE tangent matrices} \label{appendix:FE_tangent_matrices}

According to \cite{Duong:2017}, the linearization of $\delta \Pi^e_{\mathrm{int}}$ and $\delta \Pi^e_{\mathrm{ext}}$ is expressed as

\begin{equation} \label{eq:Gint_Gext_linearization_discretized}
\begin{aligned}
	\Delta \delta \Pi^e_{\mathrm{int}} &= \delta \bm{x}^T_e \left( \bm{k}^e_{\tau \tau} + \bm{k}^e_{\tau M} + \bm{k}^e_{M \tau} + \bm{k}^e_{M M} + \bm{k}^e_{\tau} + \bm{k}^e_{M} \right) \Delta \bm{x}_e, \\
	\Delta \delta \Pi^e_{\mathrm{ext}} &= \delta \bm{x}^T_e \left( \bm{k}^e_{\mathrm{ext} p} + \bm{k}^e_{\mathrm{ext} m} \right) \Delta \bm{x}_e,
\end{aligned}
\end{equation}

\noindent where the material stiffness matrices are defined by


\begin{equation} \label{eq:material_stiffness_matrices}
\begin{aligned}
	\bm{k}^e_{\tau \tau} &:= \int_{\Omega_0} c^{\alpha \beta \gamma \delta} \mathbf{N}^T_{, \alpha} \left( \bm{a}_{\beta} \otimes \bm{a}_{\gamma} \right) \mathbf{N}_{, \delta} dA, \\
	\bm{k}^e_{\tau M} &:= \int_{\Omega_0} d^{\alpha \beta \gamma \delta} \mathbf{N}^T_{, \alpha} \left( \bm{a}_{\beta} \otimes \bm{n} \right) \tilde{\mathbf{N}}_{; \gamma \delta} dA, \\
	\bm{k}^e_{M \tau} &:= \int_{\Omega_0} e^{\alpha \beta \gamma \delta} \tilde{\mathbf{N}}^T_{; \alpha \beta} \left( \bm{n} \otimes \bm{a}_{\gamma} \right) \mathbf{N}_{, \delta} dA, \\
	\bm{k}^e_{M M} &:= \int_{\Omega_0} f^{\alpha \beta \gamma \delta} \tilde{\mathbf{N}}^T_{; \alpha \beta} \left( \bm{n} \otimes \bm{n} \right) \tilde{\mathbf{N}}_{; \gamma \delta} dA
\end{aligned}	
\end{equation}

\noindent where $\mathbf{N}_{, \alpha} (\bm{\xi}) := [N_{1, \alpha} \bm{1}, N_{2, \alpha} \bm{1}, ..., N_{n, \alpha} \bm{1}]$ and

\begin{equation} \label{eq:covder_N}
	\tilde{\mathbf{N}}_{; \alpha \beta} := \mathbf{N}_{, \alpha \beta} - \Gamma^{\gamma}_{\alpha \beta} \mathbf{N}_{, \gamma},
\end{equation}

\noindent with $\mathbf{N}_{, \alpha \beta} (\bm{\xi}) := [N_{1, \alpha \beta} \bm{1}, N_{2, \alpha \beta} \bm{1}, ..., N_{n, \alpha \beta} \bm{1}]$. The geometric stiffness matrices are shown as


\begin{equation} \label{eq:geometric_stiffness_matrices}
	\bm{k}^e_M = \bm{k}^e_{M 1} + \bm{k}^e_{M 2} + \left( \bm{k}^e_{M 2} \right)^T,
\end{equation}

\noindent in which

\begin{equation} \label{eq:geometric_stiffness_terms}
\begin{aligned}
	\bm{k}^e_{M 1} &:= - \int_{\Omega_0} b_{\alpha \beta} M^{\alpha \beta}_0 a^{\gamma \delta} \mathbf{N}^T_{, \gamma} \left( \bm{n} \otimes \bm{n} \right) \mathbf{N}_{, \delta} ~dA, \\
	\bm{k}^e_{M 2} &:= \int_{\Omega_0} M^{\alpha \beta}_0 \mathbf{N}^T_{, \gamma} \left( \bm{n} \otimes \bm{a}^{\gamma} \right) \tilde{\mathbf{N}}_{; \alpha \beta} ~dA.
\end{aligned}
\end{equation}


\noindent The external moment and surface pressure components of the tangent matrix are given by \cite{Duong:2017,Sauer:2014}

\begin{equation} \label{eq:ext_moment_tangent_matrix}
\begin{aligned}
	\bm{k}_{\mathrm{ext} m} &= \int_{\partial_m \Omega^e} m_{\tau} \mathbf{N}^T_{, \alpha} \left( \nu^{\beta} \bm{n} \otimes \bm{a}^{\alpha} + \nu^{\alpha} \bm{a}^{\beta} \otimes \bm{n} \right) \mathbf{N}_{, \beta} ds - \int_{\partial_m \Omega^e} m_{\tau} \nu^{\alpha} \mathbf{N}^T_{, \alpha} \left( \bm{n} \otimes \bm{a}^{\xi} \right) \mathbf{N}_{, \xi} ~ds, \\
	\bm{k}_{\mathrm{ext} q} &= \int_{\Omega^e} q \mathbf{N}^T \left( \bm{n} \otimes \bm{a}^{\alpha} - \bm{a}^{\alpha} \otimes \bm{n} \right) \mathbf{N}_{, \alpha} ~da,
\end{aligned}
\end{equation}

\noindent where $\xi$ refers the convective coordinate of the curve $\partial_m \Omega^e$.

\end{document}